\documentclass[a4paper,fleqn,usenatbib]{mnras}

\usepackage{txfonts}

\usepackage[T1]{fontenc}
\usepackage{ae,aecompl}
\usepackage{graphics}

\usepackage{psfig}   
\usepackage{graphicx}
\usepackage{amssymb}
\usepackage{subfigure}

\title[SLR enrichment from winds and supernovae]{Short-lived radioisotope enrichment in star-forming regions from stellar winds and supernovae}

\author[R.~J.~Parker et al.]{Richard  J. Parker\thanks{E-mail: R.Parker@sheffield.ac.uk}\thanks{Royal Society Dorothy Hodgkin Fellow}$^1$, Tim Lichtenberg$^{2,3}$, Miti Patel$^{4,1}$, Cheyenne K. M. Polius$^{1}$ and \newauthor Matthew Ridsdill-Smith$^{5,1}$ \vspace*{0.1cm}\\
  $^1$Department of Physics and Astronomy, The University of Sheffield, Hicks Building, Hounsfield Road, Sheffield, S3 7RH, UK \\
  $^2$Atmospheric, Oceanic and Planetary Physics, Department of Physics, University of Oxford, Oxford, OX1 3PU, UK \\
  $^3$Kapteyn Astronomical Institute, University of Groningen, PO Box 800, 9700 AV Groningen, Netherlands \\
  $^4$School of Physics \& Astronomy, University of Leicester, University Road, Leicester, LE1 7RH, UK\\
$^5$National Astronomical Research Institute of Thailand, 260 Moo 4, T.\,\,Donkaew, A.\,\,Maerim, Chiangmai 50180, Thailand}

\begin{document}

\date{}
                             
\pagerange{\pageref{firstpage}--\pageref{lastpage}} \pubyear{2023}

\maketitle

\label{firstpage}

\begin{abstract}
The abundance of the short-lived radioisotopes $^{26}$Al and $^{60}$Fe in the early Solar system is usually explained by the Sun either forming from pre-enriched material, or the Sun's protosolar disc being polluted by a nearby supernova explosion from a massive star. Both hypotheses suffer from significant drawbacks: the former does not account for the dynamical evolution of star-forming regions, while in the latter the time for massive stars to explode as supernovae can be similar to, or even longer than, the lifetime of protoplanetary discs. In this paper, we extend the disc enrichment scenario to include the contribution of $^{26}$Al from the winds of massive stars before they explode as supernovae. We use $N$-body simulations and a post-processing analysis to calculate the amount of enrichment in each disc, and we vary the stellar density of the star-forming regions. We find that stellar winds contribute to disc enrichment to such an extent that the Solar system's $^{26}$Al/$^{60}$Fe ratio is reproduced in up to 50\,per cent of discs in dense ($\tilde{\rho} = 1000$\,M$_\odot$\,pc$^{-3}$) star-forming regions. When winds are a significant contributor to the SLR enrichment, we find that Solar system levels of enrichment can occur much earlier (before 2.5\,Myr) than when enrichment occurs from supernovae, which start to explode at later ages ($>$4\,Myr). We find that Solar system levels of enrichment all but disappear in low-density star-forming regions ($\tilde{\rho} \leq 10$\,M$_\odot$\,pc$^{-3}$), implying that the Solar system must have formed in a dense, populous star-forming region if $^{26}$Al and $^{60}$Fe were delivered directly to the protosolar disc from massive-star winds and supernovae.
  
\end{abstract}


\section{Introduction}

Placing our Solar system in the context of other planetary systems is one of the great challenges in astrophysics \citep{2020RvMP...92c0503Q}. Do planetary systems and the forming planets themselves retain signatures of their birth star-forming environments, and what are the implications for the climate diversity among extrasolar planets \citep{2021arXiv211204663W,Lichtenberg22} and the emergence of Earth-like planetary environments \citep{2021arXiv211204309M}?

Observations over the past two decades have determined that stars form in groups, often referred to as `clusters' if they are gravitationally bound/long-lived \citep{Kruijssen12b}, or `associations' if they are unbound/short-lived \citep{Wright22}. There is significant uncertainty and degeneracy when assessing which environments are either conducive or repellent to planet formation. This complication mainly arises because planets are observed to form contemporaneously with their host stars \citep{Haisch01,Brogan15,Andrews18,Alves20,SeguraCox20}, often whilst still contained within dense, obscured star-forming regions.

Indirect evidence that the Sun formed in a populous star-forming region is present in the oldest objects in the Solar system, the chondritic meteorites \citep{Lee76,Cameron77}, which contain the decay products of the short-lived radioisotopes (SLRs) $^{26}$Al and $^{60}$Fe. Whilst there are several astrophysical mechanisms for creating $^{26}$Al \citep{Lugaro18,Diehl21}, the most likely scenarios are that the Sun either formed from pre-enriched material \citep{Boss95,Boss19,Gaidos09,Gounelle12,Young14,Gounelle15,Kuffmeier16,Fujimoto18,Cote19,Forbes21} or its planet-forming disc was polluted by supernova and/or wind ejecta from massive stars \citep{Ouellette07,Ouellette10,Fatuzzo15,Parker14a,Lichtenberg16b,Nicholson17,PZ18,Fatuzzo22}.

The disc pollution scenario usually assumes that the disc is only enriched by supernovae explosions of the most massive star(s). Aside from issues around coupling hot ejecta to cold disc material \citep{Wijnen17}, even the most massive stars ($>40$\,M$_\odot$) do not explode as supernovae until after 4\,Myr \citep{Limongi06}. At these (relatively) later ages, the disc may have depleted \citep{Haisch01,Richert18}, been destroyed by encounters \citep{Vincke15} and/or photoevaporation \citep{Johnstone98,Scally01,Adams04,Nicholson19a,Winter18b,Parker21a}, or already formed planets \citep{Brogan15,Andrews18}. In the latter scenario, we might expect to see a much greater inhomogeneity in the injected $^{26}$Al than is observed. However, the massive stars that explode as supernovae also produce winds rich in $^{26}$Al during their main sequence phase \citep{Limongi06}, meaning that disc enrichment could occur much earlier than the timescales for supernovae explosions (i.e.\,\,$<<$4\,Myr).

In this paper, we determine the amount of enrichment of protoplanetary discs in $^{26}$Al and $^{60}$Fe from supernovae, which produce both SLRs, and from massive-star winds, which solely produce $^{26}$Al. From these we determine -- based on the amount of $^{26}$Al and $^{60}$Fe capture versus decay over time -- the internal radioactive heating in forming planetesimals that acts to differentiate and devolatilize them \citep{Lichtenberg19,Lichtenberg22}. We do this for star-forming regions with different initial densities, and contrast the results to the values measured in the Solar system.

Much of the recent literature has focused on $^{26}$Al and $^{60}$Fe enrichment in sequential \citep[e.g.][]{Gounelle12,Gounelle15} or extended star formation events \citep[e.g.][]{Young14,Fatuzzo22}. In this paper we incorporate the dynamical evolution of the star-forming regions, but focus on enrichment in individual star-forming regions, rather than the summed enrichment in multiple regions. \citet{Zwart19} focuses on reproducing the enrichment history of the Solar system (including wind enrichment from Wolf Rayet stars), and also examines the dynamical truncation history of the Solar system. Here, we provide more generalised simulations to infer the amount of $^{26}$Al and $^{60}$Fe enrichment from both Wolf Rayet winds and supernovae for different density star-forming regions. 

The paper is organised as follows. We describe our simulations, and our calculations of the capture of SLRs and consequent radiogenic heating in Section 2.  We present our results as cumulative distributions of the SLR ratios and heating in Section 3 and we provide a discussion in Section 4. We conclude in Section 5.

\section{Method}

In this Section we describe the set-up of the $N$-body simulations, how we calculate the enrichment of protoplanetary discs by both massive star winds and supernovae, before describing the method to derive radioactive heating values to determine whether planetary systems would experience Solar system-like internal heating of planetesimals.

\subsection{$N$-body simulations of star-forming regions}

Our star-forming regions all have $N_\star = 1500$ stars, which is roughly the expectation value ($\sim$1000\,M$_\odot$) for the mass of a star-forming region, based on the observations of Galactic star clusters and associations \citep{Lada03}. Additionally, for our chosen $N_\star$, all versions of the initial mass function (IMF) in the literature will produce some massive ($>17$\,M$_\odot$) stars that could contribute $^{26}$Al and $^{56}$Fe through both supernovae and/or winds.

We select stellar masses from the \citet{Maschberger13} IMF, which has a probability distribution function of the form
\begin{equation}
p(m) \propto \left(\frac{m}{\mu}\right)^{-\alpha}\left(1 + \left(\frac{m}{\mu}\right)^{1 - \alpha}\right)^{-\beta}.
\label{maschberger_imf}
\end{equation}
In Eqn.~\ref{maschberger_imf}  $\mu = 0.2$\,M$_\odot$ is the scale parameter, or `peak' of the IMF \citep{Bastian10,Maschberger13}, $\alpha = 2.3$ is the \citet{Salpeter55} power-law exponent for higher mass stars, and $\beta = 1.4$ describes the slope of the IMF for low-mass objects. We randomly sample this distribution in the mass range 0.1 -- 50\,M$_\odot$, such that brown dwarfs are not included in the simulations, and $^{26}$Al- and $^{60}$Fe-producing stars are present in each star-forming region.

Observations \citep{Larson81,Cartwright04,Peretto06,Sanchez09,Hacar18,Buckner19} and simulations \citep{Schmeja06,Bate09,Girichidis11,Dale15b} show that star-forming regions form with both spatial and kinematic substructure. This is likely to be an imprint of the turbulent conditions of the Giant Molecular Clouds in which stars form, although the correlation between the properties of the gas, and the stars that subsequently form, are unlikely to be linked in a trivial manner \citep{Parker15c}.

To mimic the spatial and kinematic substructure in the early phases of star formation, we set up our $N$-body simulations using a box-fractal generator \citep{Goodwin04a}, which uses a given fractal dimension, $D$, as an input to set both  the degree of spatial structure, and the degree of kinematic structure. For full details of the method, we refer the interested reader to \citet{Goodwin04a,Allison10,Lomax11,DaffernPowell20}, but we briefly summarise the method below.

The fractals are set up by placing a `parent' particle at the centre of a cube of side $N_{\rm div}$, which spawns $N_{\rm div}^3$ subcubes. Each of these subcubes contains a `child' particle at its centre, and the fractal is constructed by determining how many successive generations of `children' are produced. The likelihood of each successive generation producing their own children is given by $N_{\rm div}^{D-3}$, where $D$ is the fractal dimension.

The actual star particles are placed at the locations of the final generation of children. Fewer generations are produced when the fractal dimension is lower, leading to a less uniform appearance and hence more substructure.  Star-forming regions with a low fractal dimension (e.g.\,\,$D = 1.6$) have a high degree of substructure, whereas regions with higher values (e.g.\,\,$D = 2.0, 2.6$) have less structure and regions with $D = 3.0$ are approximately uniform.

The velocities of the parent particles are drawn from a Gaussian distribution of mean zero, and the velocities of the child particles inherit this velocity plus a small random component which scales as $1/N_{\rm div}^{g}$, where $g$ is the number of generations in the fractal and therefore the random component becomes progressively smaller with each generation. This means that physically close particles have very similar velocities, but more distant particles can have very uncorrelated velocites, similar to the observations within GMCs \citep{Larson81}.

We adopt $D = 2.0$ for all our simulations, which gives a moderate amount of spatial and kinematic substructure. We scale the velocities to a virial ratio $\alpha_{\rm vir} = T/|\Omega|$, where $T$ and $|\Omega|$ are the total kinetic and potential energies, respectively. The velocities of young stars are often observed to be subvirial along filaments \citep{Andre14,Foster15}, so we adopt a subvirial ratio ($\alpha_{\rm vir} = 0.3$) in all of our simulations.

We then scale the physical size of the fractals to produce the required stellar density. Observations of star-forming regions present a range of densities \citep{Lada03,Bressert10}, most of which are more than a factor of ten higher than the stellar density in the disc of the Milky Way \citep[$\sim$0.1\,M$_\odot$\,pc$^{-3}$,][]{Korchagin03}, and some are several 100s\,M$_\odot$\,pc$^{-3}$ \citep{King12a}. However, dynamical models postulate that the initial densities of star-forming regions may be even higher, perhaps 1000s\,M$_\odot$\,pc$^{-3}$ \citep{Marks12,Farias20,Schoettler20,Schoettler22}.

The observed present-day densities in star-forming regions are likely to be lower limits to the initial densities \citep{Parker14e}, so we set up simulations with the full range of possible initial densities for star-forming regions by adjusting the radii of the box fractals accordingly. A summary of the simulations is given in Table~\ref{simulations}.

We run ten versions of each simulation in order to gauge the amount of stochasticity in the results. We evolve the star-forming regions for 10\,Myr using the \texttt{kira} integrator in the \texttt{Starlab} package \citep{Zwart99,Zwart01}.  We do not implement mass-loss from the massive stars due to stellar evolution (although a stellar evolution module, \texttt{SeBa} is available in \texttt{Starlab}), but instead use the models from \citet{Limongi06} to determine when a massive star leaves the main sequence and explodes as a supernova.

\begin{table}
  \caption[bf]{A summary of the different initial conditions of our simulated star-forming regions. The columns show the number of stars, $N_{\rm stars}$, the initial radius of the star-forming region, $r_F$, and the initial median local stellar density, $\tilde{\rho}$. }
  \begin{center}
    \begin{tabular}{|c|c|c|}
      \hline

$N_{\rm stars}$ & $r_F$ & $\tilde{\rho}$ \\% & $r_{\disc}$  \\
\hline
1500 & 1\,pc & $1000$\,M$_\odot$\,pc$^{-3}$ \\%& 100\,au \\
1500 & 2.5\,pc & $100$\,M$_\odot$\,pc$^{-3}$ \\%& 10\,au \\
1500 & 5.5\,pc & $10$\,M$_\odot$\,pc$^{-3}$ \\%& 100\,au \\
1500 & 20\,pc & $0.1$\,M$_\odot$\,pc$^{-3}$ \\%& \citet{Eisner18} \\
%1500 & 5.5\,pc & $10$\,M$_\odot$\,pc$^{-3}$ & 100\,au  \\
%\hline
%150 & 0.75\,pc & $100$\,M$_\odot$\,pc$^{-3}$ & 100\,au \\ 

      \hline
    \end{tabular}
  \end{center}
  \label{simulations}
\end{table}

\subsection{Enrichment from massive stars}

We track the position of each $>$5\,M$_\odot$ star and the relative positions of all low-mass ($\leq$3\,M$_\odot$) stars. All low-mass stars are assigned a disc with radius $r_{\rm disc} = 100$\,au. In two sets of simulations we randomly remove discs from the enrichment calculation if a random number $\mathcal{R}$ between 0 and 1 obeys the following relation
\begin{equation}
  \mathcal{R} > e^{-0.0278t},
  \label{disc_deplete}
  \end{equation}
where $t$ is the time in Myr. This relation ensures that the discs deplete on an exponential timescale commensurate with the \citet{Richert18} fit to disc fractions in star-forming regions, which in turn is a refinement of the \citet{Haisch01} results.  

The mass of the disc is fixed, i.e.\,\,it does not deplete (apart from when the disc is removed altogether in some simulations, as described above), and is proportional to the host star mass,
\begin{equation}
m_{\rm disc} = 0.1M_\star.
\end{equation}
We assume the usual gas-to-dust ratio of 100:1, so the total dust mass, $m_{\rm dust}$, in the disc is
\begin{equation}
  m_{\rm dust} = 0.01m_{\rm disc}.
\end{equation}
The amount of $^{26}$Al is expressed in terms of the stable version of the element, $^{27}$Al, and we determine the mass of $^{27}$Al in the disc, $m_{^{27}{\rm Al}}$, from the fraction of Al in chondrites, $f_{\rm Al, CI} = 8500 \times 10^{-6}$ \citep{Lodders03}
\begin{equation}
m_{^{27}{\rm Al}} = 8500\times10^{-6} m_{\rm dust}.
\end{equation}
Similarly, the amount of $^{60}$Fe is expressed in terms of its stable version, $^{56}$Fe, and we determine the mass of  $^{56}$Fe in the disc, $m_{^{56}{\rm Fe}}$, from the fraction of Fe in chondrites, $f_{\rm Fe, CI} = 1828 \times 10^{-4}$ \citep{Lodders03}
\begin{equation}
m_{^{56}{\rm Fe}} = 1828\times10^{-4} m_{\rm dust}.
\end{equation}
    
In order to calculate both the location of, and the $^{26}$Al and $^{60}$Fe yields from, supernovae in the star-forming regions, for each massive star we perform a linear interpolation using its mass to estimate the time at which it explodes as a supernova, and its SLR yields, based on the models in \citet{Limongi06}.

For each $<$3\,M$_\odot$ star at each snapshot in time, we calculate the geometric cross section of the disc for capturing material from a supernova thus
\begin{equation}
\eta_{\rm SN} = \frac{\pi r_{\rm disc}^2}{4\pi d^2}{\rm cos}\theta,
  \end{equation}
where $d$ is the distance to the massive star(s) at the instant of supernova \citep[from the models of][]{Limongi06} and $\theta$ is the inclination of the disc. Following \citet{Lichtenberg16b}, for each star we adopt $\theta = 60^\circ$ as the likely average inclination between the disc and the ejecta for a random distribution.

Similarly, to calculate the SLR yield from the massive star winds, we use the main sequence lifetimes from \citet{Limongi06} as well as the total yield from the winds, to determine the yield per Myr. In Appendix~\ref{appendix} we calculate the yields using more recent models \citep{Limongi18}, but find no major differences compared with the models from \citet{Limongi06}, especially when realistic estimates of stellar rotation are included in the calculations \citep{deMink13}.

The cross section for capture of wind material is calculated as the volume of material swept out by a low-mass star as it traverses a distance $\Delta r_\star$ through a wind bubble of radius $r_{\rm bub}$
\begin{equation}
\eta_{\rm wind} = \frac{3}{4}\frac{\pi r_{\rm disc}^2\Delta r_\star}{\pi r_{\rm bub}^3}.
\end{equation}
We implement two different regimes for the density of the wind bubbles. First, we assume a very compact bubble around each massive star with a radius $r_{\rm bub} = 0.1$\,pc. Secondly, we assume the bubble(s) disperse rapidly, and the bubble has a radius $r_{\rm bub} = 2r_{1/2}$, where $r_{1/2}$ is the half-mass radius of the star-forming region (i.e.\,\,the radius within which half the total stellar mass in the region is enclosed). Previous work \citep[e.g.][]{Parker12a,Schoettler19} has shown this to be a reasonable estimate of the extent of these simulated star-forming regions without including distant outliers and/or ejected stars.

We use the total mass of $^{26}$Al and $^{60}$Fe captured/swept up by each low mass star and divide this by the mass in stable isotopes, $m_{^{27}{\rm Al}}$ and  $m_{^{56}{\rm Fe}}$, to determine the $^{26}$Al and $^{60}$Fe ratios:
\begin{equation}
Z_{\rm Al} = \frac{m_{^{26}{\rm Al}}}{m_{^{27}{\rm Al}}},
\end{equation}
and
\begin{equation}
Z_{\rm Fe} = \frac{m_{^{60}{\rm Fe}}}{m_{^{56}{\rm Fe}}}.
\end{equation}
For each low-mass star, we calculate $Z_{\rm Al}$ three times; for supernovae only, for supernovae and local winds ($r_{\rm bub} = 0.1$\,pc), and then finally for supernovae and dispersed winds  ($r_{\rm bub} = 2r_{1/2}$). The contribution of $^{60}$Fe comes from supernovae only.

We then compare each of the calculated $Z_{\rm Al}$, and the calculated $Z_{\rm Fe}$ to the respective measured Solar system values, $Z_{\rm Al, SS} = 5.85 \times 10^{-5}$ \citep{Thrane06} and $Z_{\rm Fe, SS} = 1.15 \times 10^{-8}$ \citep{Tang12} or $Z_{\rm Fe, SS} = 1 \times 10^{-6}$ \citep{Mishra16}. Different literature sources are in reasonable agreement on the measurement of $Z_{\rm Al}$ \citep[see also][who find a slightly lower value, around $Z_{\rm Al, SS} \sim 5 \times 10^{-5}$]{Jacobsen08,Kita13}, but the measurement of $^{60}$Fe is far more controversial \citep[see][for recently quoted values toward the lower end of the range]{Trappitsch18,Kodolanyi22}, and so we will compare our results to the full range of values quoted in the literature.

\subsection{Internal radioactive heating in planetesimals}

We then use the $Z_{\rm Al}$ and $Z_{\rm Fe}$ ratios to calculate the potential radioactive heating in forming planetesimals \citep{Moskovitz11,Lichtenberg16b} due to these SLRs. We calculate the radiogenic heating for each star, $Q(t)$ at each snapshot in time $t$, taking into account the radioactive decay of both $^{26}$Al and $^{60}$Fe from the current and previous snapshots using
\begin{equation}
  Q(t) = f_{\rm Al,CI}Z_{\rm Al}\frac{E_{\rm Al}}{\tau_{\rm Al}}e^{-t/\tau_{\rm Al}} + f_{\rm Fe,CI}Z_{\rm Fe}\frac{E_{\rm Fe}}{\tau_{\rm Fe}}e^{-t/\tau_{\rm Fe}},
  \label{ss_heating}
  \end{equation}
where $f_{\rm Al,CI}$ is the fraction of Al in chondrites \citep{Lodders03}, as defined earlier, $E_{\rm Al} = 3.12$\,MeV is the decay energy of $^{26}$Al, $\tau_{\rm Al} = 0.717$\,Myr is the radioactive half-live of $^{26}$Al  \citep{CastilloRogez09}. Similarly, $f_{\rm Fe,CI}$ is the fraction of Fe in chondrites \citep{Lodders03}, $E_{\rm Fe} = 2.712$\,MeV is the decay energy of $^{60}$Fe \citep{CastilloRogez09} and  $\tau_{\rm Fe} = 2.6$\,Myr is the half-life of $^{60}$Fe \citep{Wallner15}. The initial Solar system heating value is calculated from these values to be $Q_{\rm SS} = 3.4 \times 10^{-7}$\,W\,kg$^{-1}$.

\section{Results}

In this Section we present the short-lived isotope ratios, $Z_{\rm Al} = {\rm ^{26}Al}/{\rm^{27}\rm Al}$ and $Z_{\rm Al} = {\rm ^{60}Fe}/{\rm ^{56}Fe}$ for star-forming regions with different initial densities. For each density regime, we also use the SLR ratios to calculate the long-term internal heating of the planetesimals that form in the discs, $Q(t)$.

\subsection{High stellar density}

We first calculate the SLR ratios and internal heating of planetesimals for discs in inititally very dense ($\tilde{\rho} \sim 1000$M$_\odot$\,pc$^{-3}$) star-forming regions. The details of the evolution of these regions is documented elsewhere \citep{Allison10,Parker14b,Lichtenberg16b}, but we briefly outline it again here.

The star-forming regions are initially out of equilibrium. First, the spatial and kinematic substructure is erased before the star-forming regions undergo violent relaxation \citep{LyndenBell67,Spitzer69} in which the stars fall into the potential well of the star-forming region, leading to the formation of a smooth, centrally concentrated star cluster.

This violent relaxation facilitates mass segregation, whereby the most massive stars become more spatially concentrated than the average mass stars in the cluster \citep{McMillan07,Allison10,Parker14b}. In some cases the massive star(s) are ejected \citep{Schoettler19}, but are replaced in the central region by the next most massive star(s) \citep{Parker16c}. In these dense regions we expect low-mass stars to be in close proximity to massive stars \citep{Parker21a}, thereby maximising the likelihood of enrichment occurring \citep{Parker14a}.

\begin{figure*}
  \begin{center}
\setlength{\subfigcapskip}{10pt}
\hspace*{-1.5cm}\subfigure[Supernovae only -- 10\,Myr]{\label{high_density-Al-a}\rotatebox{270}{\includegraphics[scale=0.27]{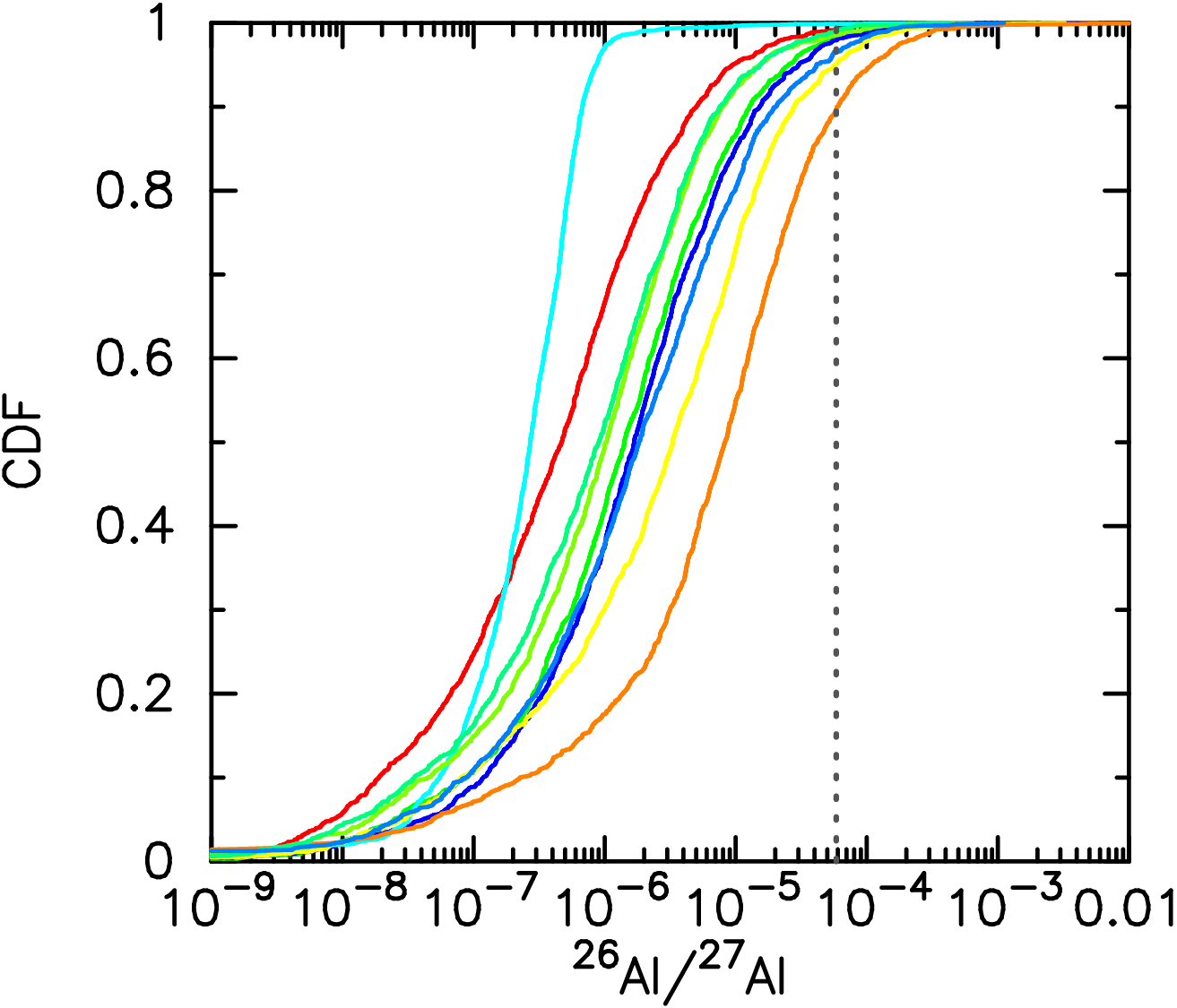}}}
\hspace*{0.3cm} 
\subfigure[Supernovae and local wind bubbles -- 10\,Myr]{\label{high_density-Al-b}\rotatebox{270}{\includegraphics[scale=0.27]{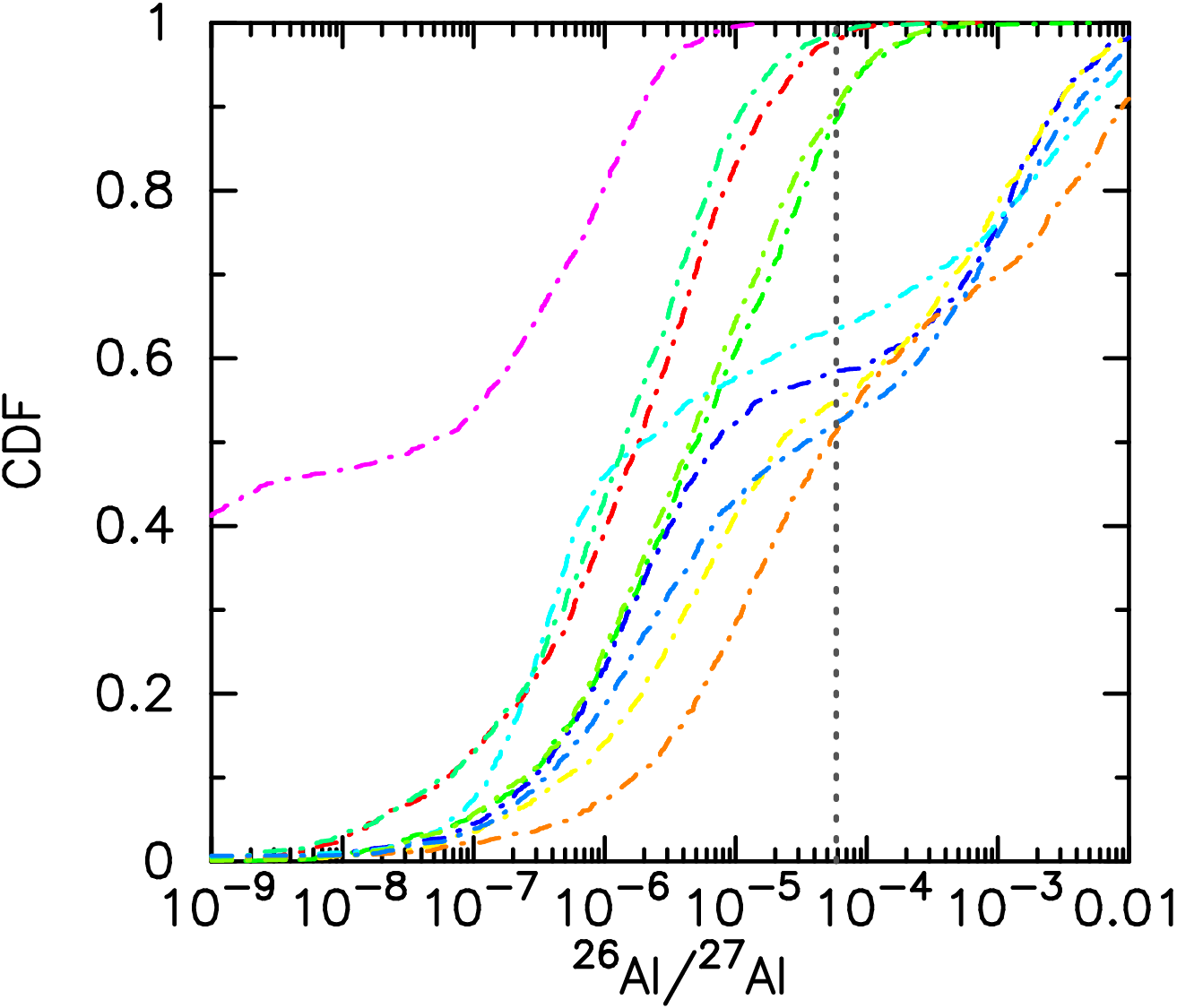}}}
\hspace*{0.3cm} 
\subfigure[Supernovae and global wind bubbles -- 10\,Myr]{\label{high_density-Al-c}\rotatebox{270}{\includegraphics[scale=0.27]{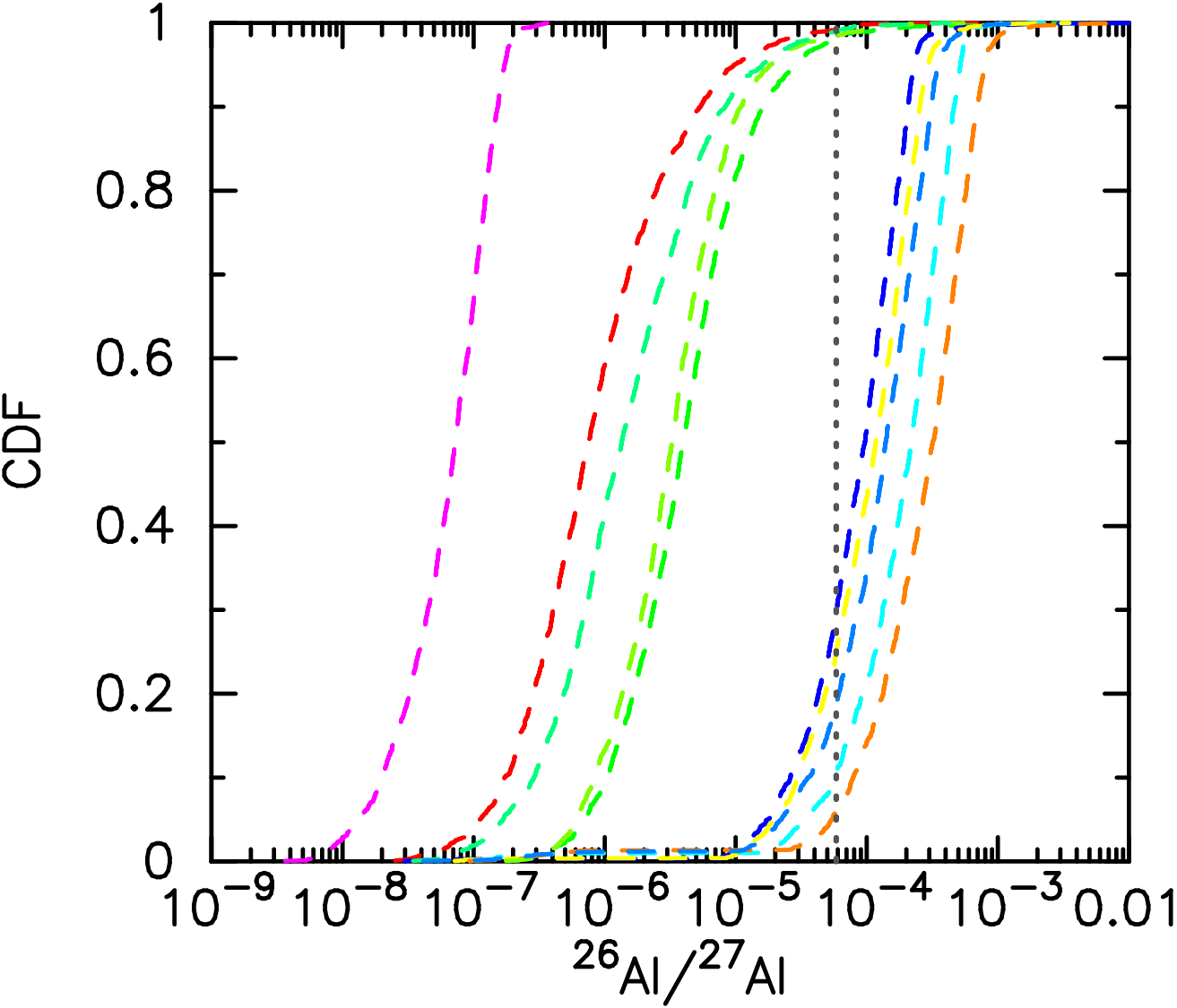}}} 
\hspace*{-1.5cm}\subfigure[All three line sets -- 1\,Myr]{\label{high_density-Al-d}\rotatebox{270}{\includegraphics[scale=0.27]{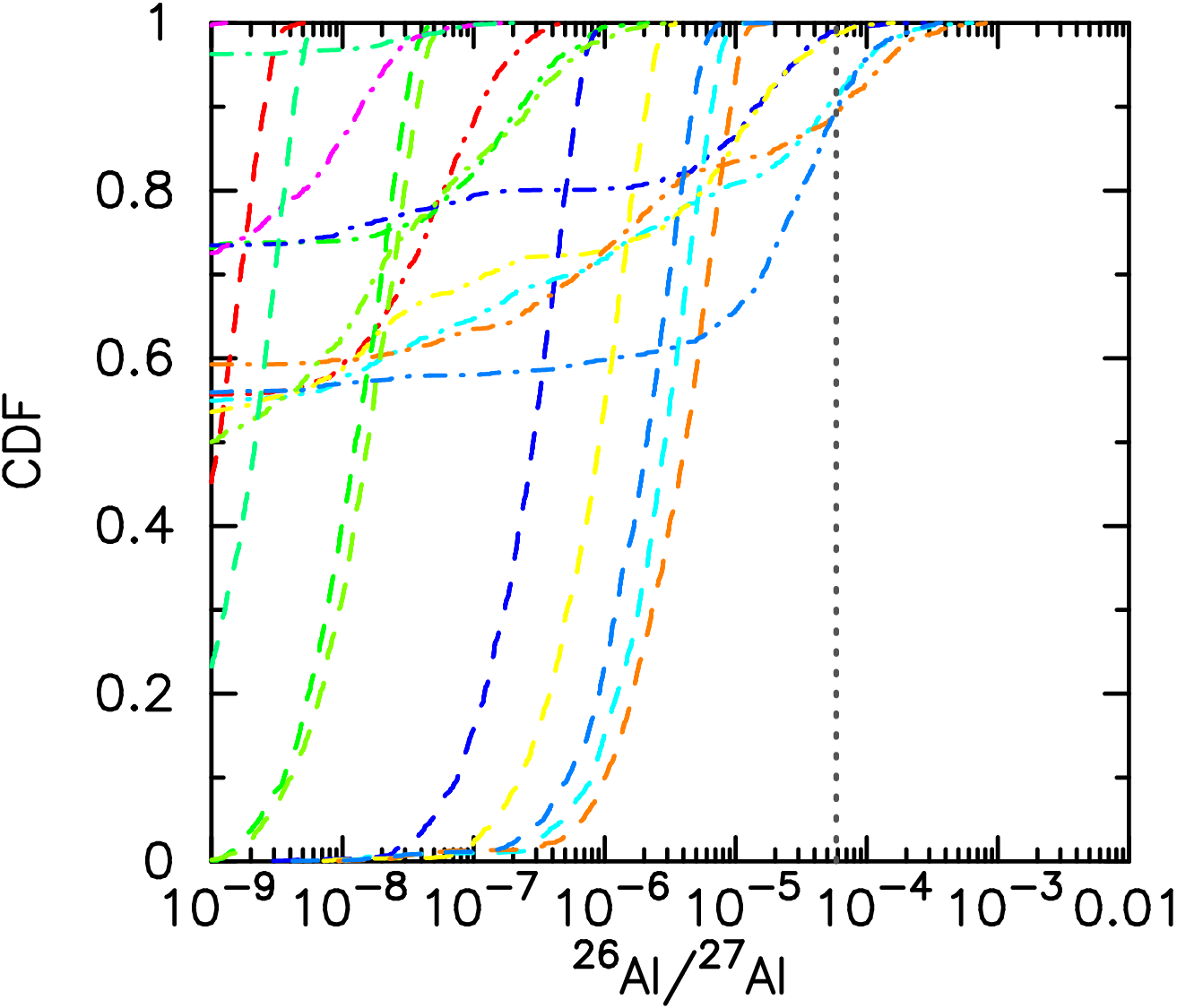}}}
\hspace*{0.3cm} 
\subfigure[All three line sets -- 2.5\,Myr]{\label{high_density-Al-e}\rotatebox{270}{\includegraphics[scale=0.27]{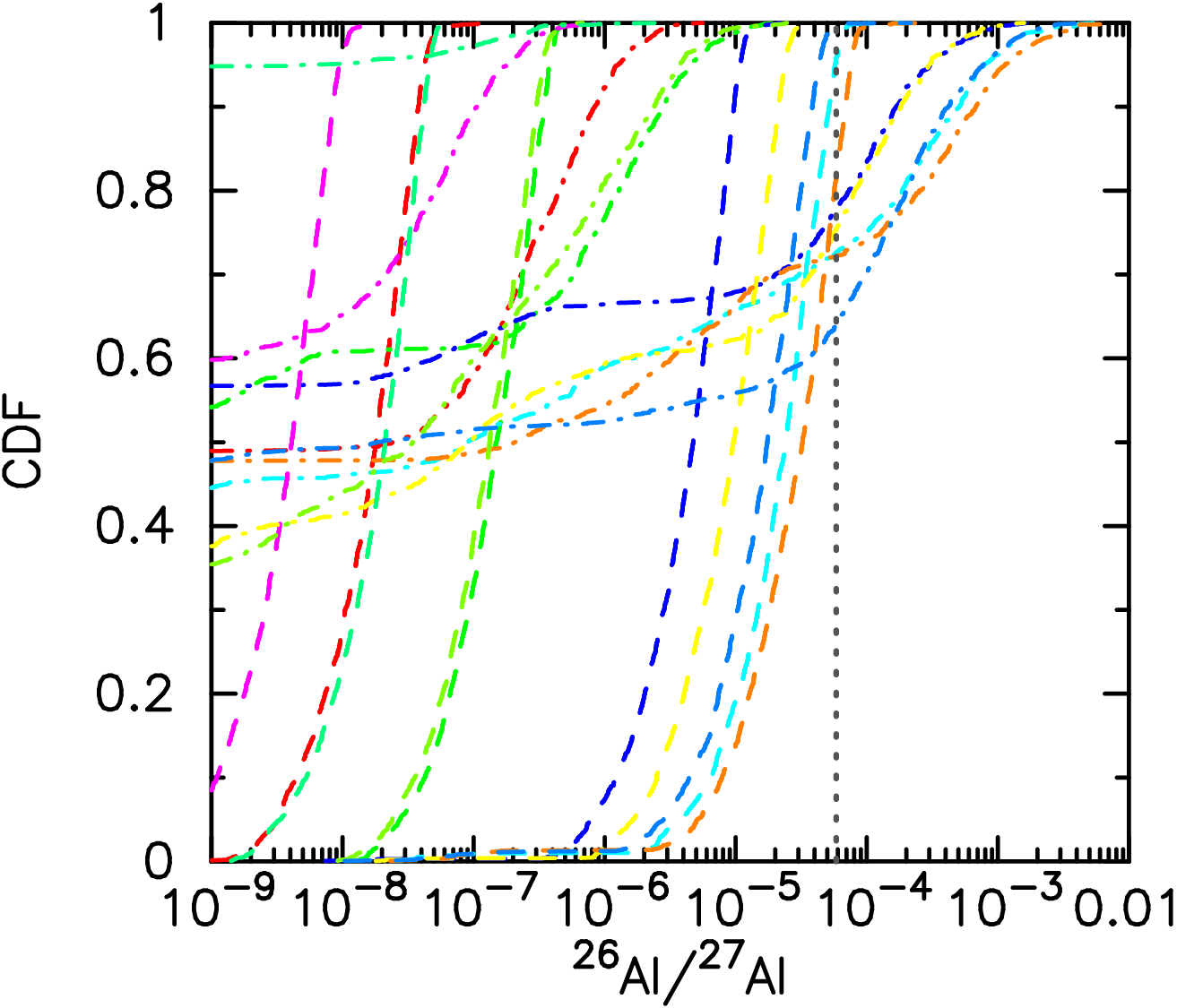}}}
\hspace*{0.3cm} 
\subfigure[All three line sets -- 5\,Myr]{\label{high_density-Al-f}\rotatebox{270}{\includegraphics[scale=0.27]{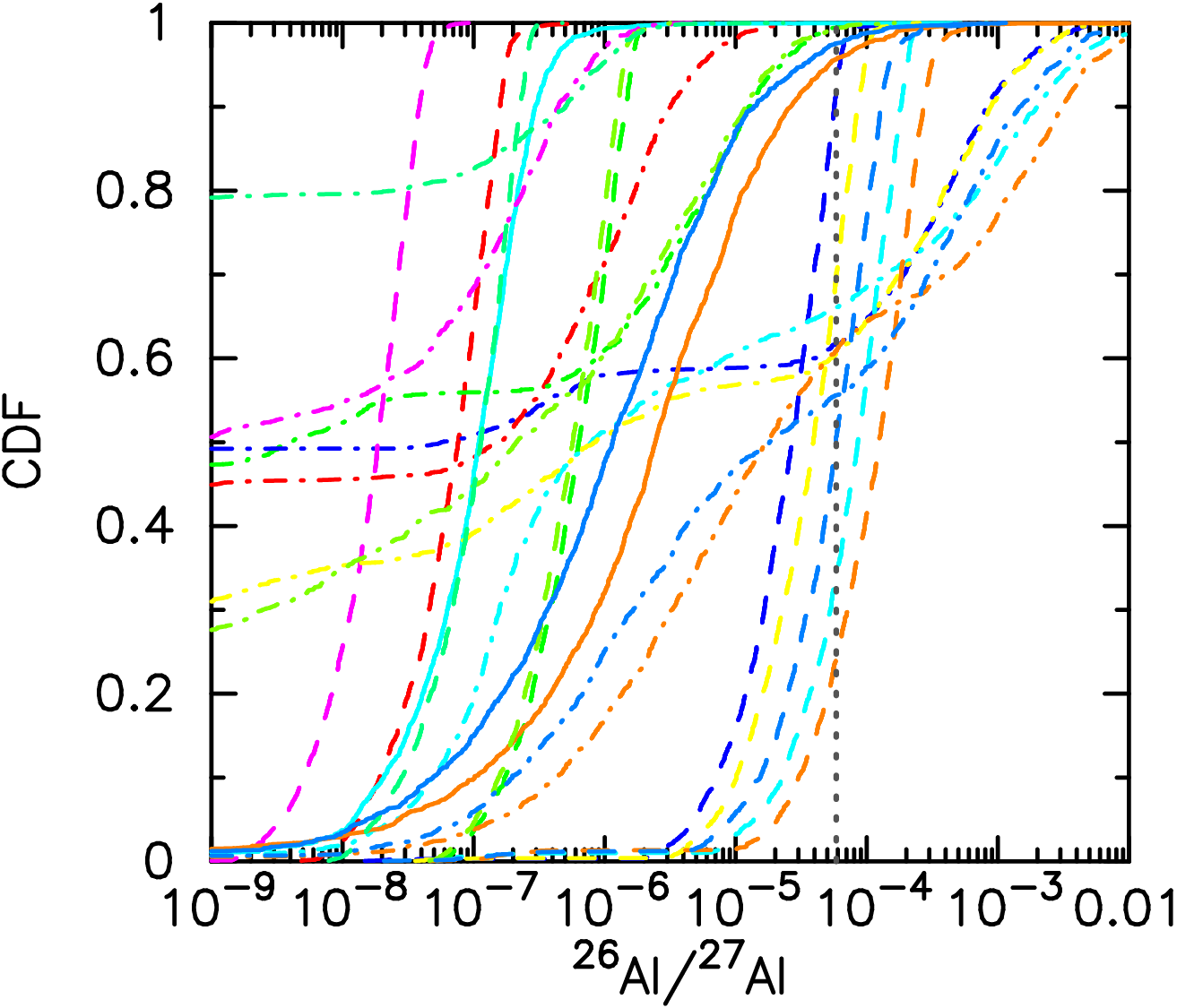}}} 
\caption[bf]{Cumulative Distribution  Functions (CDFs) of the SLR ratios $Z_{\rm Al}$ in a star-forming region with an initially high stellar density ($\tilde{\rho} \sim 1000$M$_\odot$\,pc$^{-3}$). The initial disc radii are all $r_{\rm disc} = 100$\,au, and the discs do not evolve with time.  Solid lines are ratios calculated from supernovae only. Dashed lines are global, or dispersed wind bubbles ($r_{\rm bub} = 2r_{1/2}$) plus any supernovae, dot-dashed lines are local wind bubbles ($r_{\rm bub} = 0.1$\,pc)  plus any supernovae. We show $^{26}$Al/$^{27}$Al ratios calculated at 10\,Myr (top row), then 1, 2.5 \& 5\,Myr (bottom row). The vertical dotted lines indicate the measured Solar system value, $Z_{\rm Al,SS} = 5.85 \times 10^{-5}$ \citep{Thrane06}.}
\label{high_density-Al}
  \end{center}
\end{figure*}

In Fig.~\ref{high_density-Al} we show the enrichment calculations for these high density regions, assuming that all the protoplanetary discs have a fixed radius of 100\,au. In the top row we show the $^{26}$Al abundance ratios, $Z_{\rm Al}$, calculated after 10\,Myr for supernovae only (panel a), supernovae and dispersed winds (i.e.\,where the winds fill the volume of the clusters, panel b) and supernovae and localised winds (i.e.\,where the wind material remains in a small [$r_{\rm bub} = 0.1$\,pc] volume around the massive stars, panel c). Each individual line is one simulation (there are ten versions of the same simulation, identical apart from the random number seed used to set up the initial conditions).

In the bottom row we show all three different $Z_{\rm Al}$ in each panel, but calculated at 1, 2.5 and 5\,Myr (panels d, e, and f, respectively). At 1 and 2.5\,Myr, no supernovae have occurred yet, and only the most massive stars ($>$40\,M$_\odot$) have exploded after 5\,Myr, so the winds dominate the enrichment budget.

Our first notable result is that the amount of $^{26}$Al enrichment from supernovae alone (Fig.~\ref{high_density-Al-a}) is only at Solar system levels \citep[or higher, i.e.\,\,$\geq 5.85 \times 10^{-5}$][]{Thrane06} for a few stars (less than 10\,per cent) in the majority of simulations. However, when discs are also enriched by wind material, the fraction of stars with Solar system levels of enrichment is more than 50\,per cent in some regions (Figs.~\ref{high_density-Al-b}~and~\ref{high_density-Al-c}). In five simulations (the three blue lines, the yellow line, the brown line in Fig.~\ref{high_density-Al-c}) the dispersed winds provide a level of enrichment far in excess of Solar system levels. For the same simulations, more localised winds also provide significant enrichment, with higher $Z_{\rm Al}$ ratios possible (Fig.~\ref{high_density-Al-b}).

Secondly, the $Z_{\rm Al}$ ratios at early ages (1\,Myr, panel d) can reach Solar system values for the localised wind bubbles, and after 2.5 (panel e) and 5\,Myr (panel f) the enrichment ratios increase further (recall that this is before most supernovae have ocurred).

\begin{figure*}
  \begin{center}
\setlength{\subfigcapskip}{10pt}
\subfigure[Supernovae -- 5\,Myr]{\label{high_density-Fe-a}\rotatebox{270}{\includegraphics[scale=0.27]{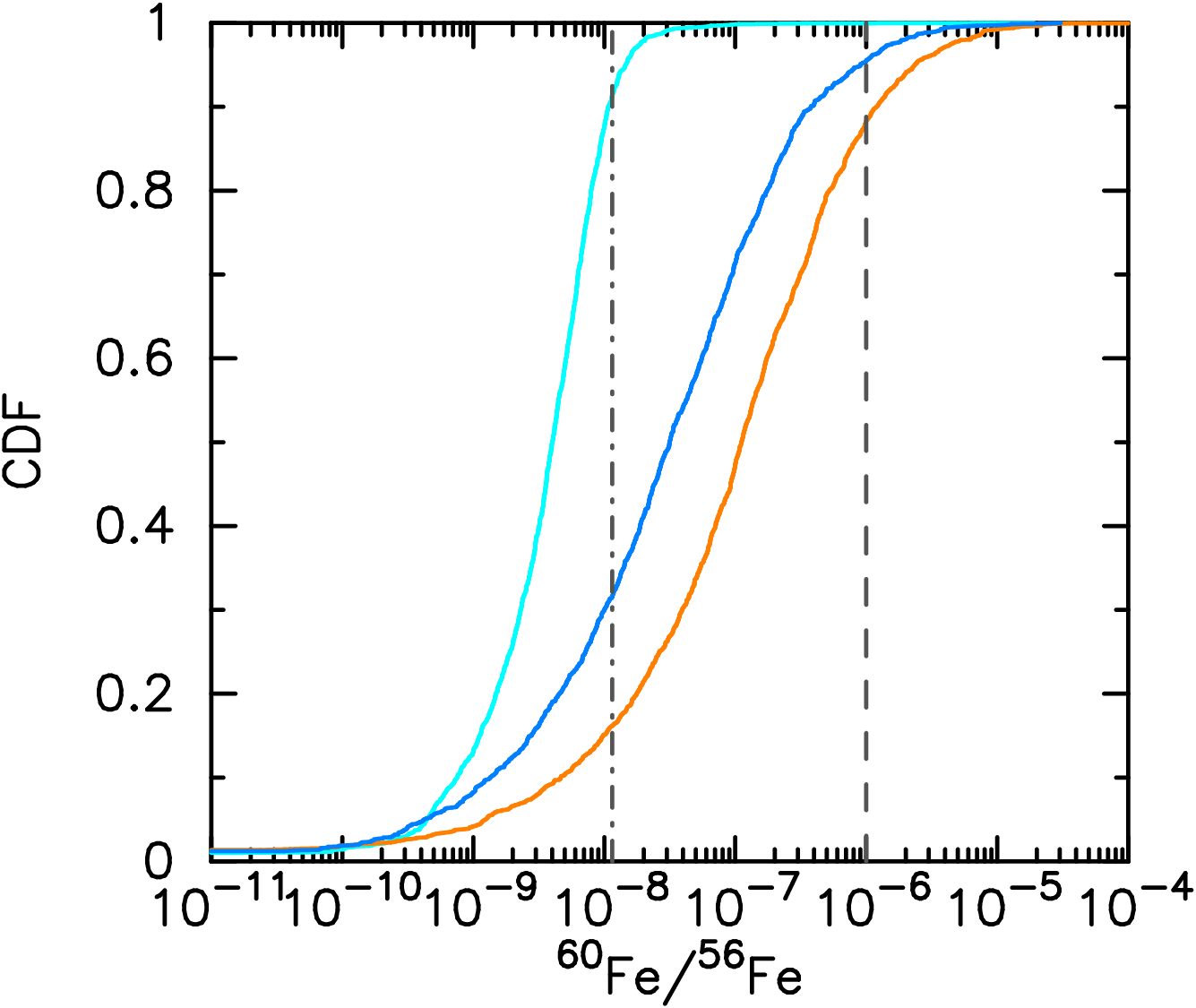}}}
\hspace*{0.3cm} 
\subfigure[Supernovae -- 10\,Myr]{\label{high_density-Fe-b}\rotatebox{270}{\includegraphics[scale=0.27]{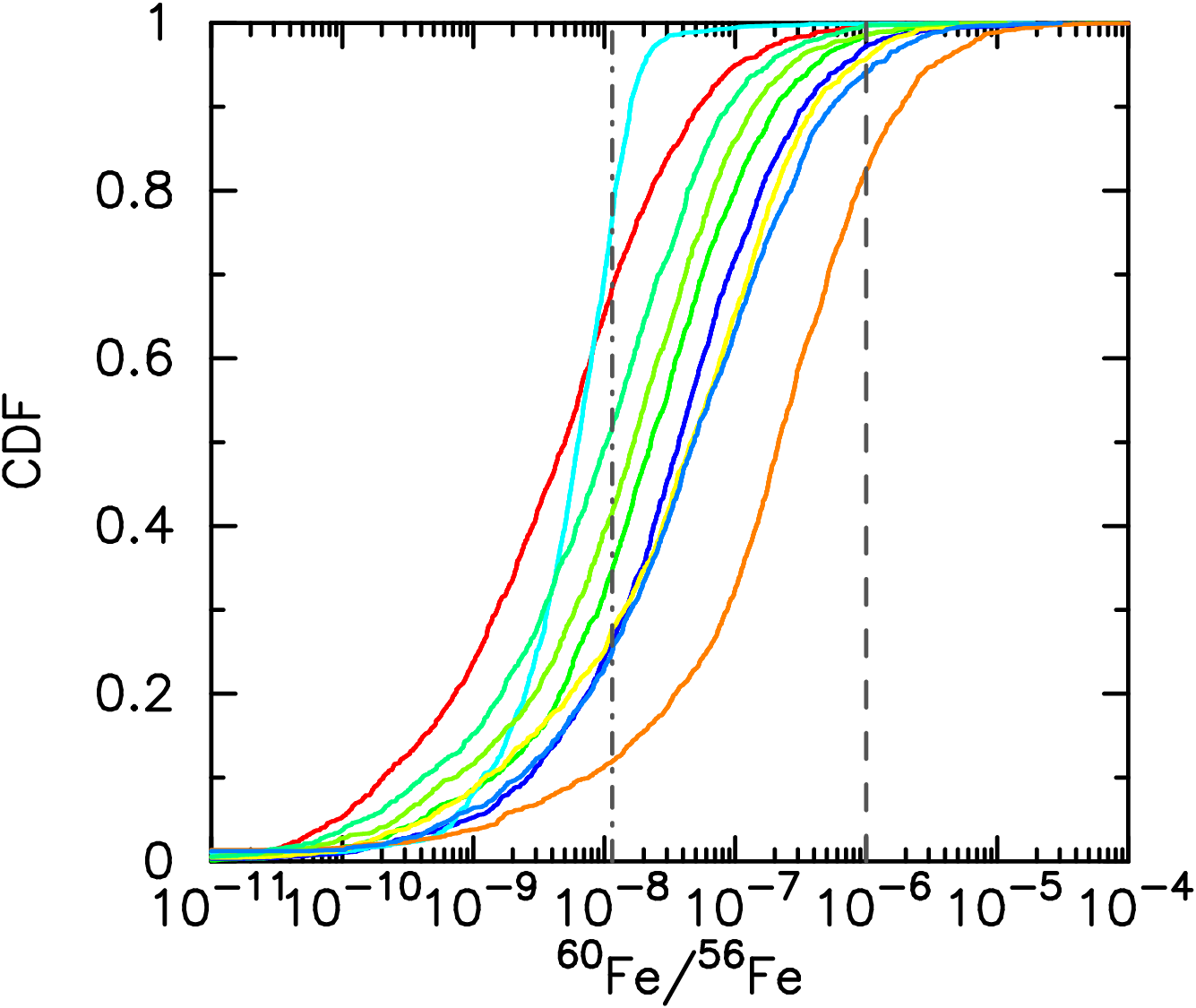}}} 
\caption[bf]{Cumulative Distribution  Functions (CDFs) of the SLR ratios $Z_{\rm Fe}$ in a star-forming region with an initially high stellar density ($\tilde{\rho} \sim 1000$M$_\odot$\,pc$^{-3}$). The initial disc radii are all $r_{\rm disc} = 100$\,au, and the discs do not evolve with time.  We show the $^{60}$Fe/$^{56}$Fe ratios (which only come from supernovae ejecta) calculated at 5 and 10\,Myr. The vertical dot-dashed line indicates the measured Solar system value, $Z_{\rm Fe,SS} = 1.15 \times 10^{-8}$ \citep{Tang12} and the vertical dashed line indicates the alternative measurement, $Z_{\rm Fe,SS} = 1 \times 10^{-6}$ \citep{Mishra16}.}
\label{high_density-Fe}
  \end{center}
\end{figure*}

In Fig.~\ref{high_density-Fe} we show the $^{60}$Fe abundance ratios at 5\,Myr (panel (a)) and after 10\,Myr (panel b). Again, after 5\,Myr few stars have exploded as supernovae, and in fact this only occurs in three out of ten simulations (Fig~\ref{high_density-Fe-a}). However, after 10\,Myr (Fig~\ref{high_density-Fe-b}) all simulations have stars that produce $^{60}$Fe on exploding as supernovae, and the values lie within the range of measured $Z_{\rm Fe}$ ratios for the Solar system \citep{Tang12,Mishra16}.

\begin{figure*}
  \begin{center}
\setlength{\subfigcapskip}{10pt}
\hspace*{-1.5cm}\subfigure[Supernovae only]{\label{high_density-Qheat-a}\rotatebox{270}{\includegraphics[scale=0.25]{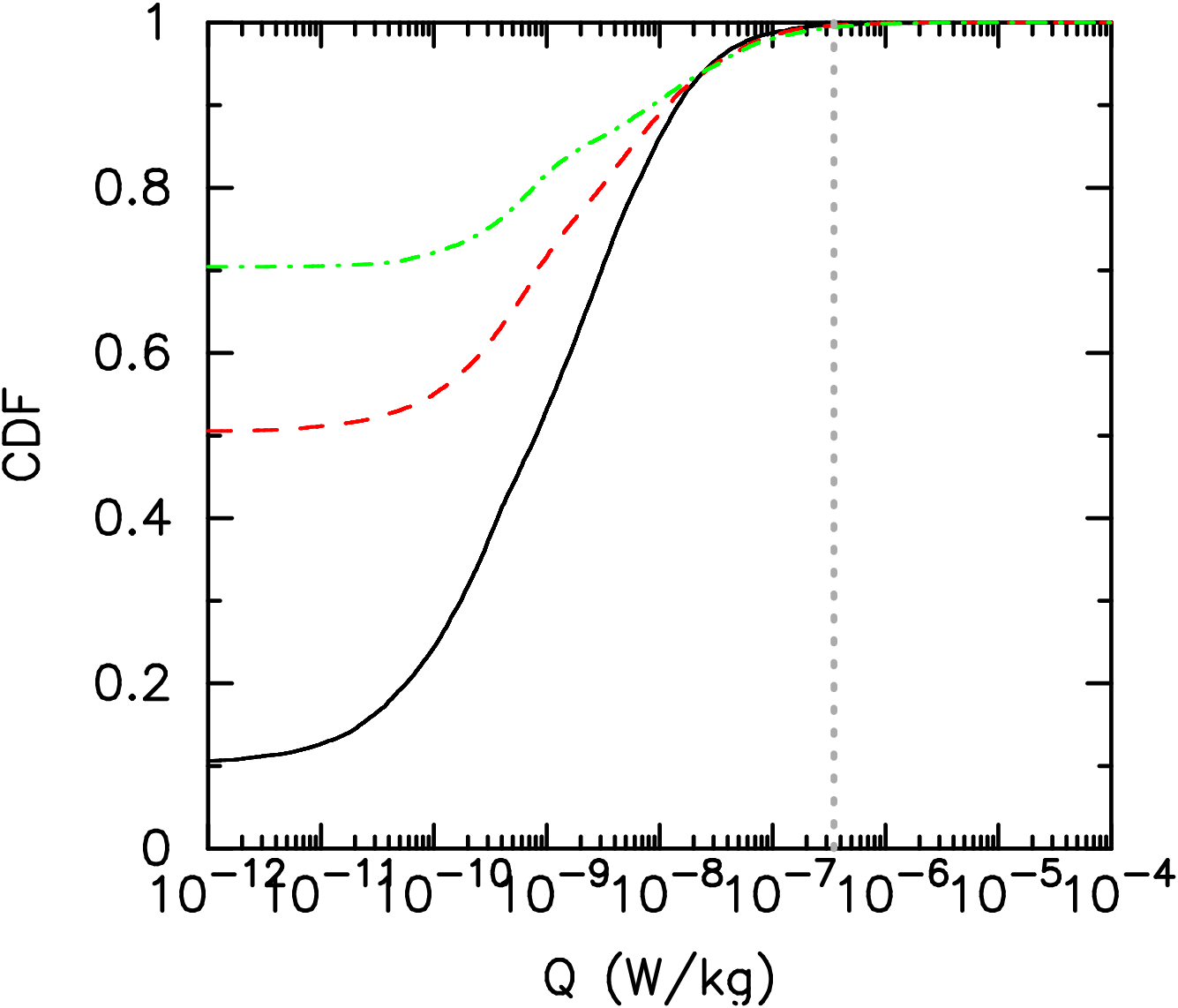}}}
\hspace*{0.3cm} 
\subfigure[Supernovae and local wind bubbles]{\label{high_density-Qheat-b}\rotatebox{270}{\includegraphics[scale=0.25]{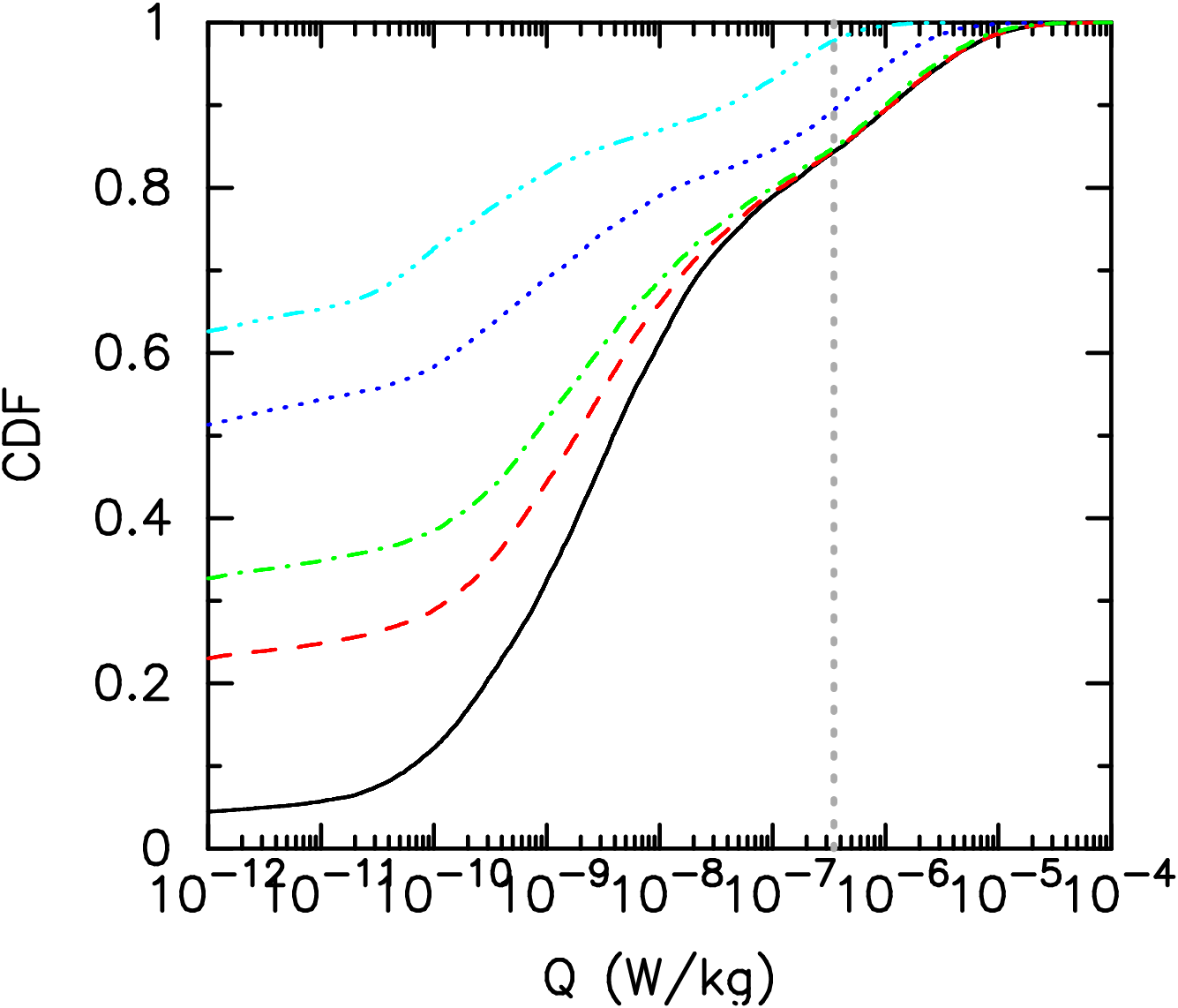}}}
\hspace*{0.3cm} 
\subfigure[Supernovae and despersed wind bubbles]{\label{high_density-Qheat-c}\rotatebox{270}{\includegraphics[scale=0.25]{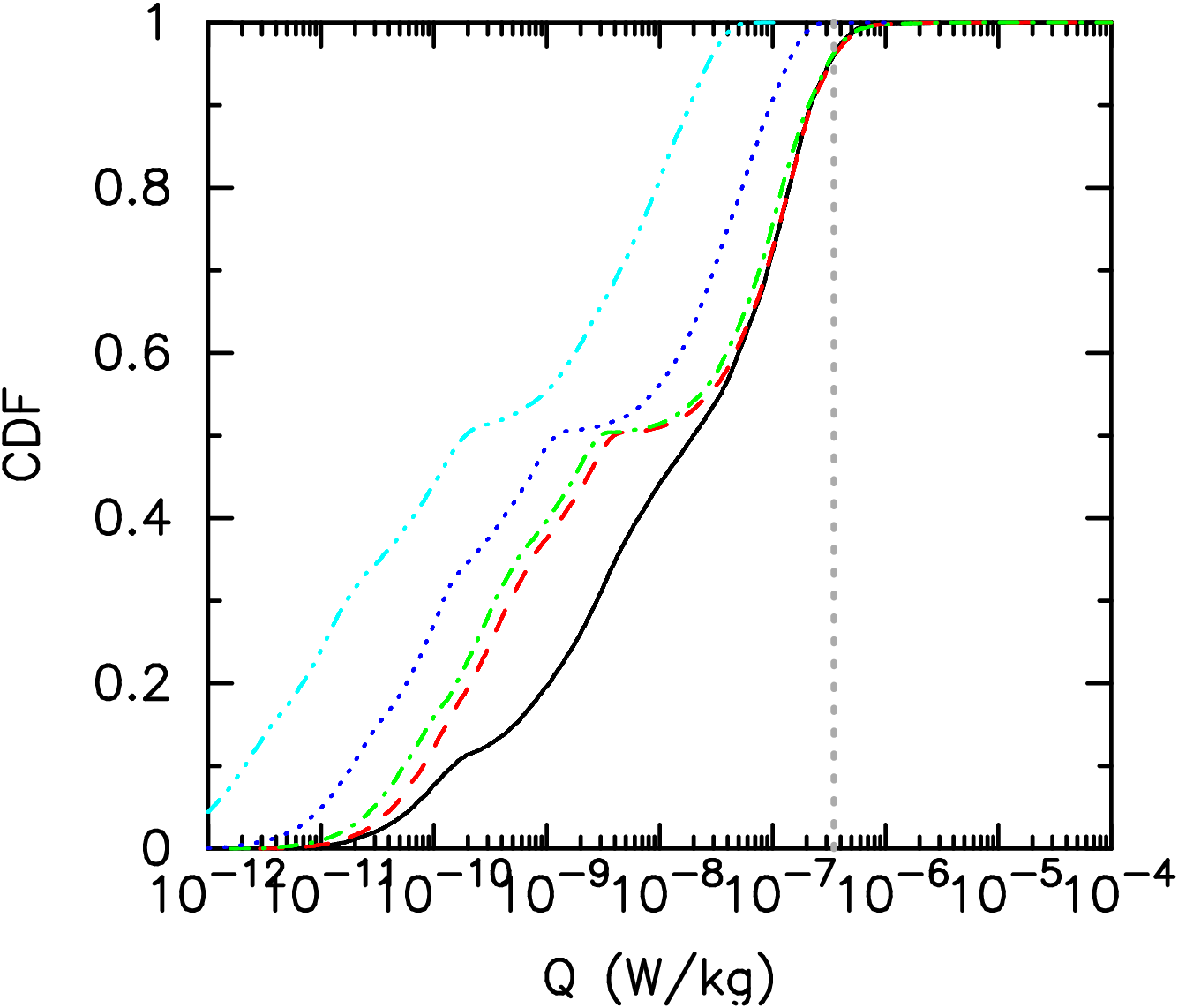}}} 
\caption[bf]{Cumulative Distribution  Functions (CDFs) of the radioactive heating of planetesimals from both $^{26}$Al and $^{60}$Fe in initially high-density star-forming regions ($\tilde{\rho} \sim 1000$M$_\odot$\,pc$^{-3}$). Panel (a) shows the heating from supernovae ejecta alone, panel (b) shows the heating from supernovae and local wind bubbles ($r_{\rm bub} = 0.1$\,pc) and panel (c) shows the heating from supernovae and dispersed wind bubbles ($r_{\rm bub} = 2r_{1/2}$\,pc). The heating is calculated at 10\,Myr (solid black lines), 7.5\,Myr (the dashed red lines), 5\,Myr (dot-dashed green lines), 2.5\,Myr (dotted blue lines) and 1\,Myr (dot-dot-dashed cyan lines). The vertical dotted lines indicate the initial heating for the Solar system, $Q_{\rm SS} = 3.4 \times 10^{-7}$\,W\,kg$^{-1}$, calculated from Eqn.~\ref{ss_heating}.}
\label{high_density-Qheat}
  \end{center}
\end{figure*}

We now plot the radioactive heating of planetesimals from these SLRs in Fig.~\ref{high_density-Qheat}, following the method in \citet{Moskovitz11}. We show the heating from supernovae only (Fig.~\ref{high_density-Qheat-a}), from supernovae and dispersed winds (Fig.~\ref{high_density-Qheat-b}) and from supernovae and local winds (Fig.~\ref{high_density-Qheat-c}). Each line represents the distribution of $Q$ values calculated at 10\,Myr (the solid black line), 7.5\,Myr (dashed red line), 5\,Myr (dot-dashed green line), 2.5\,Myr (dotted blue line) and at 1\,Myr (dot-dot-dashed cyan line). The latter two lines are not present in Fig.~\ref{high_density-Qheat-a} because there are no supernovae before $\sim$4\,Myr. The value calculated for the Solar system is shown by the vertical dotted line. Clearly, very few systems obtain Solar system-levels of heating from supernovae alone, or to frame it in terms independent of the assumptions about disc radius/evolution, enrichment is enhanced by the presence of winds, especially if the material remains in localised bubbles.

\subsection{Moderate stellar densities}

\begin{figure*}
  \begin{center}
\setlength{\subfigcapskip}{10pt}
\hspace*{-1.5cm}\subfigure[Supernovae only]{\label{moderate_density-Qheat-a}\rotatebox{270}{\includegraphics[scale=0.25]{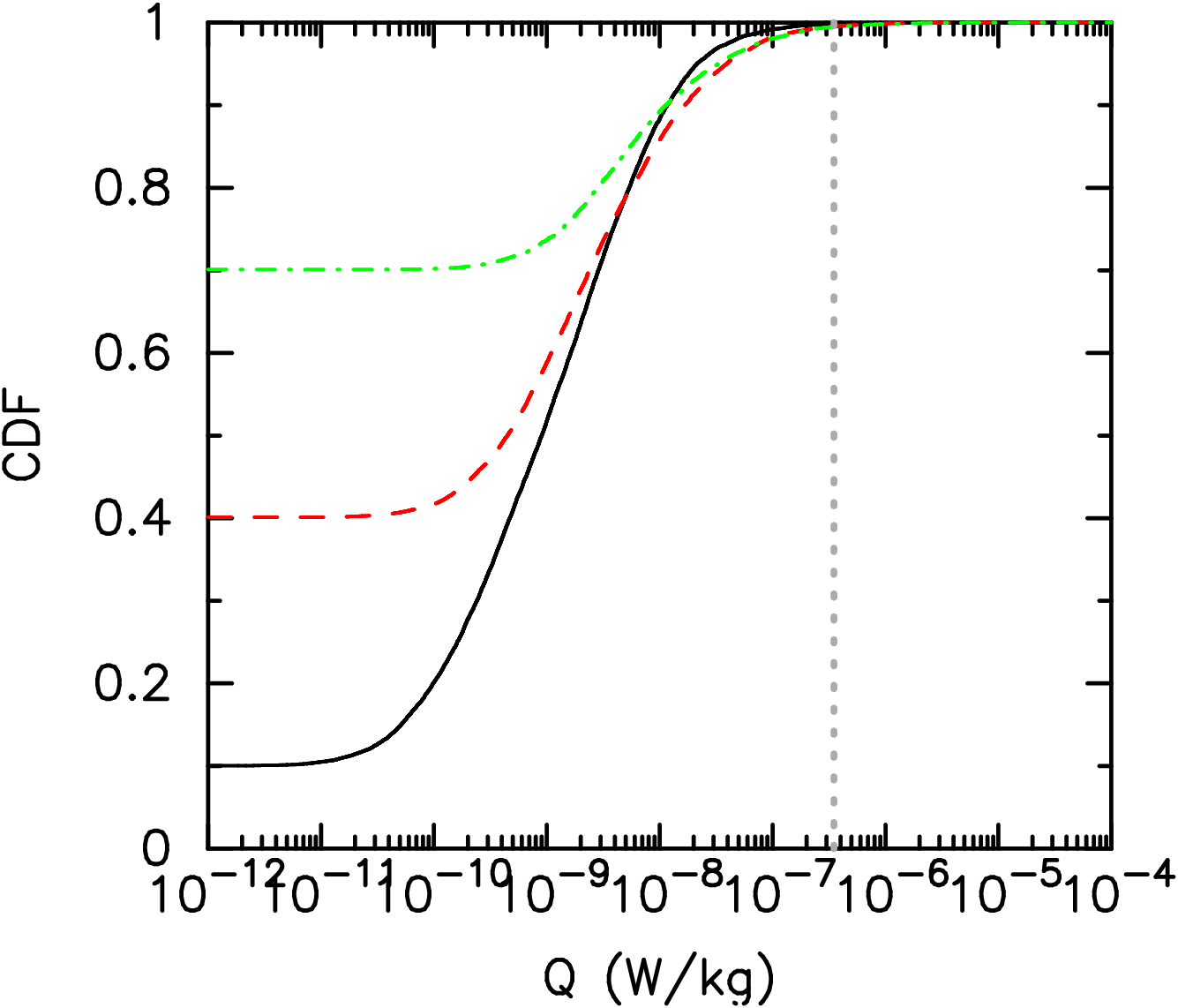}}}
\hspace*{0.3cm} 
\subfigure[Supernovae and local wind bubbles]{\label{moderate_density-Qheat-b}\rotatebox{270}{\includegraphics[scale=0.25]{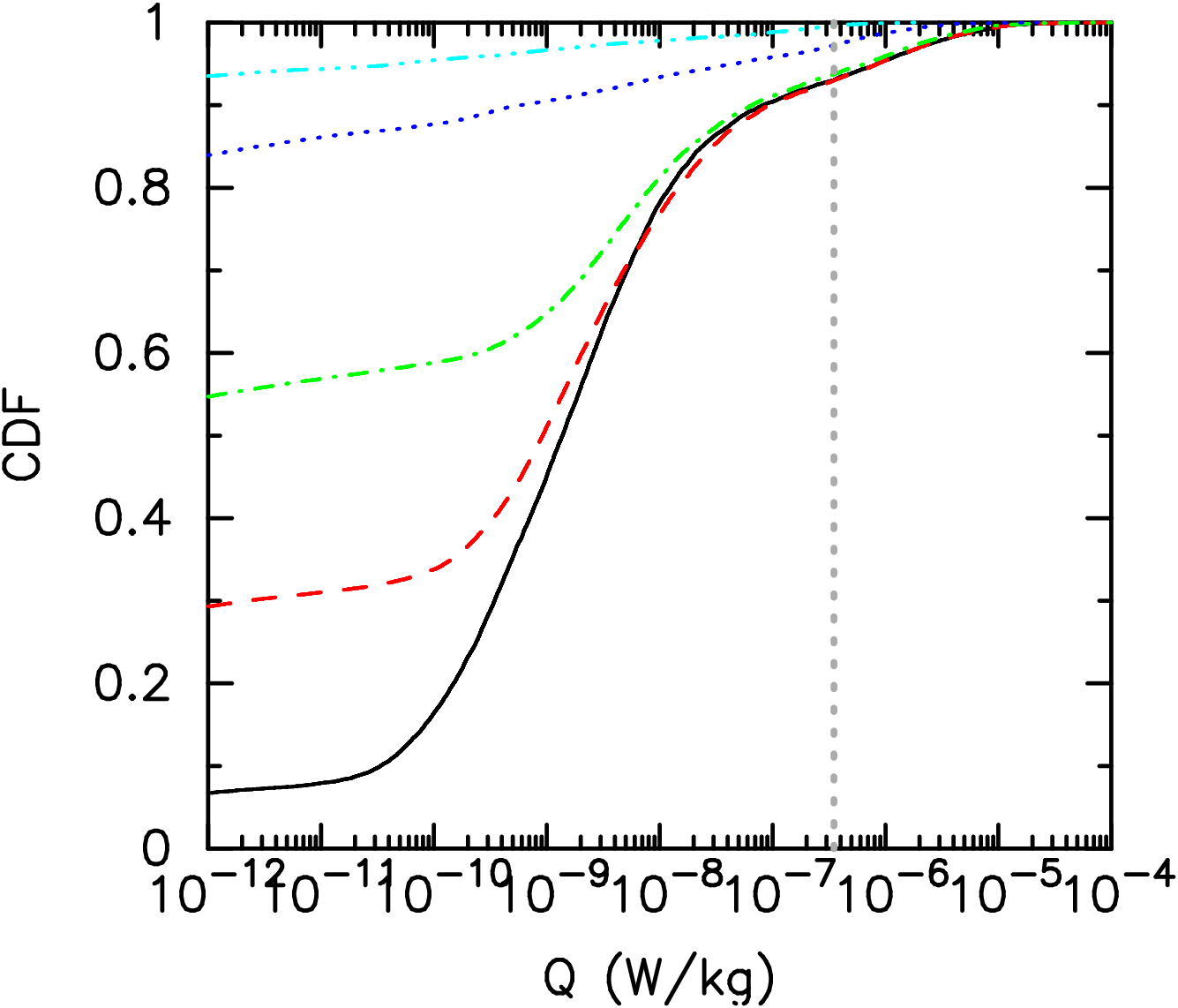}}}
\hspace*{0.3cm} 
\subfigure[Supernovae and dispersed wind bubbles]{\label{moderate_density-Qheat-c}\rotatebox{270}{\includegraphics[scale=0.25]{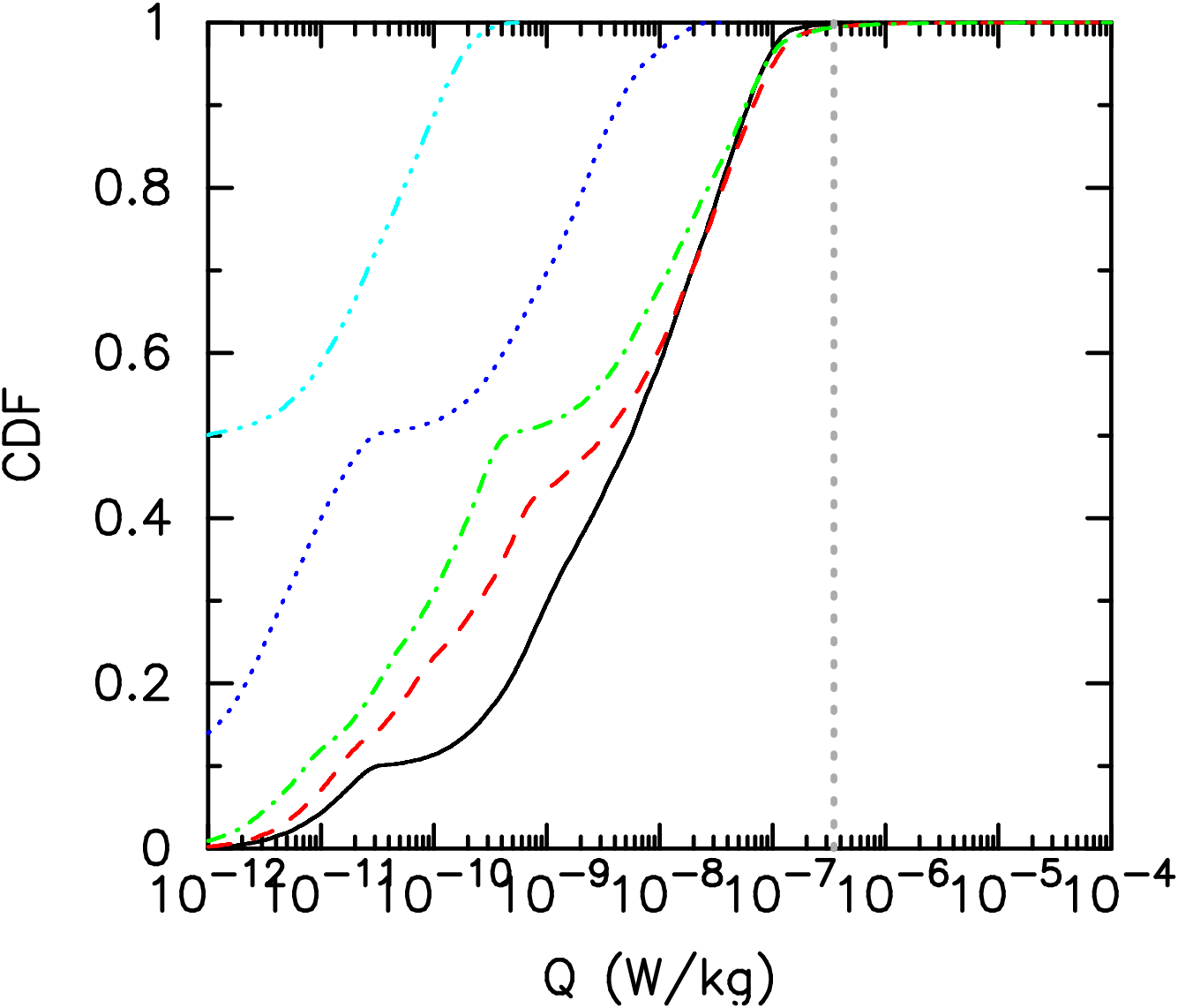}}} 
\caption[bf]{Cumulative Distribution  Functions (CDFs) of the radioactive heating of planetesimals from both $^{26}$Al and $^{60}$Fe in moderate-density star-forming regions ($\tilde{\rho} \sim 100$M$_\odot$\,pc$^{-3}$). Panel (a) shows the heating from supernovae ejecta alone, panel (b) shows the heating from supernovae and local wind bubbles ($r_{\rm bub} = 0.1$\,pc) and panel (c) shows the heating from supernovae and dispersed wind bubbles ($r_{\rm bub} = 2r_{1/2}$\,pc). The heating is calculated at 10\,Myr (solid black lines), 7.5\,Myr (the dashed red lines), 5\,Myr (dot-dashed green lines), 2.5\,Myr (dotted blue lines) and 1\,Myr (dot-dot-dashed cyan lines). The vertical dotted lines indicate the initial heating for the Solar system, $Q_{\rm SS} = 3.4 \times 10^{-7}$\,W\,kg$^{-1}$, calculated from Eqn.~\ref{ss_heating}.}
\label{moderate_density-Qheat}
  \end{center}
\end{figure*}

The high stellar density star-forming regions shown in Figs.~\ref{high_density-Al}--\ref{high_density-Qheat} are likely to be representative of some regions, e.g. the ONC \citep{Allison11,Parker14e,Schoettler20}, Carina \citep{Reiter19} and NGC\,2264 \citep{Schoettler22,Parker22a}, but not all. Many star-forming regions local to the Sun have much lower postulated initial densities, likely to be in the range 100 -- 1\,M$_\odot$\,pc$^{-3}$ \citep{Lada03,Bressert10,Parker17a}.

When we reduce the stellar density, the amount of enrichment (unsurprisingly) decreases. At moderate stellar densities ($\tilde{\rho} \sim 100$M$_\odot$\,pc$^{-3}$), depending on the amount of enrichment from winds, up to $\sim$50\,per cent of stars attain Solar system-like (or higher) $Z_{\rm Al}$ ratios after 10\,Myr, and 10\,per cent of stars attain Solar sytem-like (or higher) $Z_{\rm Fe}$ ratios.

In terms of the internal heating of planetesimals, however, only the simulations in which the $^{26}$Al is captured from localised stellar winds reach Solar system-levels of heating (the middle panel of Fig.~\ref{moderate_density-Qheat-b}).

\subsection{Low- and Field-like stellar densities}

\begin{figure*}
  \begin{center}
\setlength{\subfigcapskip}{10pt}
\hspace*{-1.5cm}
\subfigure[$Z_{\rm Al}$ -- supernovae and local wind bubbles]{\label{low_density-a}\rotatebox{270}{\includegraphics[scale=0.27]{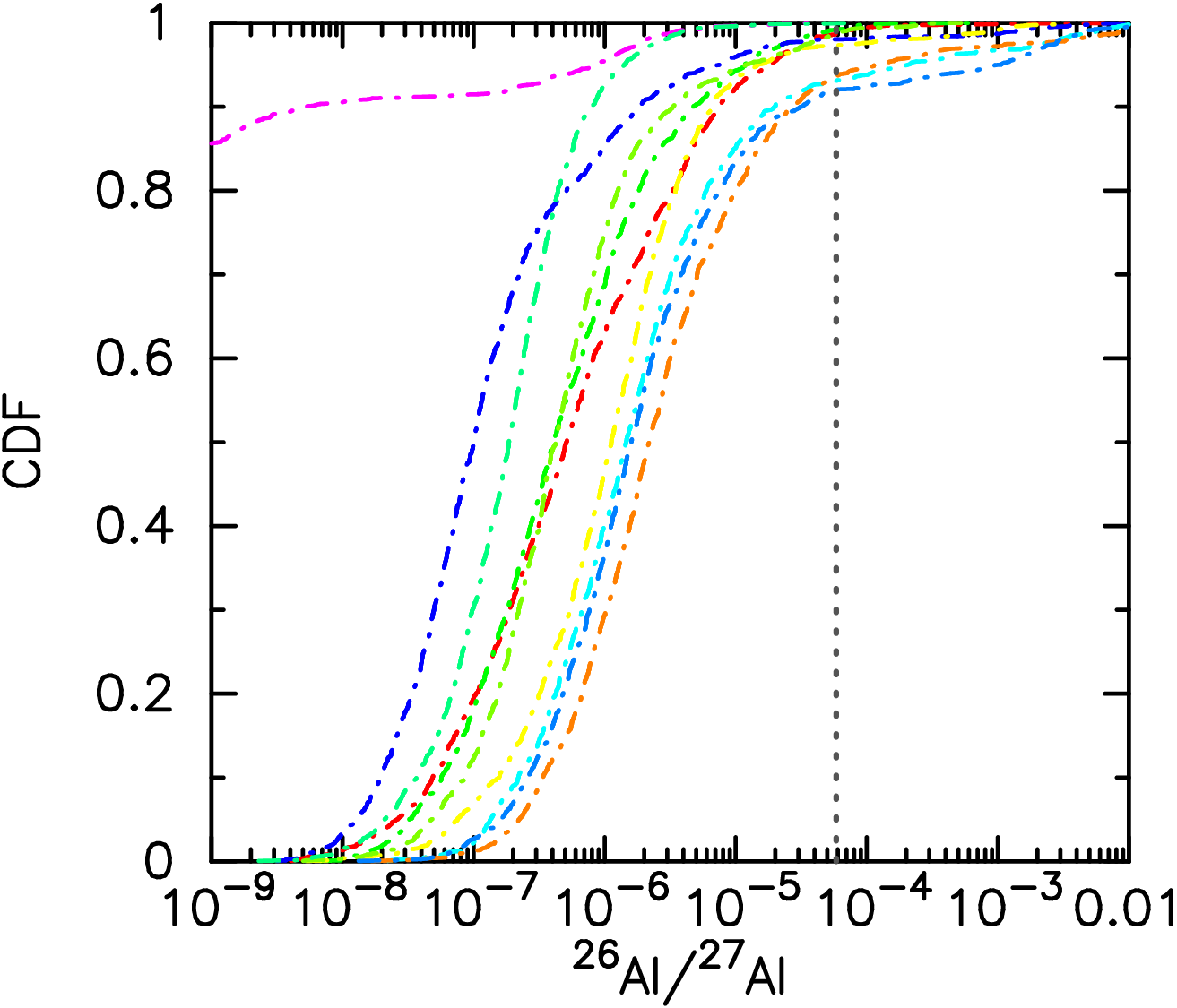}}}
\hspace*{0.3cm} 
\subfigure[$Z_{\rm Al}$ -- supernovae and dispersed wind bubbles]{\label{low_density-b}\rotatebox{270}{\includegraphics[scale=0.27]{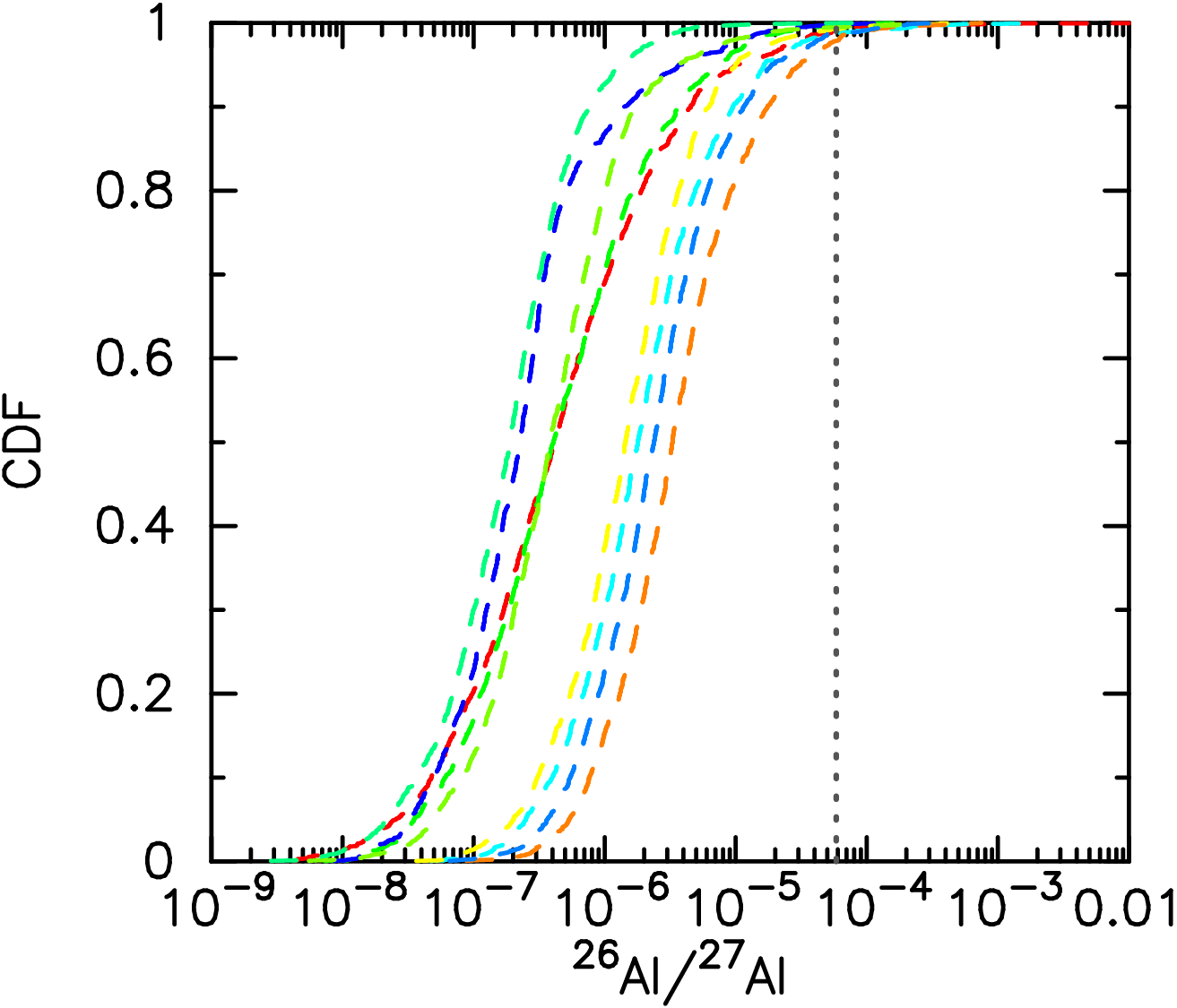}}}
\subfigure[$Z_{\rm Fe}$ -- supernovae only]{\label{low_density-c}\rotatebox{270}{\includegraphics[scale=0.27]{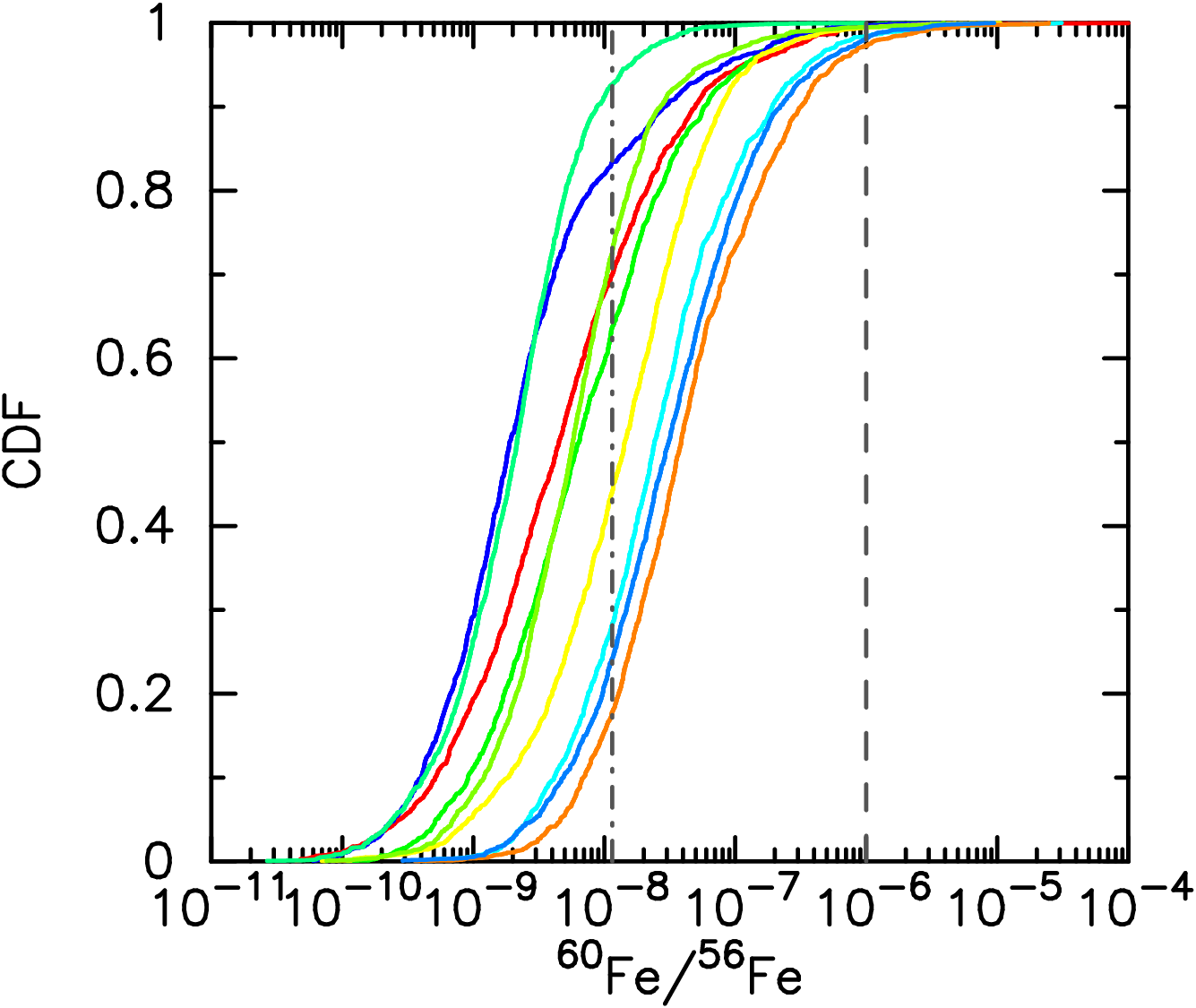}}}
\hspace*{0.3cm} 
\caption[bf]{Cumulative Distribution  Functions (CDFs) of the SLR ratios $Z_{\rm Al}$ and $Z_{\rm Fe}$ in a star-forming region with an initially low stellar density ($\tilde{\rho} \sim 10$M$_\odot$\,pc$^{-3}$). The initial disc radii are all $r_{\rm disc} = 100$\,au, and the discs do not evolve with time.  Panel (a) shows the $Z_{\rm Al}$ ratios for supernovae and local wind bubbles ($r_{\rm bub} = 0.1$\,pc), panel (b)  shows the $Z_{\rm Al}$ ratios for supernovae and dispersed wind bubbles ($r_{\rm bub} = 2r_{1/2}$), and panel (c) shows the $Z_{\rm Fe}$ ratios (for supernovae only). All $Z$ ratios are calculated at 10\,Myr. In panels (a) and (b) the vertical dotted lines indicate the measured Solar system value for Al, $Z_{\rm Al,SS} = 5.85 \times 10^{-5}$ \citep{Thrane06}. In panel (c) the vertical dot-dashed line indicates the measured Solar system value for Fe, $Z_{\rm Fe,SS} = 1.15 \times 10^{-8}$ \citep{Tang12} and the vertical dashed line indicates the alternative measurement, $Z_{\rm Fe,SS} = 1 \times 10^{-6}$ \citep{Mishra16}.}
\label{low_density}
  \end{center}
\end{figure*}

If we reduce the density further, so that the regions have low densities ($\tilde{\rho} \leq 10$M$_\odot$\,pc$^{-3}$) then the amount of enrichment begins to fall below Solar system levels for all stars. In Fig.~\ref{low_density} we show the $Z_{\rm Al}$ and $Z_{\rm Fe}$ ratios for these low-density regions at 10\,Myr. Fig.~\ref{low_density-a} shows the $Z_{\rm Al}$ ratios for the combination of supernovae and localised winds, and some stars do attain Solar system-like values. However, when the winds are more dispersed the distribution of $Z_{\rm Al}$ ratios falls short of Solar system-values (Fig.~\ref{low_density-b}), as do the $Z_{\rm Fe}$ ratios (Fig.~\ref{low_density-c}).

\begin{figure*}
  \begin{center}
\setlength{\subfigcapskip}{10pt}
\hspace*{-1.5cm}\subfigure[Supernovae only]{\label{low_density-Qheat-a}\rotatebox{270}{\includegraphics[scale=0.25]{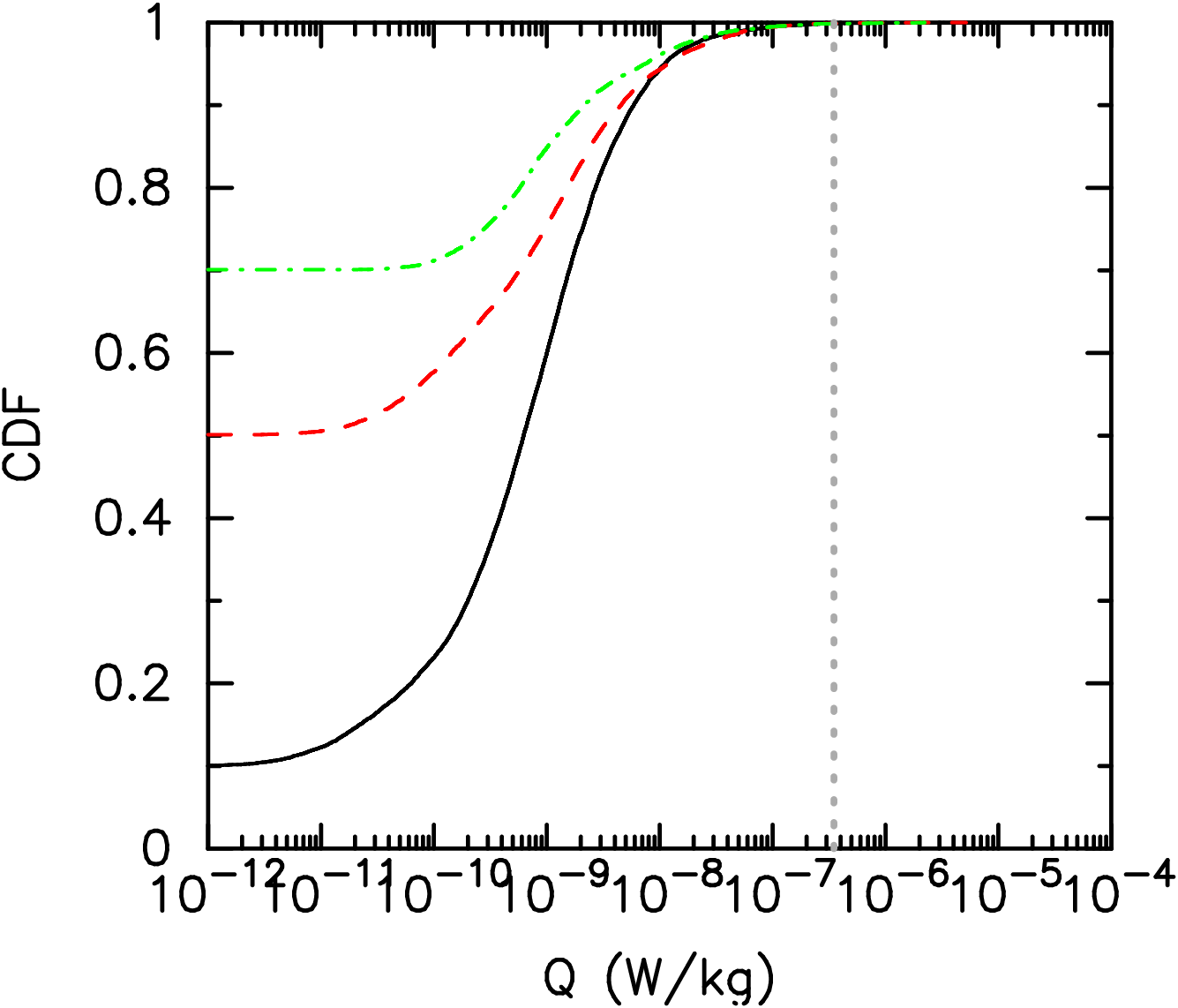}}}
\hspace*{0.3cm} 
\subfigure[Supernovae and local wind bubbles]{\label{low_density-Qheat-b}\rotatebox{270}{\includegraphics[scale=0.25]{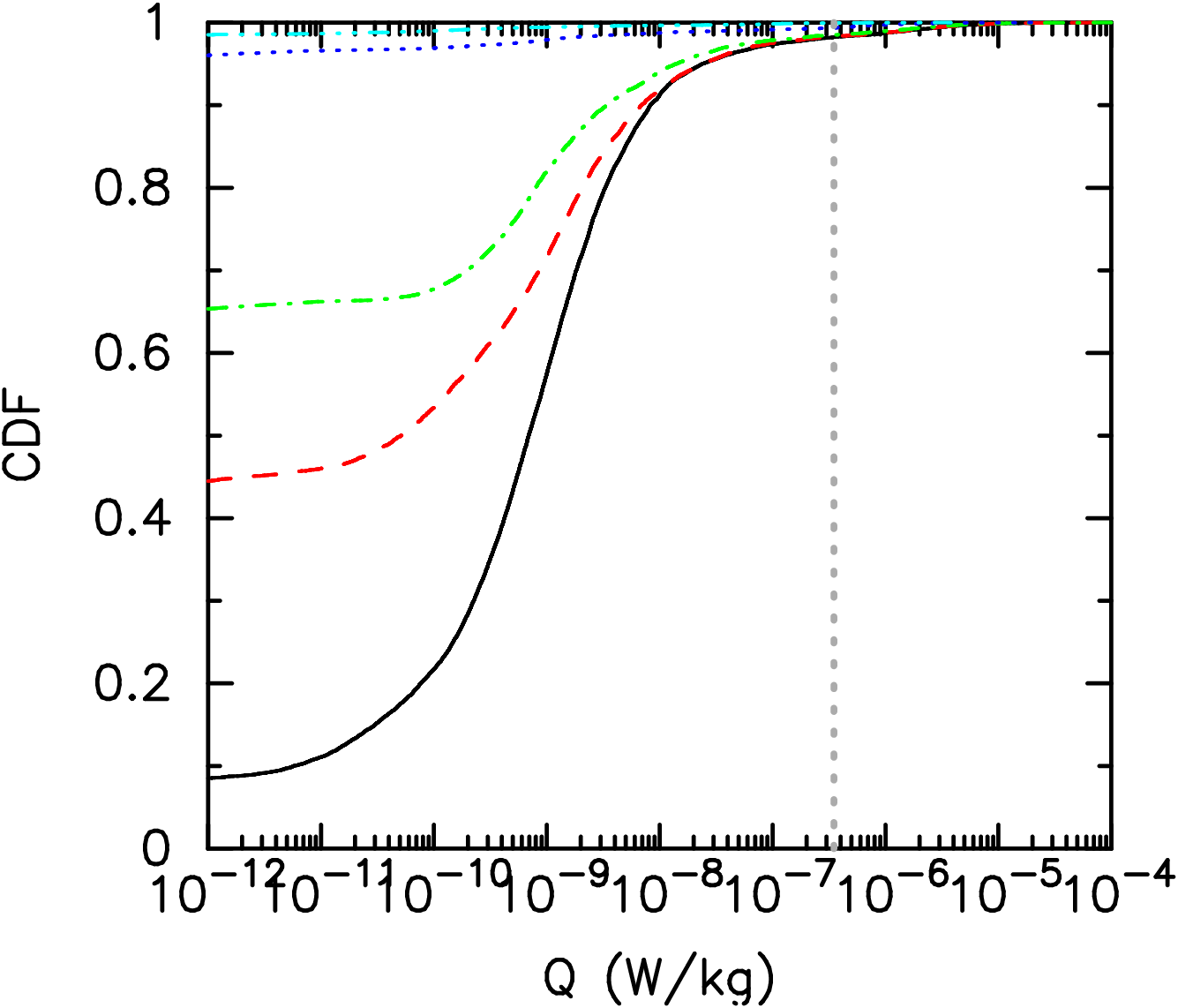}}}
\hspace*{0.3cm} 
\subfigure[Supernovae and despersed wind bubbles]{\label{low_density-Qheat-c}\rotatebox{270}{\includegraphics[scale=0.25]{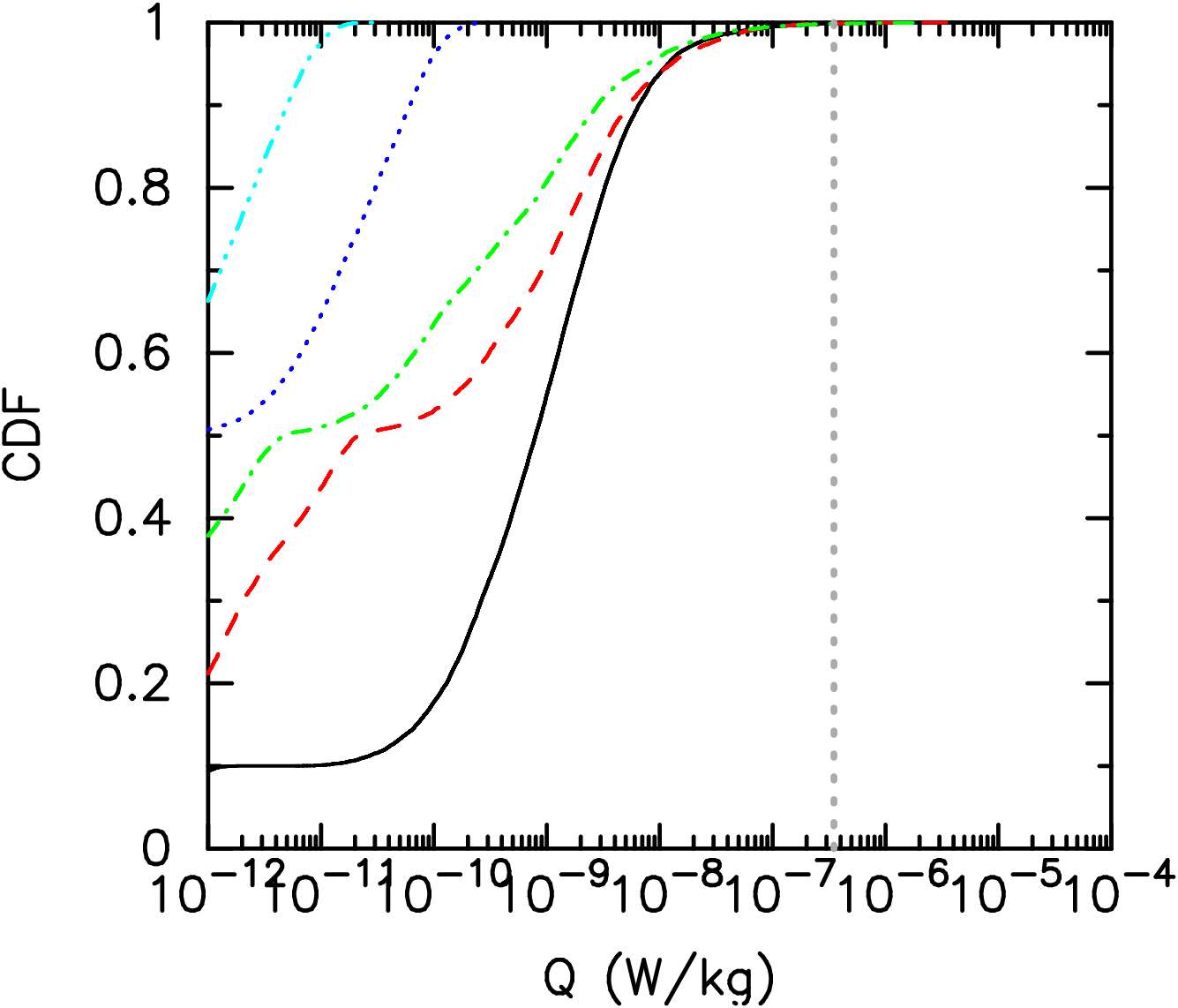}}} 
\caption[bf]{Cumulative Distribution  Functions (CDFs) of the radioactive heating of planetesimals from both $^{26}$Al and $^{60}$Fe in low-density star-forming regions ($\tilde{\rho} \sim 10$M$_\odot$\,pc$^{-3}$). Panel (a) shows the heating from supernovae ejecta alone, panel (b) shows the heating from supernovae and local wind bubbles ($r_{\rm bub} = 0.1$\,pc) and panel (c) shows the heating from supernovae and dispersed wind bubbles ($r_{\rm bub} = 2r_{1/2}$\,pc). The heating is calculated at 10\,Myr (solid black lines), 7.5\,Myr (the dashed red lines), 5\,Myr (dot-dashed green lines), 2.5\,Myr (dotted blue lines) and 1\,Myr (dot-dot-dashed cyan lines). The vertical dotted lines indicate the initial heating for the Solar system, $Q_{\rm SS} = 3.4 \times 10^{-7}$\,W\,kg$^{-1}$, calculated from Eqn.~\ref{ss_heating}.}
\label{low_density-Qheat}
  \end{center}
\end{figure*}

When we use the $Z_{\rm Al}$ and $Z_{\rm Fe}$ ratios to calculate the SLR heating, we find that almost no stars have values commensurate with the early Solar system (Fig.~\ref{low_density-Qheat}).

If we reduce the density further, to values similar to the Sun's present-day location \citep[0.1\,M$_\odot$\,pc$^{-3}$][]{Korchagin03}, then no stars have Solar system levels of enrichment. 

%\begin{figure*}
%  \begin{center}
%\setlength{\subfigcapskip}{10pt}
%\hspace*{-1.5cm}\subfigure[Supernovae only]{\label{field_density-Qheat-a}\rotatebox{270}{\includegraphics[scale=0.25]{Plot_q_heat_CDF_Or_C0p3F2p20pSmFS10_100Xf_SN.ps}}}
%\hspace*{0.3cm} 
%\subfigure[Supernovae and local wind bubbles]{\label{field_density-Qheat-b}\rotatebox{270}{\includegraphics[scale=0.25]{Plot_q_heat_CDF_Or_C0p3F2p20pSmFS10_100Xf_w2.ps}}}
%\hspace*{0.3cm} 
%\subfigure[Supernovae and despersed wind bubbles]{\label{field_density-Qheat-c}\rotatebox{270}{\includegraphics[scale=0.25]{Plot_q_heat_CDF_Or_C0p3F2p20pSmFS10_100Xf_w1.ps}}} 
%\caption[bf]{The radioactive heating of planetesimals from both $^{26}$Al and $^{60}$Fe in field-density star-forming regions ($\tilde{\rho} \sim 0.1$M$_\odot$\,pc$^{-3}$). Panel (a) shows the heating from supernovae ejecta alone, panel (b) shows the heating from supernovae and local wind bubbles ($r_{\rm bub} = 0.1$\,pc) and panel (c) shows the heating from supernovae and dispersed wind bubbles ($r_{\rm bub} = 2r_{1/2}$\,pc). The heating is calculated at 10\,Myr (solid black lines), 7.5\,Myr (the dashed red lines), 5\,Myr (dot-dashed green lines), 2.5\,Myr (dotted blue lines) and 1\,Myr (dot-dot-dashed cyan lines).}
%\label{field_density-Qheat}
%  \end{center}
%\end{figure*}

\subsection{Enrichment during disc depletion}

Up to this point we have assumed that the discs remain extant for the duration of the simulation (10\,Myr), and as this is roughly the main sequence lifetime of many of the enriching (massive) stars, the discs can capture material from the winds of these stars, and then their supernovae ejecta when they explode.

We have not allowed the discs to expand through viscous evolution \citep{Hartmann98,Lichtenberg16a}, but also we have not factored in any disc destruction mechanisms, such as truncation from encounters and/or photoevaporation from the massive stars \citep{Johnstone98,Scally01,Adams04,Vincke15,Nicholson19a,Parker21a}.

We have also not accounted for planet formation in the discs, which although does not remove material from the disc, it may mean the composition of the disc (and distribution of SLRs within) is no longer homogeneous.

\begin{figure*}
  \begin{center}
\setlength{\subfigcapskip}{10pt}
\hspace*{-1.5cm}\subfigure[Supernovae only -- 10\,Myr]{\label{high_density-disappear-Al-a}\rotatebox{270}{\includegraphics[scale=0.27]{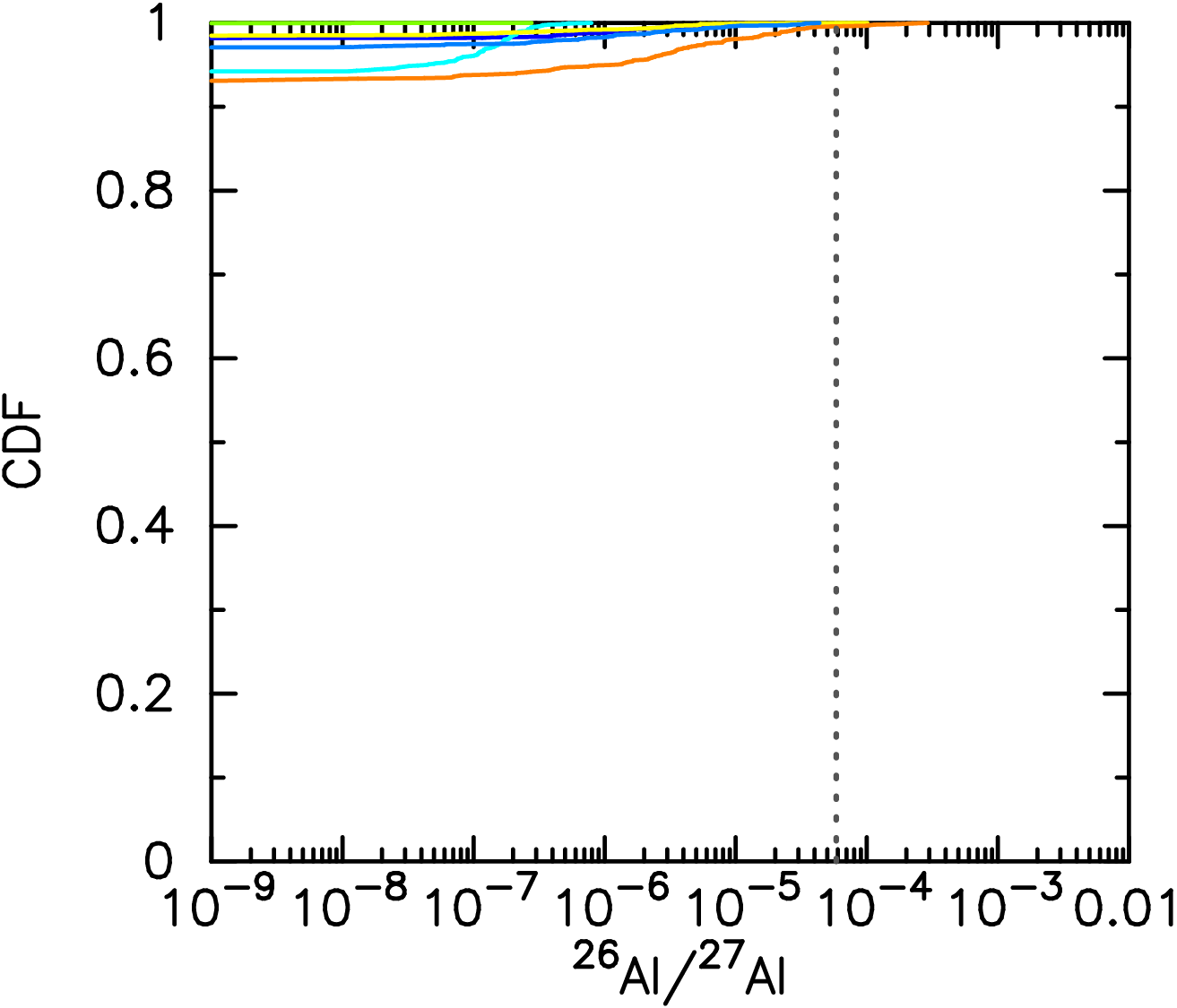}}}
\hspace*{0.3cm} 
\subfigure[Supernovae and local wind bubbles -- 10\,Myr]{\label{high_density-disappear-Al-b}\rotatebox{270}{\includegraphics[scale=0.27]{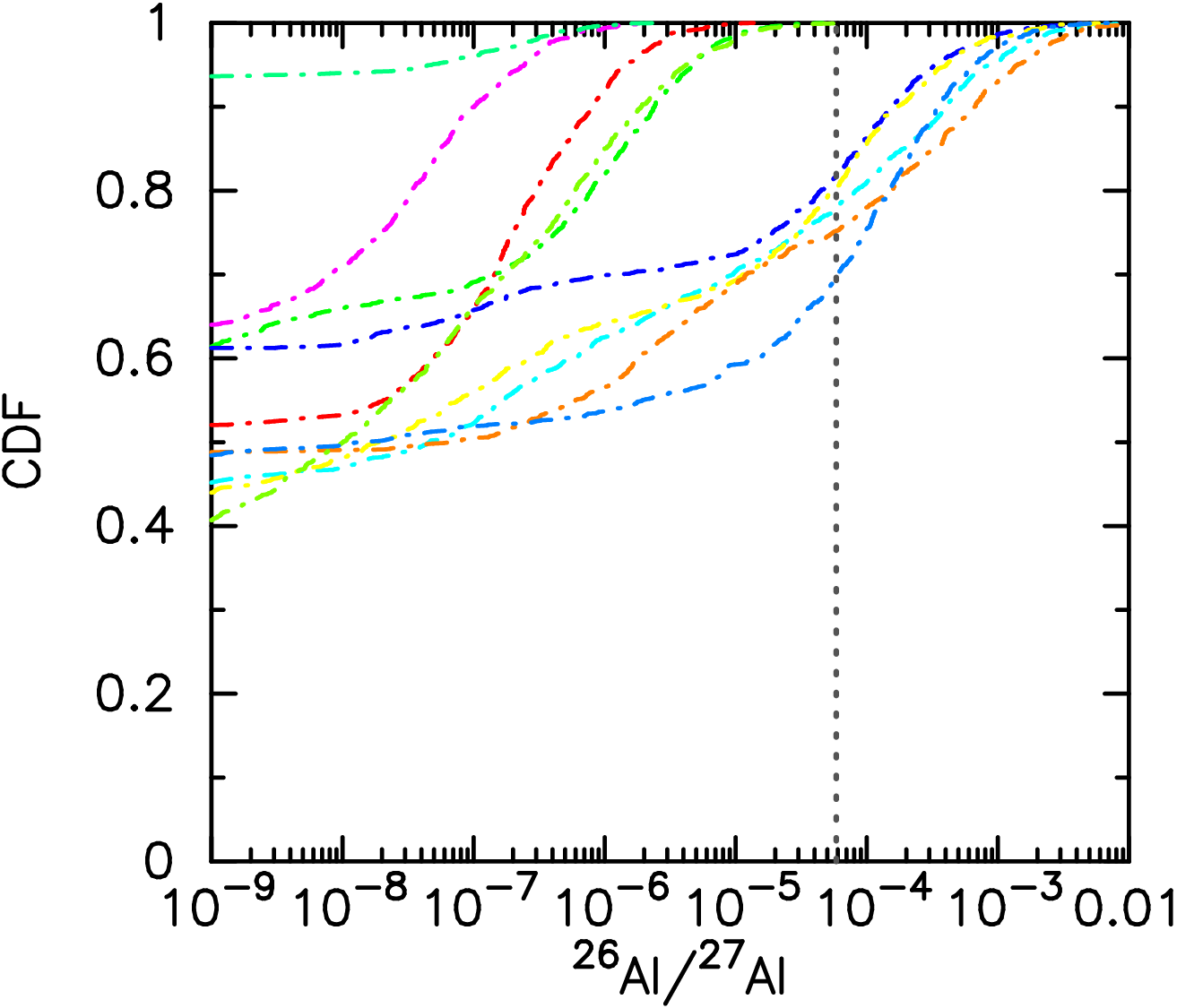}}}
\hspace*{0.3cm} 
\subfigure[Supernovae and global wind bubbles -- 10\,Myr]{\label{high_density-disappear-Al-c}\rotatebox{270}{\includegraphics[scale=0.27]{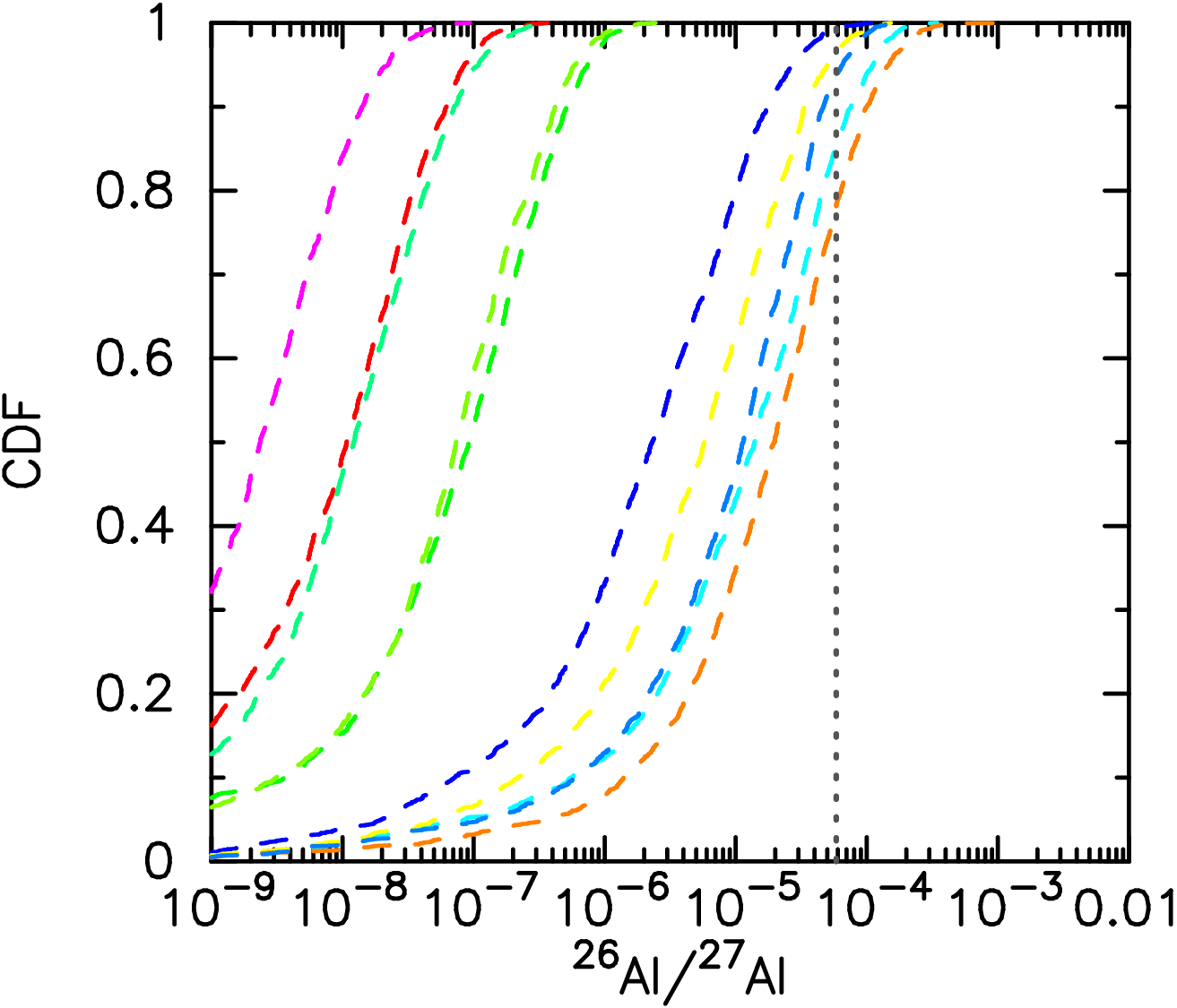}}} 
\hspace*{-1.5cm}\subfigure[All three line sets -- 1\,Myr]{\label{high_density-disappear-Al-d}\rotatebox{270}{\includegraphics[scale=0.27]{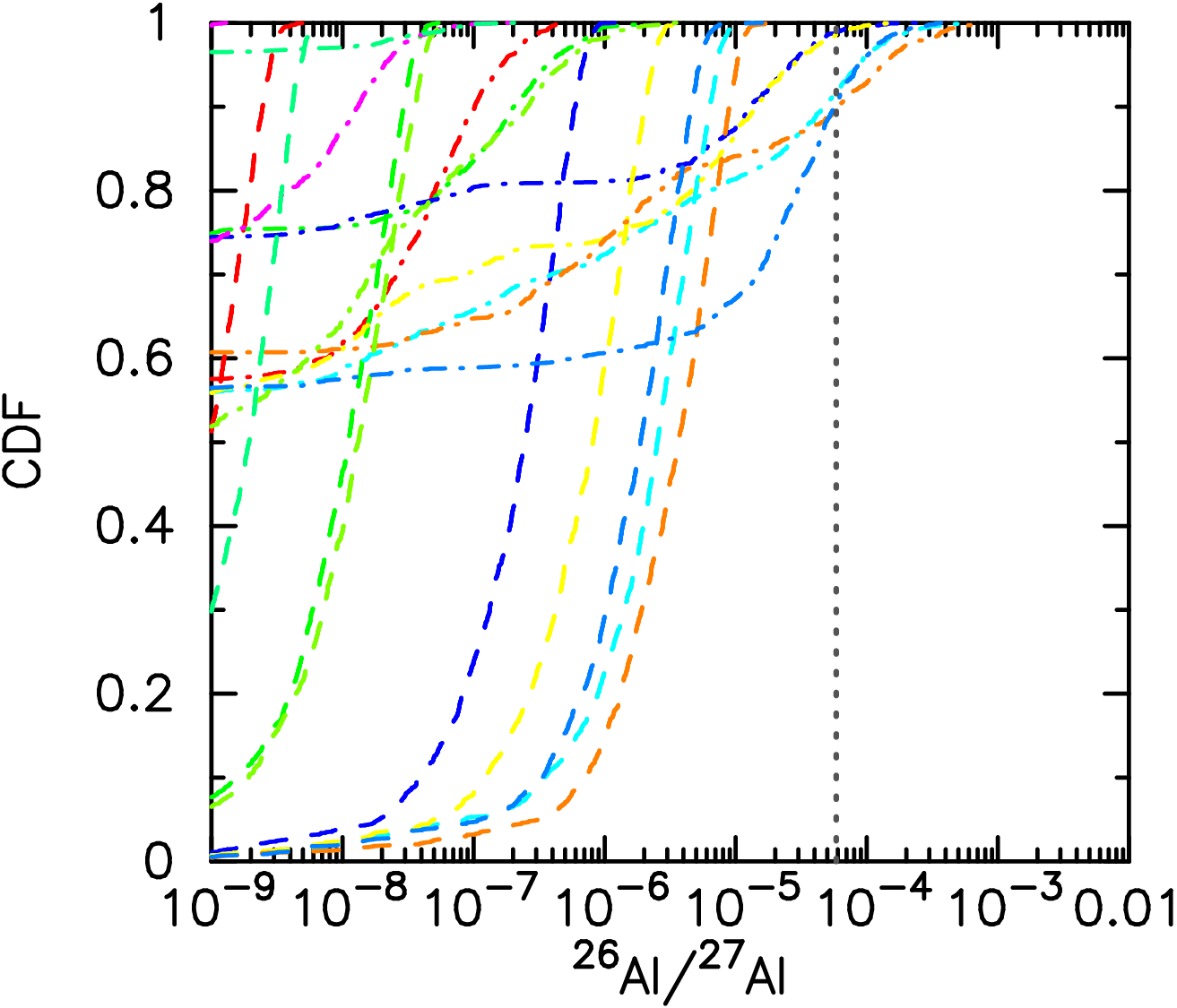}}}
\hspace*{0.3cm} 
\subfigure[All three line sets -- 2.5\,Myr]{\label{high_density-disappear-Al-e}\rotatebox{270}{\includegraphics[scale=0.27]{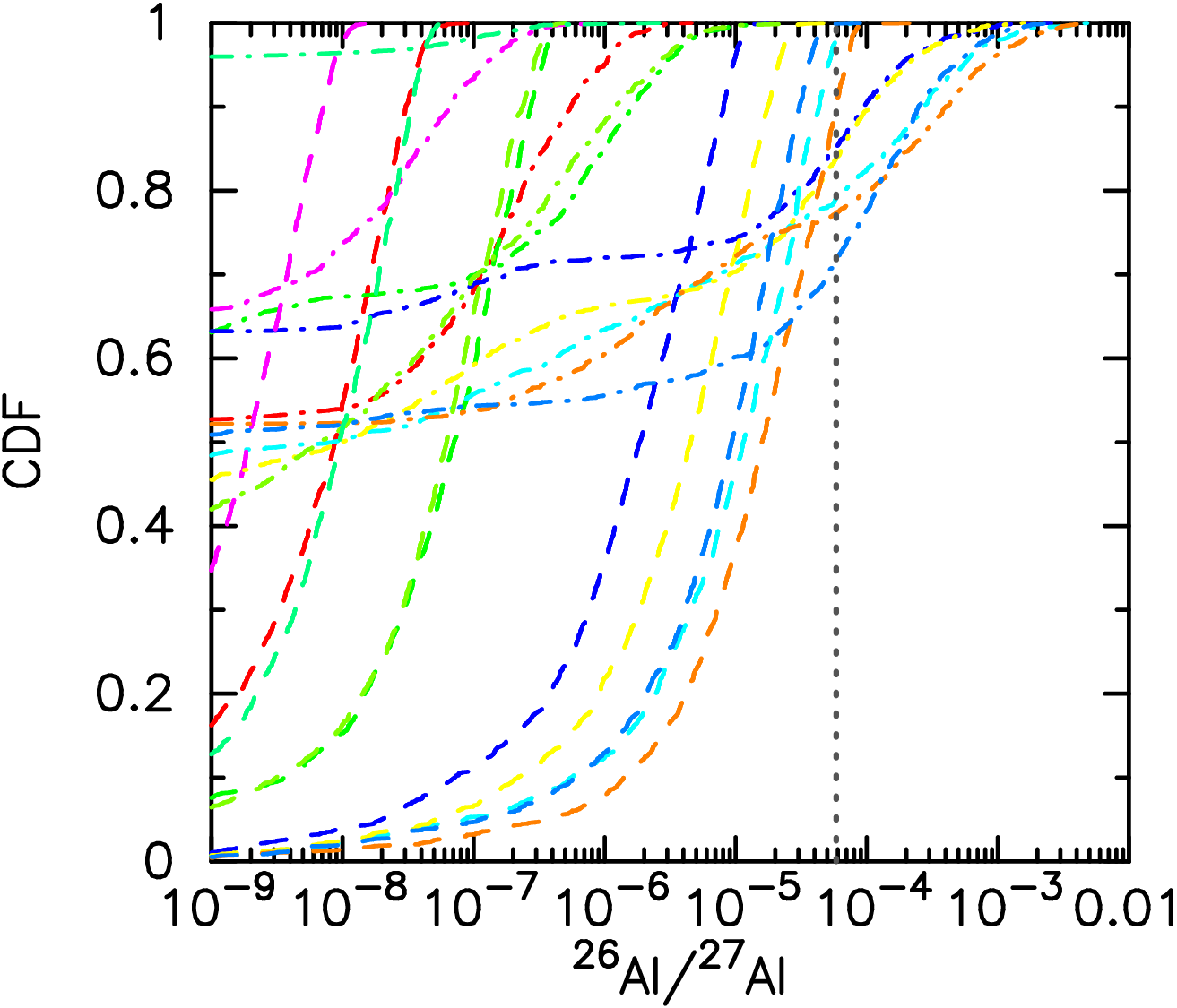}}}
\hspace*{0.3cm} 
\subfigure[All three line sets -- 5\,Myr]{\label{high_density-disappear-Al-f}\rotatebox{270}{\includegraphics[scale=0.27]{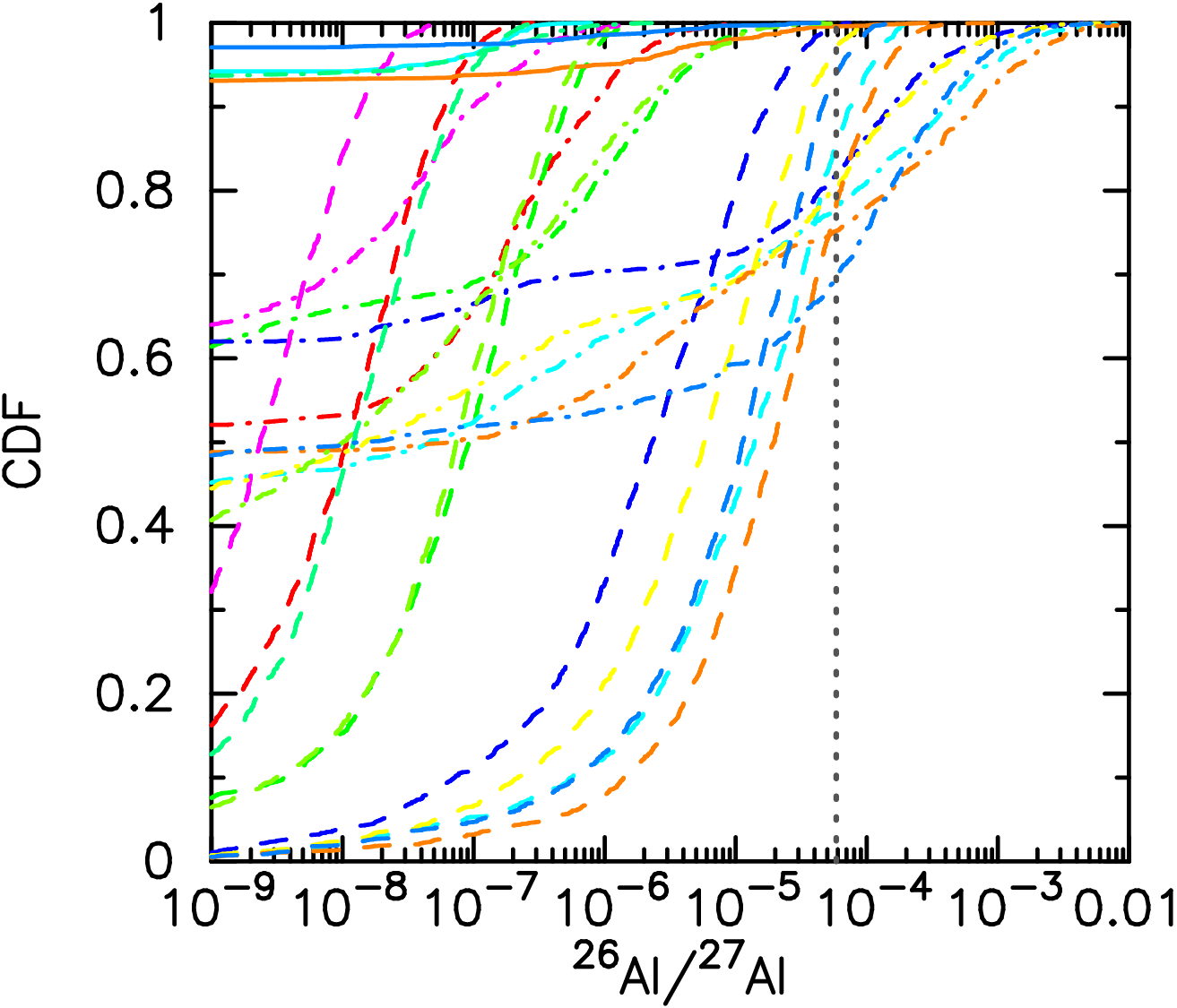}}} 
\caption[bf]{Cumulative Distribution  Functions (CDFs) of the SLR ratios $Z_{\rm Al}$ in a star-forming region with an initially high stellar density ($\tilde{\rho} \sim 1000$M$_\odot$\,pc$^{-3}$) where the discs deplete in line with the observations of \citet{Haisch01} and \citet{Richert18}. The initial disc radii are all $r_{\rm disc} = 100$\,au, and the discs do not evolve with time.  Solid lines are ratios calculated from supernovae only. Dashed lines are global, or dispersed wind bubbles ($r_{\rm bub} = 2r_{1/2}$) plus any supernovae, dot-dashed lines are local wind bubbles ($r_{\rm bub} = 0.1$\,pc)  plus any supernovae. We show $^{26}$Al/$^{27}$Al ratios calculated at 10\,Myr (top row), then 1, 2.5 \& 5\,Myr (bottom row). The vertical dotted lines indicate the measured Solar system value, $Z_{\rm Al,SS} = 5.85 \times 10^{-5}$ \citep{Thrane06}.}
\label{high_density-disappear-Al}
  \end{center}
\end{figure*}

\begin{figure*}
  \begin{center}
\setlength{\subfigcapskip}{10pt}
\subfigure[Supernovae -- 5\,Myr]{\label{high_density-disappear-Fe-a}\rotatebox{270}{\includegraphics[scale=0.27]{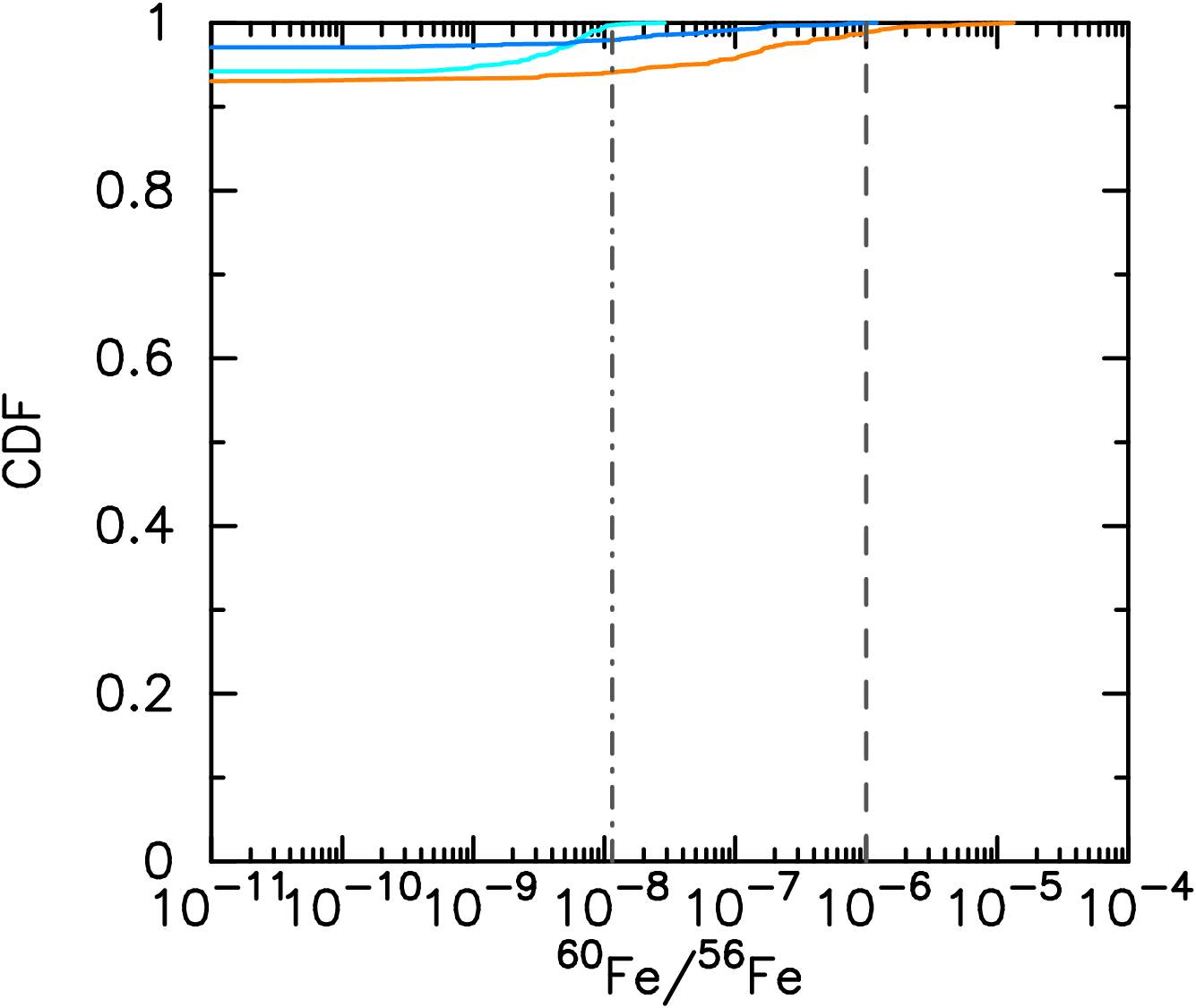}}}
\hspace*{0.3cm} 
\subfigure[Supernovae -- 10\,Myr]{\label{high_density-disappear-Fe-b}\rotatebox{270}{\includegraphics[scale=0.27]{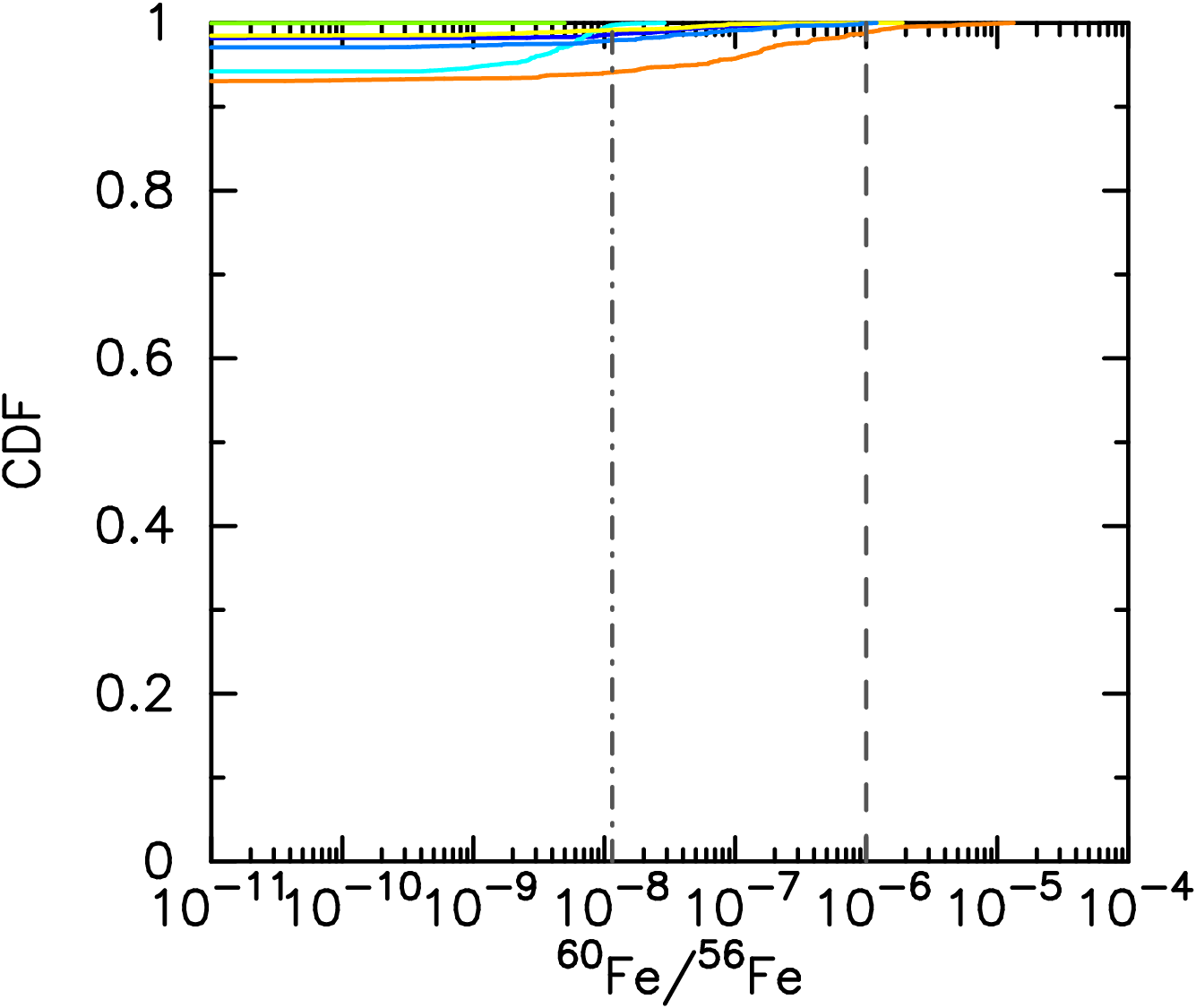}}} 
\caption[bf]{Cumulative Distribution  Functions (CDFs) of the SLR ratios $Z_{\rm Fe}$ in a star-forming region with an initially high stellar density ($\tilde{\rho} \sim 1000$M$_\odot$\,pc$^{-3}$) where the discs deplete in line with the observations of \citet{Haisch01} and \citet{Richert18}. The initial disc radii are all $r_{\rm disc} = 100$\,au, and the discs do not evolve with time.  We show the $^{60}$Fe/$^{56}$Fe ratios (which only come from supernovae ejecta) calculated at 5 and 10\,Myr. The vertical dot-dashed line indicates the measured Solar system value for Fe, $Z_{\rm Fe,SS} = 1.15 \times 10^{-8}$ \citep{Tang12} and the vertical dashed line indicates the alternative measurement, $Z_{\rm Fe,SS} = 1 \times 10^{-6}$ \citep{Mishra16}.}
\label{high_density-disappear-Fe}
  \end{center}
\end{figure*}

\begin{figure*}
  \begin{center}
\setlength{\subfigcapskip}{10pt}
\hspace*{-1.5cm}\subfigure[Supernovae only]{\label{high_density-disappear-Qheat-a}\rotatebox{270}{\includegraphics[scale=0.25]{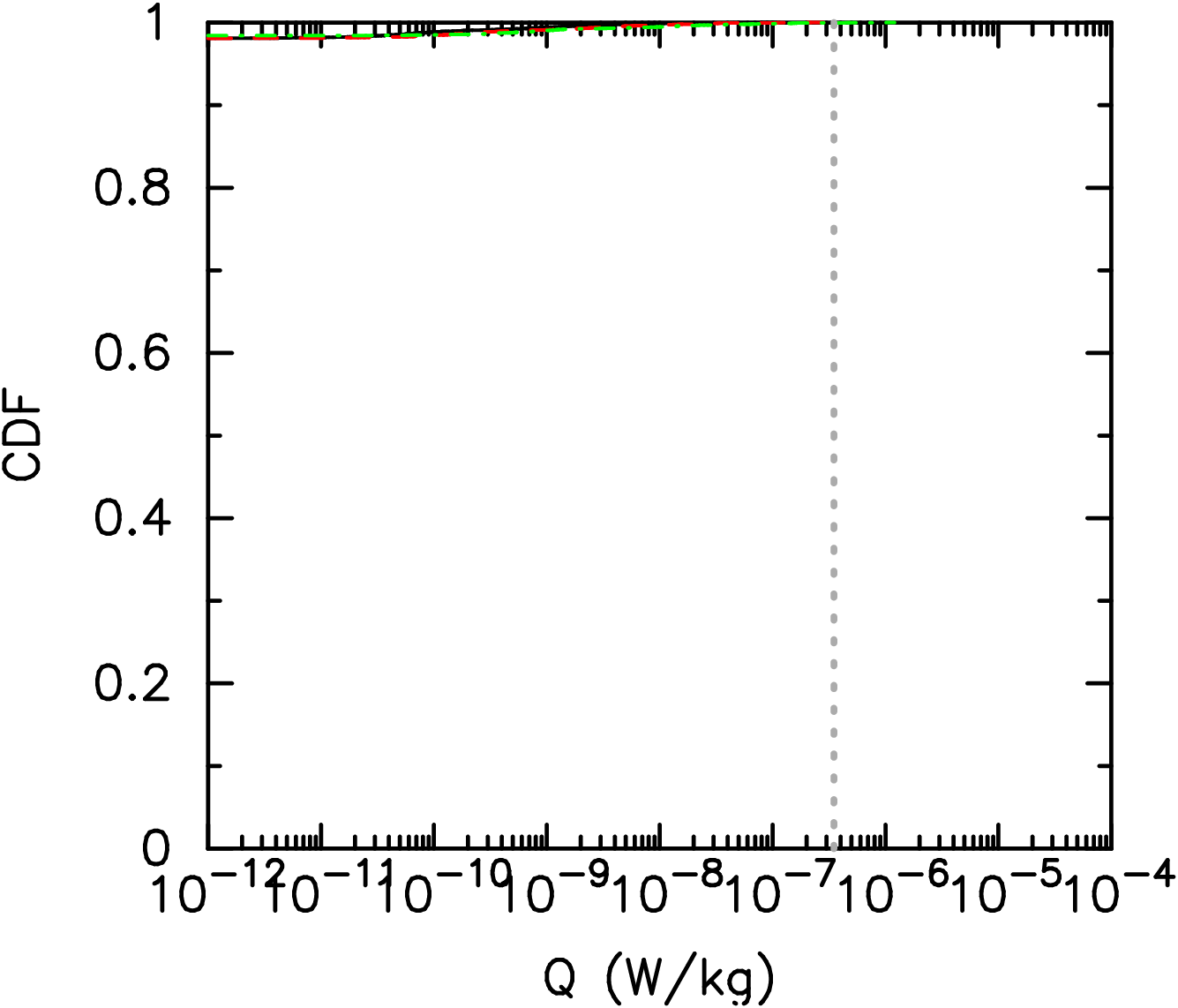}}}
\hspace*{0.3cm} 
\subfigure[Supernovae and local wind bubbles]{\label{high_density-disappear-Qheat-b}\rotatebox{270}{\includegraphics[scale=0.25]{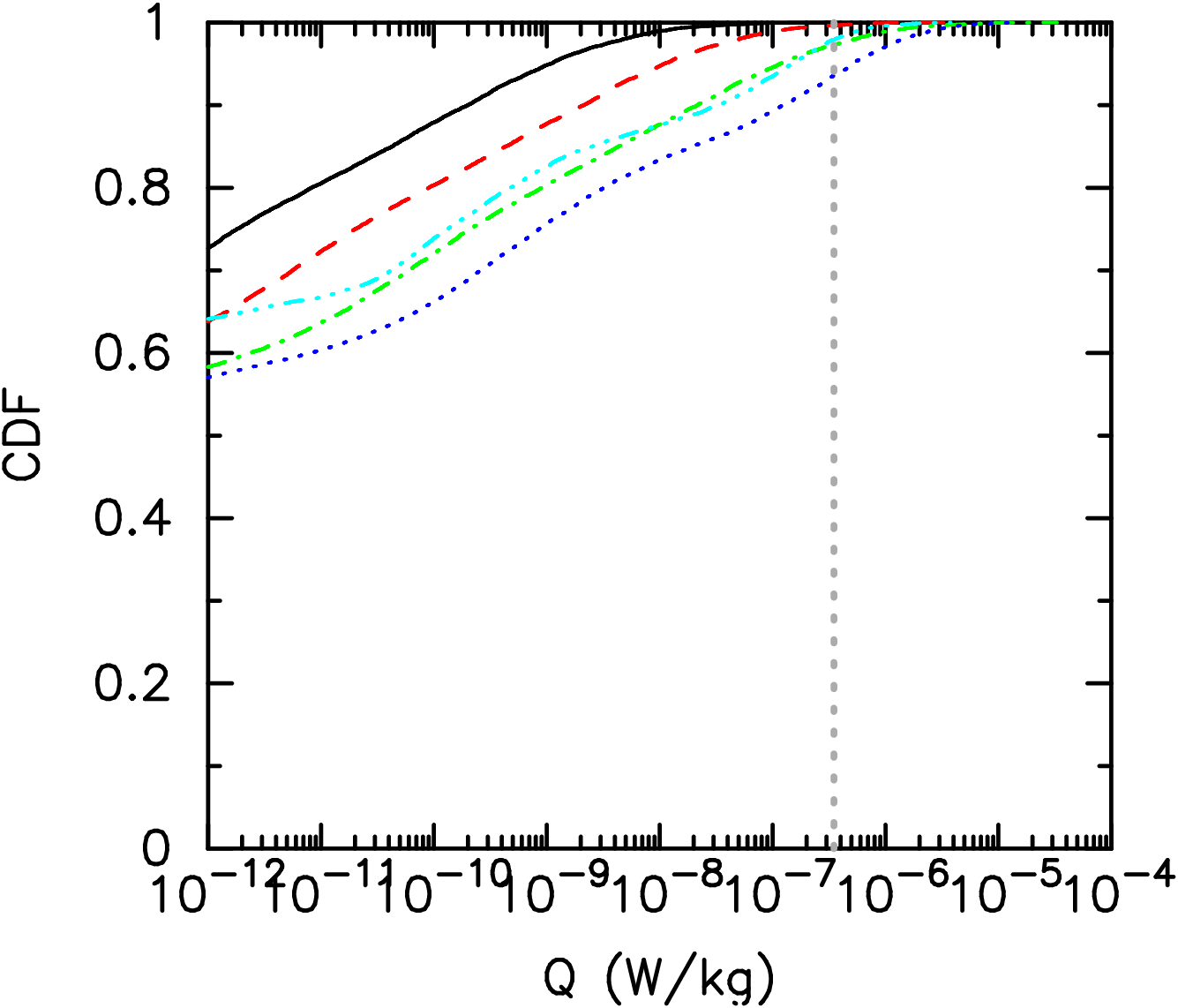}}}
\hspace*{0.3cm} 
\subfigure[Supernovae and despersed wind bubbles]{\label{high_density-disappear--Qheat-c}\rotatebox{270}{\includegraphics[scale=0.25]{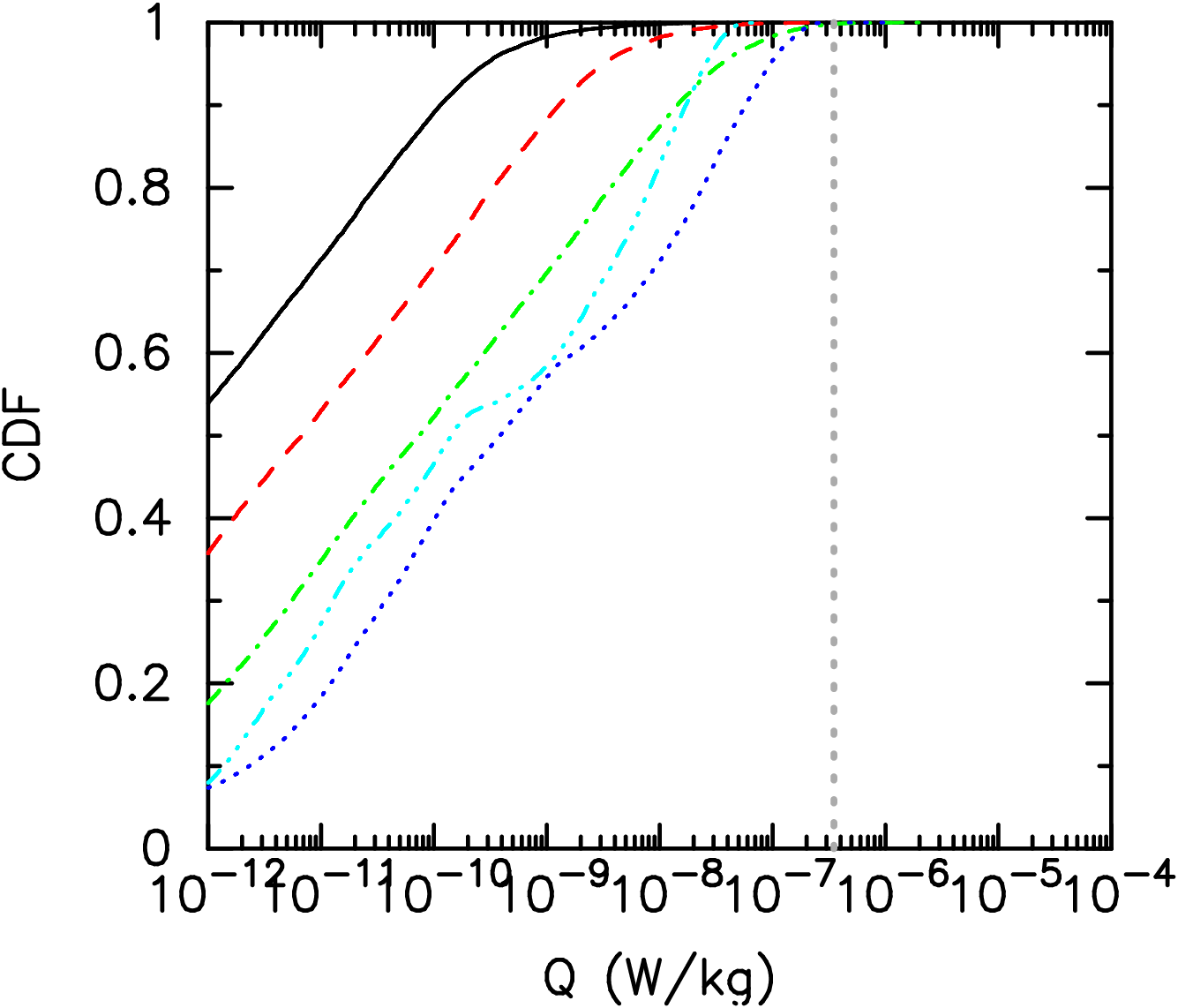}}} 
\caption[bf]{Cumulative Distribution  Functions (CDFs) of the radioactive heating of planetesimals from both $^{26}$Al and $^{60}$Fe in initially high-density star-forming regions ($\tilde{\rho} \sim 1000$M$_\odot$\,pc$^{-3}$). In these simulations the discs deplete in line with the observations of \citet{Haisch01,Richert18}. Panel (a) shows the heating from supernovae ejecta alone, panel (b) shows the heating from supernovae and local wind bubbles ($r_{\rm bub} = 0.1$\,pc) and panel (c) shows the heating from supernovae and dispersed wind bubbles ($r_{\rm bub} = 2r_{1/2}$\,pc). The heating is calculated at 10\,Myr (solid black lines), 7.5\,Myr (the dashed red lines), 5\,Myr (dot-dashed green lines), 2.5\,Myr (dotted blue lines) and 1\,Myr (dot-dot-dashed cyan lines). The vertical dotted lines indicate the initial heating for the Solar system, $Q_{\rm SS} = 3.4 \times 10^{-7}$\,W\,kg$^{-1}$, calculated from Eqn.~\ref{ss_heating}.}
\label{high_density-disappear-Qheat}
  \end{center}
\end{figure*}

In Figs.~\ref{high_density-disappear-Al}--\ref{high_density-disappear-Qheat} we repeat the calculations for the high density star-forming regions ($\tilde{\rho} \sim 1000$M$_\odot$\,pc$^{-3}$) but `deplete' our discs using the exponential function in Eqn.~\ref{disc_deplete}. At a given time, if a random number exceeds the value in Eqn.~\ref{disc_deplete} we do not add any further $^{26}$Al or $^{60}$Fe to the disc (either from winds, or supernovae).

The effect of this is clearly demonstrated in Fig.~\ref{high_density-disappear-Al}. In panel (a) we show the $Z_{\rm Al}$ ratios for discs for enrichment from supernovae only. Only three out of ten simulations have any discs remaining at this age, and very few of these discs have any notable enrichment.

In the simulations where wind enrichment occurs (panels (b) and (c)), then after 10\,Myr five simulations have enrichment commensurate with the Solar system if the wind enrichment occurs in small, localised bubbles. Fewer simulations (three) have Solar system levels of enrichment if the winds expand into large bubbles.

However, despite the early depletion of discs \citep[the discs have a `half-life' of 2\,Myr,][]{Richert18}, enrichment does occur early on, as shown in panels (c)--(f) in Fig.~\ref{high_density-disappear-Al}. Here, the enrichment is dominated by winds, and the majority occurs in the first few Myr when the star-forming regions are more dense (the dashed, and dot-dashed lines, representing enrichment from dispersed, and more local winds, respectively, creep rightwards in these plots over time, but many stars obtain most of their enrichment in the first 2\,Myr).

We find a similar pattern in the $Z_{\rm Fe}$ ratios, as shown in Fig.~\ref{high_density-disappear-Fe}. Recall that the enrichment in $^{60}$Fe comes only from supernovae, and the majority of the discs have depleted by the time of the supernovae. Panel (a) in this figure shows the  $Z_{\rm Fe}$ ratios at 5\,Myr, where there are more discs, but fewer massive stars have exploded as supernovae, whereas panel (b) shows the $Z_{\rm Fe}$ ratios at 10\,Myr, where there are fewer discs, but more supernovae have occurred. Clearly, the $Z_{\rm Fe}$ ratios are low, or non-existent, for most stars at both times.

Finally, we plot the amount of heating due to $^{26}$Al and $^{60}$Fe when the discs are depleted and the results are shown in Fig.~\ref{high_density-disappear-Qheat}. We see that if enrichment occurs from supernovae alone (panel a), or from dispersed winds (panel c) then planetesimals will not attain Solar system-levels of heating. However, panel (b) shows that if the winds are entrained in smaller, localised bubbles, then Solar system levels of heating can still occur. This is because the  $^{26}$Al and $^{60}$Fe are added to the discs earlier; in this plot the distribution with the highest $Q$ values is at 2.5\,Myr (the dark blue dotted line), whereas in similar simulations when we do not deplete the discs (Fig.~\ref{high_density-Qheat}) the heating after 5 -- 10\,Myr is greatest.

\subsection{Relative abundances of $^{26}$Al to $^{60}$Fe}

A further issue with the disc enrichment from supernovae scenario is that the relative abundance of $^{26}$Al to $^{60}$Fe is usually too low in simulations, at around unity. The ratio of $^{26}$Al/$^{60}$Fe is affected by the uncertainty in the measurements of $^{60}$Fe, but for example, \citet{Gounelle12} quote $^{26}$Al/$^{60}$Fe = 8.2, and \citet{Lugaro18} quote $^{26}$Al/$^{60}$Fe $\approx$ 3--25. This ratio is sensitively dependent on the measured value of $^{60}$Fe.

\begin{figure*}
  \begin{center}
\setlength{\subfigcapskip}{10pt}
\hspace*{-1.5cm}\subfigure[Supernovae only]{\label{high_density_al_fe_ratios-a}\rotatebox{270}{\includegraphics[scale=0.25]{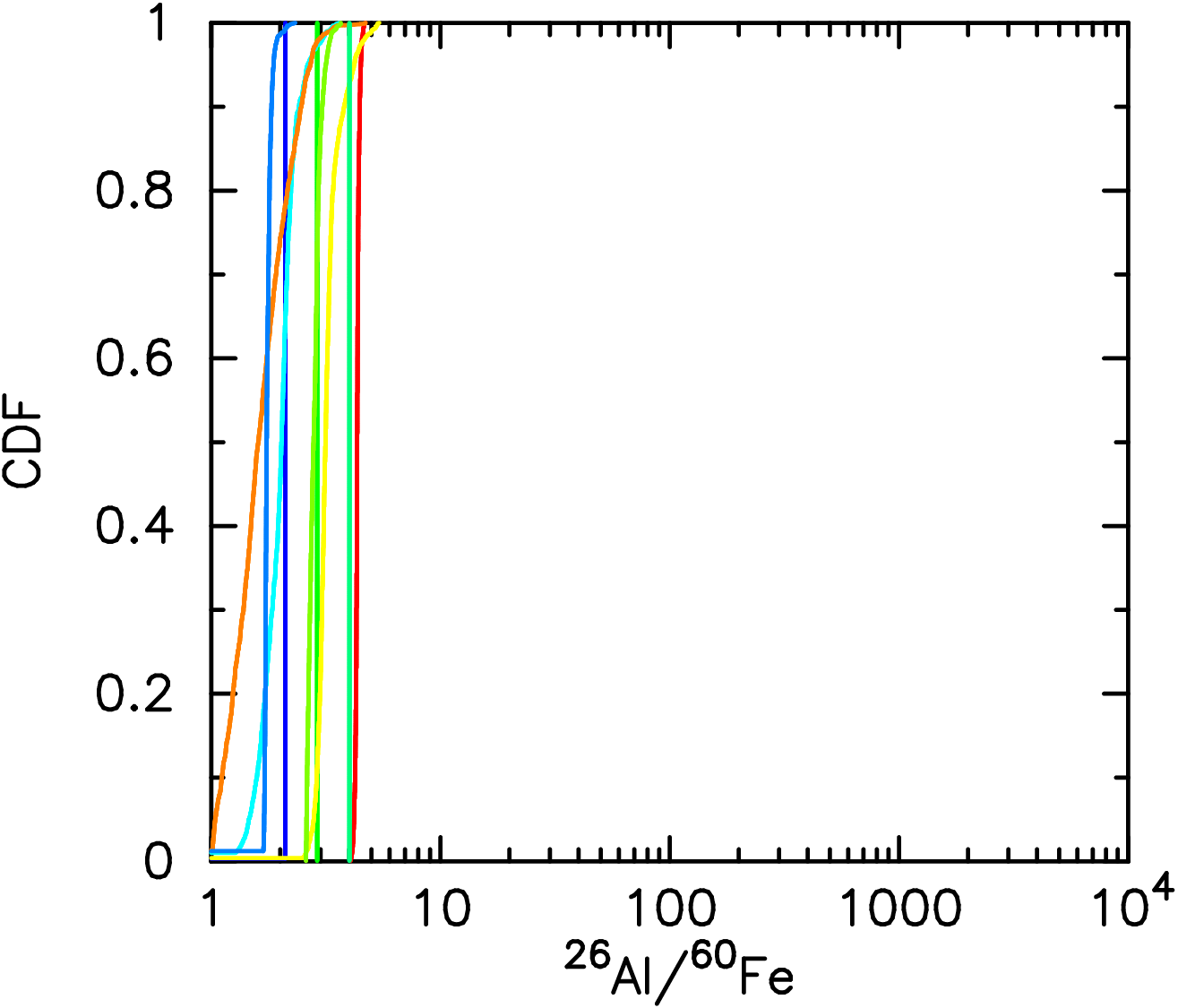}}}
\hspace*{0.3cm} 
\subfigure[Supernovae and local wind bubbles]{\label{high_density_al_fe_ratios-b}\rotatebox{270}{\includegraphics[scale=0.25]{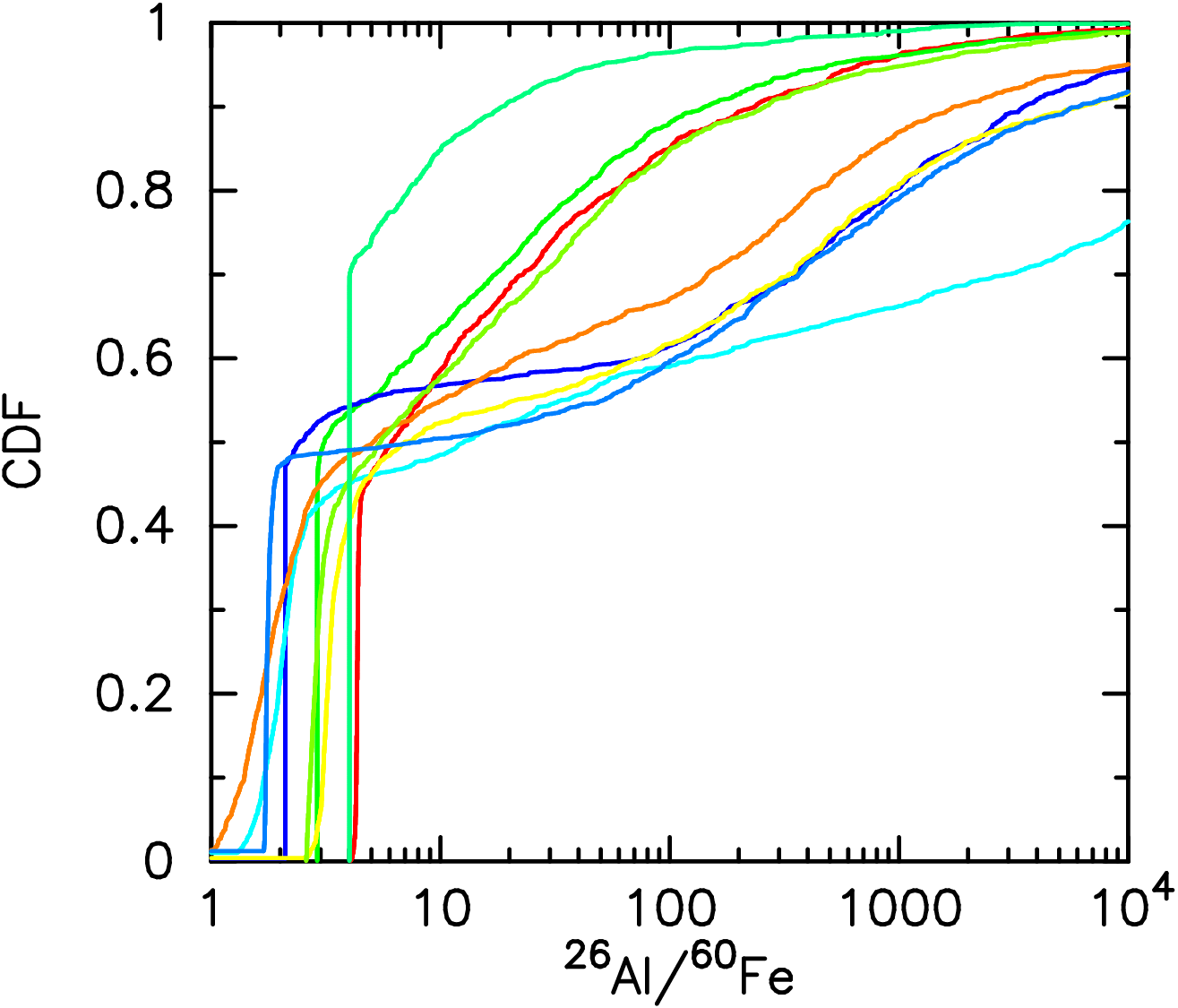}}}
\hspace*{0.3cm} 
\subfigure[Supernovae and despersed wind bubbles]{\label{high_density_al_fe_ratios-c}\rotatebox{270}{\includegraphics[scale=0.25]{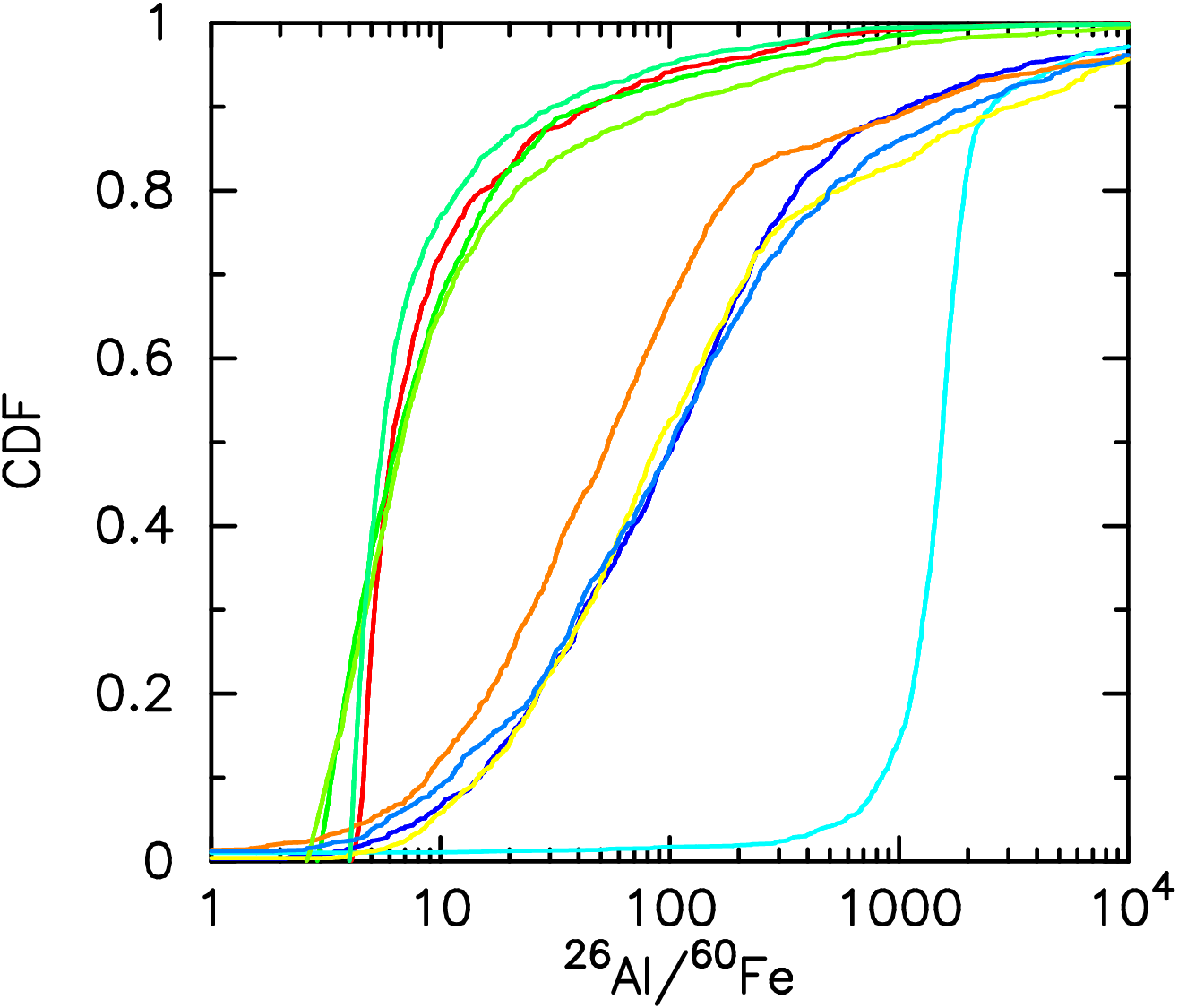}}} 
\caption[bf]{Cumulative Distribution  Functions (CDFs) of the ratios of the two SLRs $^{26}$Al to $^{60}$Fe in initially high-density star-forming regions ($\tilde{\rho} \sim 1000$M$_\odot$\,pc$^{-3}$) where the discs do not deplete. Panel (a) shows the $^{26}$Al/$^{60}$Fe ratio from supernovae ejecta alone, panel (b) shows the $^{26}$Al/$^{60}$Fe ratio from supernovae and local wind bubbles ($r_{\rm bub} = 0.1$\,pc) and panel (c) shows the $^{26}$Al/$^{60}$Fe ratio from supernovae and dispersed wind bubbles ($r_{\rm bub} = 2r_{1/2}$\,pc). The ratios are calculated at 10\,Myr and the different lines are for individual simulations.}
\label{high_density_al_fe_ratios}
  \end{center}
\end{figure*}

We plot the $^{26}$Al/$^{60}$Fe ratio from our dense simulations ($\tilde{\rho} = 1000$M$_\odot$\,pc$^{-3}$) in Fig.~\ref{high_density_al_fe_ratios}. In panel (a) we show the ratios for simulations where the $^{26}$Al comes from supernovae alone, in panel (b) we show the ratios where the $^{26}$Al also comes from winds that disperse through the star cluster, and in panel (c) we show the ratios where the $^{26}$Al also comes from winds that are entrained in local bubbles.

Clearly, when the $^{26}$Al comes from supernovae alone, the $^{26}$Al/$^{60}$Fe ratio is close to unity in many cases, but when the $^{26}$Al comes from winds this ratio can span a much larger range in values and is within the measured values for the Solar system. In strongly wind-enriched systems, the $^{26}$Al/$^{60}$Fe can even exceed the likely range in the Solar system by more than an order of magnitude.

\section{Discussion}

Our results show that protoplanetary discs can be enriched in the SLRs $^{26}$Al and $^{60}$Fe to Solar system levels and beyond by capturing the material ejected in the winds of massive stars.

As one might expect, the amount of enrichment is dependent on the initial stellar density, with more enrichment, to higher levels, occurring in dense star-forming regions where the average stellar density is $\tilde{\rho} \sim 1000$\,M$_\odot$\,pc$^{-3}$ compared to moderately dense ($\tilde{\rho} \sim 100$\,M$_\odot$\,pc$^{-3}$) star-forming regions.

While Solar system-like levels of $^{26}$Al enrichment can be reproduced in an appreciable number of cases, we find a high degree of $^{26}$Al variability in young planetary systems. This suggests that internal radiogenic heating can be anticipated to vary substantially between planetary systems, with potentially significant implications for the distribution of atmosphere-forming volatile elements, such as water \citep{Lichtenberg19} and carbon compounds \citep{LichtenbergKrijt21}. Enrichment on a level of a few per cent to a few tens of per cent is consistent with the amount of differentiated debris in polluted white dwarfs \citep{Bonsor20,Curry22}, which suggests that internal heating by short-lived radioactive isotopes chemically differentiates and thus alters a significant fraction of planetary bodies during planetary formation \citep{JuraYoung14,Lichtenberg22}.

However, at densities $\tilde{\rho} \le 10$\,M$_\odot$\,pc$^{-3}$ the amount of enrichment does not result in planetessimal heating to Solar system levels. It is worth highlighting that many star-forming regions, from Taurus \citep{Guedel07} to Cyg OB2 \citep{Wright14}, have densities of this order, and if they were not more dense in the past, it is difficult to see how these regions could enrich protoplanetary discs to Solar system levels.

When we implement a disc depletion algorithm, many of the discs are gone before the stars explode as supernovae, and so the enrichment is dominated by the massive stars' winds. If -- as appears to be the case in the Solar system -- enrichment happens early on, then our results suggest this can only happen in the most dense star-forming regions.

We have also demonstrated that the relative abundance of $^{26}$Al/$^{60}$Fe is consistent with a contribution of $^{26}$Al from massive star winds, and negates the need for multiple phases, or sequential star formation, to be required for solar system formation \citep[as postulated by][]{Gounelle12,Young14}. 

The most significant caveat in our work is the omission of disc evolution, both internal viscous evolution and mass-loss due to external photoevaporation due to Far/Extreme Ultraviolet (FUV/EUV) radiation from massive stars. Discs undergo significant spreading due to viscous evolution \citep{ConchaRamirez19}, which would increase the cross section of the discs and potentially lead to higher levels of enrichment. It is likely that the implementation of viscous evolution in the models of \citet{Lichtenberg16b} is the reason why some of their simulations display higher $Z$ ratios and $Q$ heating values.

However, our discs could also be significantly altered by photoevaporation and the subsequent inward evolution of the discs. Whilst external photoevaporation is thought to remove very little of the dust content of discs \citep{Haworth18b}, it may drive the dust radii inwards \citep{Sellek20}, which -- if dominant over viscous spreading -- would reduce the cross section of the disc for capturing wind and supernovae ejecta.

Furthermore, whilst little dust is lost from discs due to photoevaporation, significant amounts of the gas -- sometimes all -- is removed on very short ($<$1\,Myr) timescales \citep{Scally01,Winter18b,Nicholson19a,Parker21a,ConchaRamirez19}. This means that gas giant planet formation can be hindered or suppressed altogether. In a follow-up paper, we will determine how many Solar system analogues can be enriched yet retain gas in their discs in these star-forming regions, and compare the destruction of discs due to FUV radiation with those that may be enriched but truncated by dynamical encounters, e.g.\,\,\citet{Zwart19}.

A further caveat to our results is that we have only modelled subvirial star-forming regions undergoing collapse to form a star cluster. These initial conditions probably optimise the amount of enrichment, as opposed to e.g.\,\,a supervirial region undergoing expansion \citep[though we note that these regions also develop dense (sub)clusters which are very similar in both appearance and physical processes to the subvirial clusters and do not preclude enrichment,][]{Parker14a,Parker14b,Rate20}.

We have also not varied the degree of spatial substructure, the number of stars \citep[enrichment can occur in low-$N$ star-forming regions if there are massive stars present][]{Nicholson17} or included primordial binaries in our calculations.

\section{Conclusions}

We present $N$-body simulations of star-forming regions and employ a post-processing analysis to determine the amount of enrichment in the short-lived radioisotopes (SLRs) $^{26}$Al and $^{60}$Fe that occurs in protoplanetary discs. Previous work on the dynamical evolution of young star-forming regions has usually only considered the contribution of $^{26}$Al from the supernovae explosions of massive stars, but in this work we have implemented enrichment from the winds of these massive stars. We then use the amount of  $^{26}$Al and $^{60}$Fe to calculate the radioactive internal heating of planetesimals due to these SLRs. Our simulations results suggest the following:

(i) The contribution of $^{26}$Al from stellar winds is significant, and for many stars is the difference between enrichment at lower levels than the Solar system, and levels equal to or higher than the Solar system.

(ii) The amount of enrichment is slightly sensitive to the dispersal rate of the wind material; if the material disperses throughout the whole star-forming region, then more stars are enriched, but fewer at Solar system levels or higher. If the winds are entrained in local bubbles, then fewer stars overall are enriched, but more of those that are have Solar system or higher levels of enrichment.

(iii) The stellar density is the most important variable in determining the number of stars that are enriched. In the most dense regions ($\tilde{\rho} \sim 1000$\,M$_\odot$\,pc$^{-3}$), up to 50\,per cent of stars can attain Solar system levels of enrichment, whereas in regions with much lower densities ($\tilde{\rho} \sim 10$\,M$_\odot$\,pc$^{-3}$, similar to OB associations observed today), almost no enrichment occurs.

(iv) Our results are very sensitive to the lifetimes of the protoplanetary discs. In a set of simulations where we forbid enrichment if the disc has either been destroyed or already formed planets, enrichment can only occur in the most dense star-forming regions.

(v) When winds make a significant contribution to the enrichment, this occurs much earlier than in previous simulations that solely relied on supernovae (Solar system-levels of enrichment from winds occur from 2.5\,Myr, rather than from beyond the times of the first supernovae at around 4\,Myr).

(vi) Our simulations reproduce the amount of internal heating calculated for planetesimals in the early Solar system, and also produce a wide spread in the relative abundance of $^{26}$Al/$^{60}$Fe.

However, for simplicity, our simulations do not include photoevaporation of the gas component of the discs due to the FUV/EUV radiation fields of massive stars (the same massive stars that are enriching the discs), nor are the radii of the protoplanetary discs evolving due to photoevaporation or viscous speading. In a follow-up paper we will implement both mechanisms to establish whether our Solar system could form, be enriched early, and retain a gas component in its disc that would enable Jupiter and Saturn to form. 

\section*{Acknowledgements}

We thank the anonymous referee for their comments and suggestions, which have improved the manuscript. RJP acknowledges support from the Royal Society in the form of a Dorothy Hodgkin Fellowship. TL acknowledges support from the Simons Foundation (SCOL Award No.~611576), and the Branco Weiss foundation. This work benefitted from information exchange within the program `Alien Earths' (NASA grant No.\,\,80NSSC21K0593) for NASA's Nexus for Exoplanet System Science (NExSS) research coordination network, and the AEThER project, funded by the Alfred P. Sloan Foundation under grant No. G202114194. We are grateful to Marco Limongi for clarifying an aspect of the models from \citet{Limongi18} during the refereeing process.

\section*{Data availability statement}

The data used to produce the plots in this paper will be shared on reasonable request to the corresponding author. 

\bibliographystyle{mnras}  
\bibliography{general_ref}

\appendix

\section{Comparison with different yield models}
\label{appendix}

The stellar evolution models we use to calculate the yields are from \citet{Limongi06}, which assume that stars $>$25\,M$_\odot$ produce $^{26}$Al and $^{60}$Fe (in addition to other heavy elements) in supernovae. However, the most recent models \citep[e.g.][]{Limongi18} assume that $>$25\,M$_\odot$ stars collapse directly to a black hole and these new models produce different yields, especially for the supernovae.

\begin{figure}
\begin{center}
\rotatebox{270}{\includegraphics[scale=0.4]{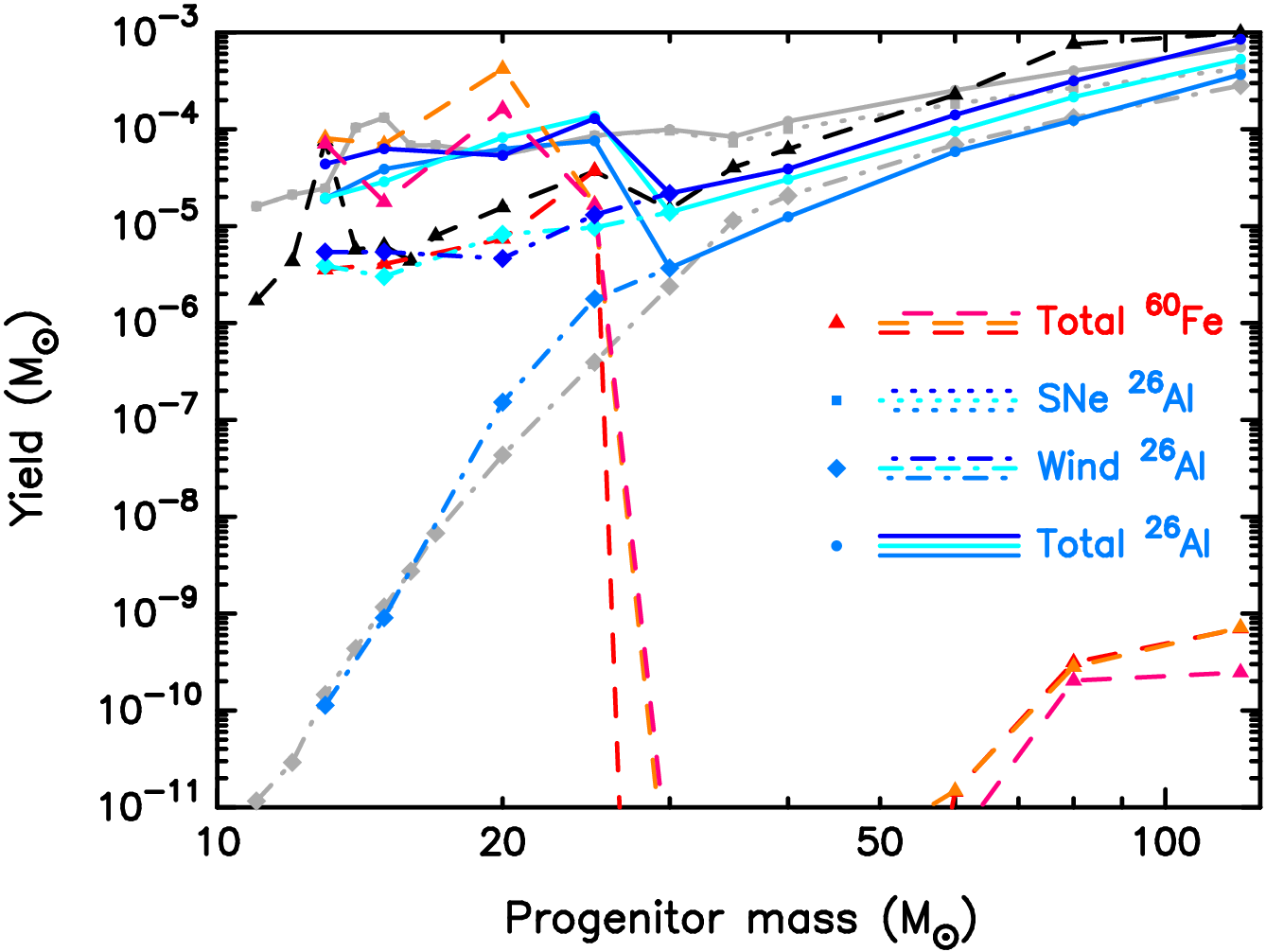}}
\caption[bf]{Theoretical yields of $^{26}$Al and $^{60}$Fe from two different sets of stellar evolution models. Solid lines and circles indicate the total yield (wind plus supernovae) of $^{26}$Al, dot-dashed lines and diamonds indicate the $^{26}$Al contribution from massive star winds, and dotted lines and squares indicate the $^{26}$Al yield from supernovae only.  The dashed lines and triangles indicate the yield of $^{60}$Fe from supernovae and winds, although the contribution from the latter is negligible for stars less massive than 50\,M$_\odot$. The data from the models in \citet{Limongi06} are shown by the grey lines ($^{26}$Al) and the black lines ($^{60}$Fe). The data for $^{26}$Al from the models in \citet{Limongi18} are shown by the blue lines, with the different rotation velocities of the massive stars (0, 150 and 300\,km\,s$^{-1}$) indicated by the powder blue, cyan and royal blue lines, respectively. The data for $^{60}$Fe from the models in \citet{Limongi18} are shown by the red lines, with the different rotation velocities of the massive stars (0, 150 and 300\,km\,s$^{-1}$) indicated by the crimson, raspberry and orange lines, respectively. The drastic drop-off in $^{60}$Fe yields is by virtue of stars with masses $>$25\,M$_\odot$ not exploding as supernovae in these updated models. }

\label{yield_models}
  \end{center}
\end{figure}

\begin{figure*}
  \begin{center}
\setlength{\subfigcapskip}{10pt}
\hspace*{-1.5cm}\subfigure[$^{26}$Al, supernovae only -- 10\,Myr]{\label{high_density-Al-LC18_r1-a}\rotatebox{270}{\includegraphics[scale=0.31]{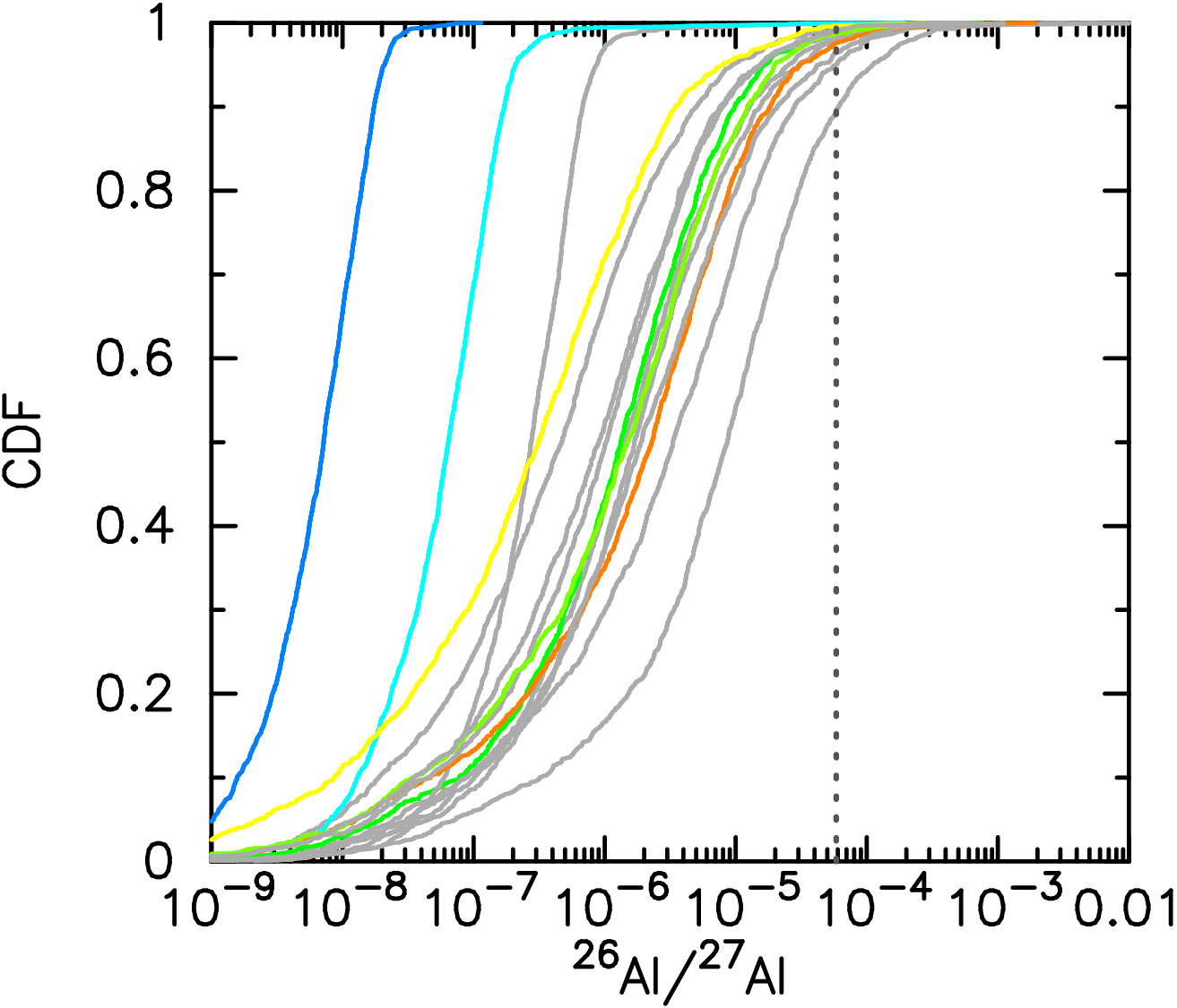}}}
\hspace*{0.3cm} 
\subfigure[$^{26}$Al, supernovae and local wind bubbles -- 10\,Myr]{\label{high_density-Al-LC18_r1-b}\rotatebox{270}{\includegraphics[scale=0.31]{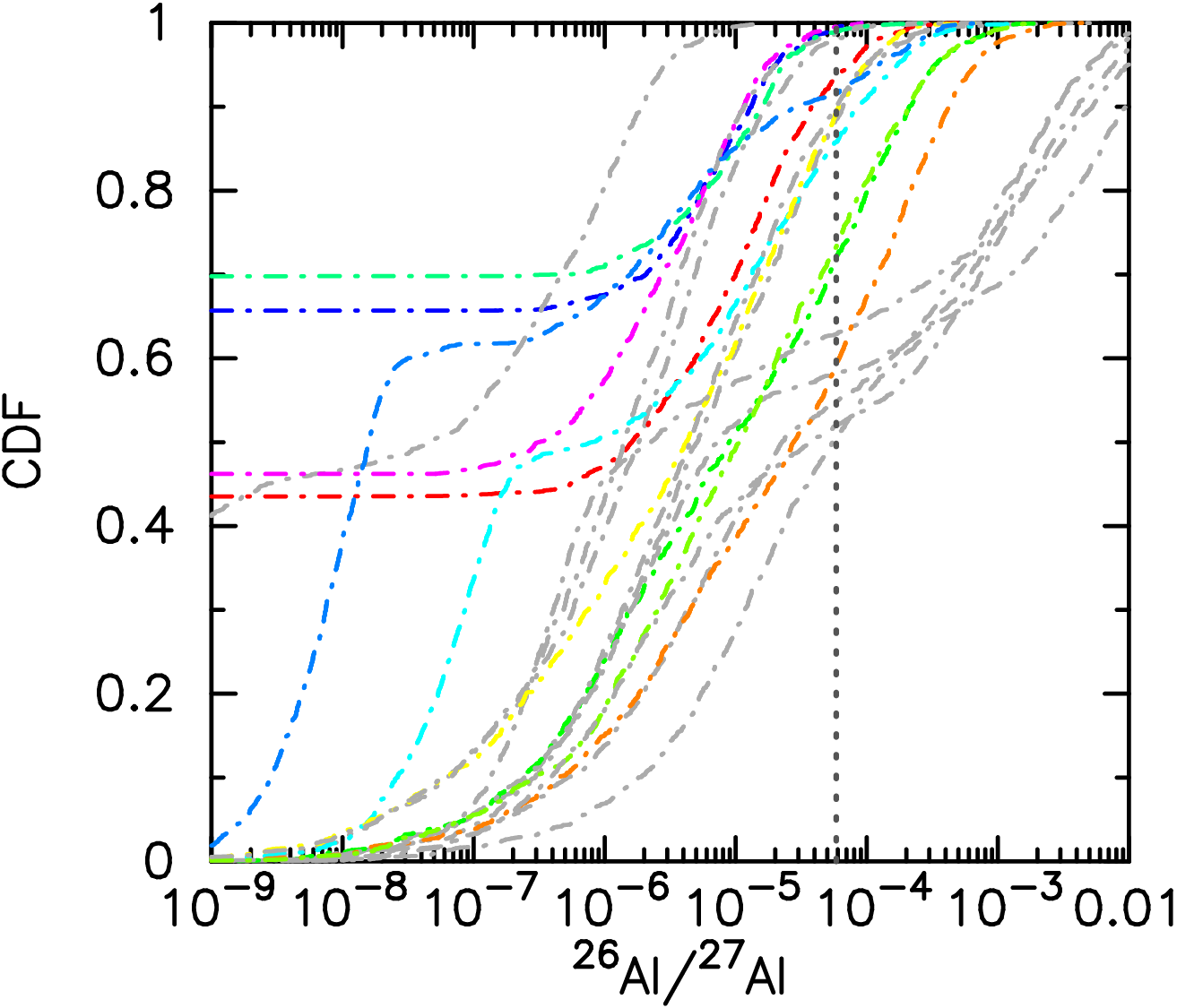}}}
\hspace*{-1.5cm} 
\subfigure[$^{26}$Al, supernovae and global wind bubbles -- 10\,Myr]{\label{high_density-Al-LC18_r1-c}\rotatebox{270}{\includegraphics[scale=0.31]{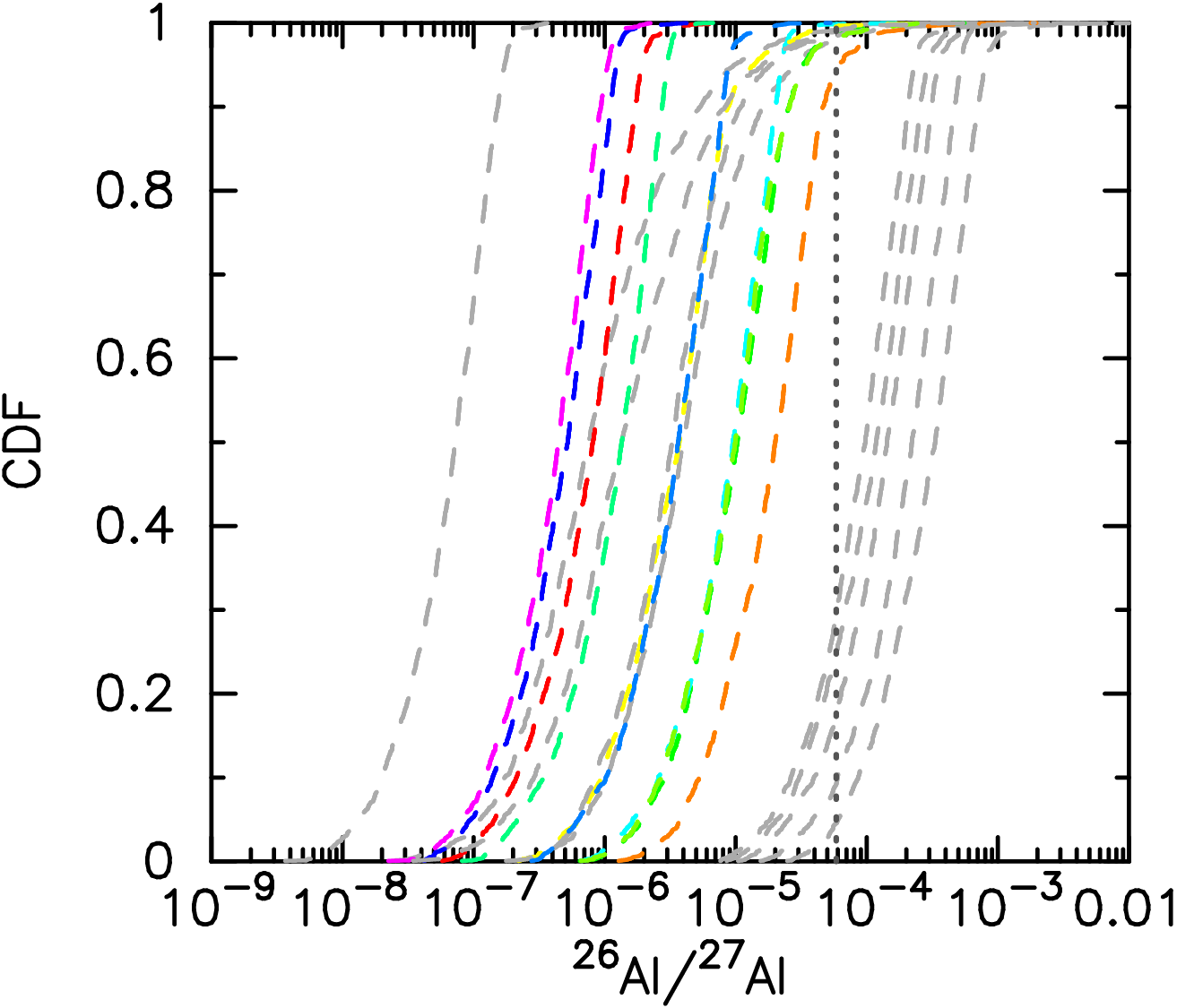}}}
\hspace*{0.3cm}\subfigure[$^{60}$Fe, supernovae only -- 10\,Myr]{\label{high_density-Al-LC18_r1-d}\rotatebox{270}{\includegraphics[scale=0.31]{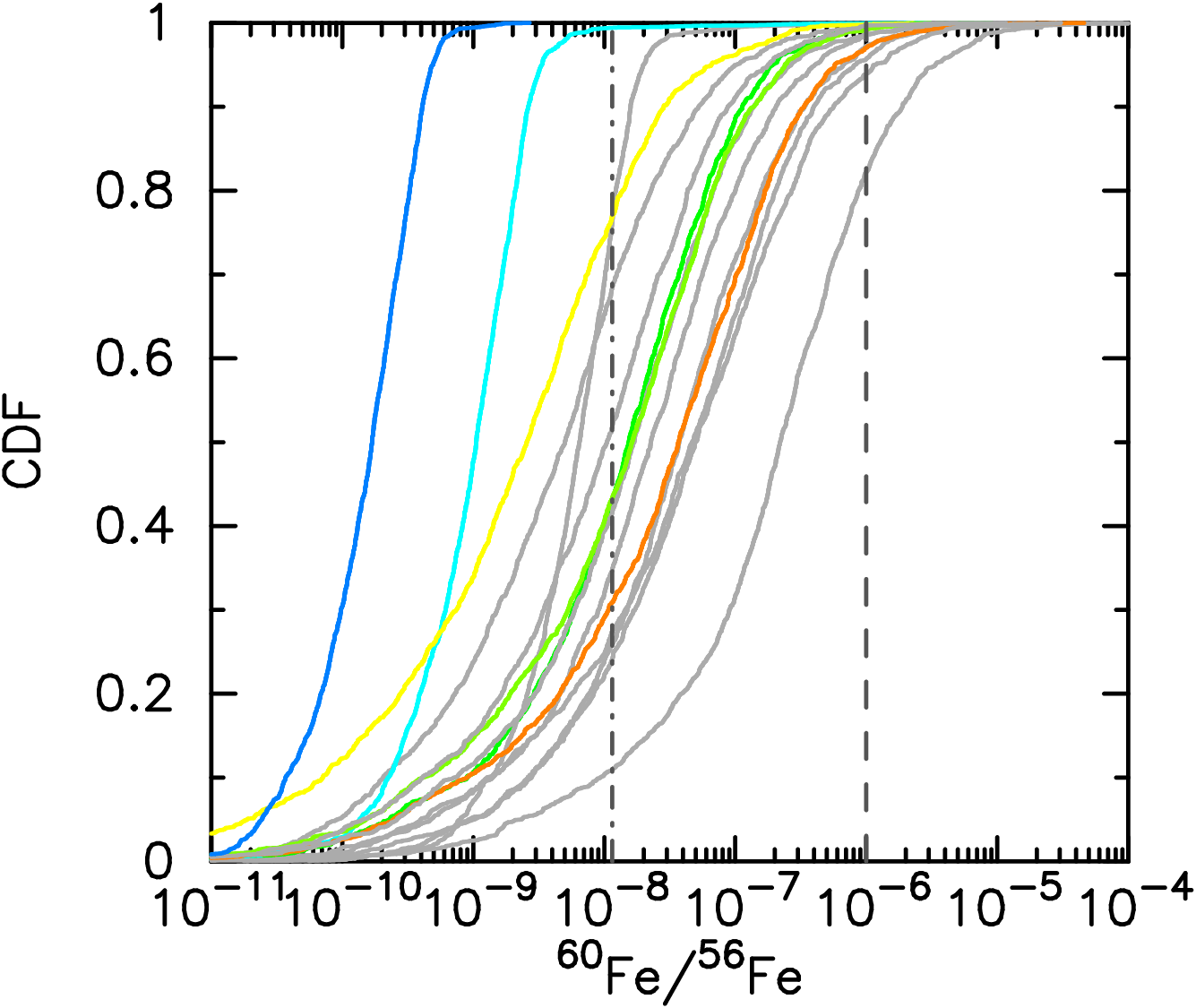}}}
\caption[bf]{Comparison between the SLR ratios $Z_{\rm Al, Fe}$ calculated after 10\,Myr in our simulations with the \citet{Limongi06} models, versus the \citet{Limongi18} models. We show the Cumulative Distribution  Functions (CDFs) of the SLR ratios $Z_{\rm Al, Fe}$ in a star-forming region with an initially high stellar density ($\tilde{\rho} \sim 1000$M$_\odot$\,pc$^{-3}$). The initial disc radii are all $r_{\rm disc} = 100$\,au, and the discs do not evolve with time. The coloured lines in all panels are the yields calculated using the non-rotating (0\,km\,s$^{-1}$) models from \citet{Limongi18}, whereas the grey lines are the yields calculated using the models in \citet{Limongi06}, which we show in Figs.~\ref{high_density-Al}~and~\ref{high_density-Fe}. Solid lines are ratios calculated from supernovae only. Dashed lines are global, or dispersed wind bubbles ($r_{\rm bub} = 2r_{1/2}$) plus any supernovae, dot-dashed lines are local wind bubbles ($r_{\rm bub} = 0.1$\,pc)  plus any supernovae. We show the $^{26}$Al/$^{27}$Al and  $^{60}$Fe/$^{56}$Fe ratios calculated at 10\,Myr. The vertical dotted lines in panels (a)--(c) indicate the measured Solar system value, $Z_{\rm Al,SS} = 5.85 \times 10^{-5}$ \citep{Thrane06}, whereas the vertical dot-dashed line  in panel (d) indicates the Solar system value for Fe, $Z_{\rm Fe,SS} = 1.15 \times 10^{-8}$ \citep{Tang12} and the vertical dashed line indicates the alternative measurement, $Z_{\rm Fe,SS} = 1 \times 10^{-6}$ \citep{Mishra16}.}
\label{high_density-Al-LC18_r1}
  \end{center}
\end{figure*}

In Fig.~\ref{yield_models} we show the yields from \citet{Limongi06} by the grey lines ($^{26}$Al, for supernovae only, winds only, and total yield) and the black dashed line ($^{60}$Fe).  The new models from \citet{Limongi18} are shown by the coloured lines (different shades of blue for $^{26}$Al and different shades of red for $^{60}$Fe). The different shades of colour indicate different rotation velocities for the massive stars; for $^{26}$Al the powder blue, cyan and royal blue lines show yields for rotation velocities of 0, 150 and 300\,km\,s$^{-1}$, respectively, and for $^{60}$Fe the crimson, raspberry and orange lines show yields for rotation velocities of 0, 150 and 300\,km\,s$^{-1}$, respectively.

\begin{figure*}
  \begin{center}
\setlength{\subfigcapskip}{10pt}
\hspace*{-1.5cm}\subfigure[$^{26}$Al, supernovae only -- 10\,Myr]{\label{high_density-Al-LC18_r3-a}\rotatebox{270}{\includegraphics[scale=0.31]{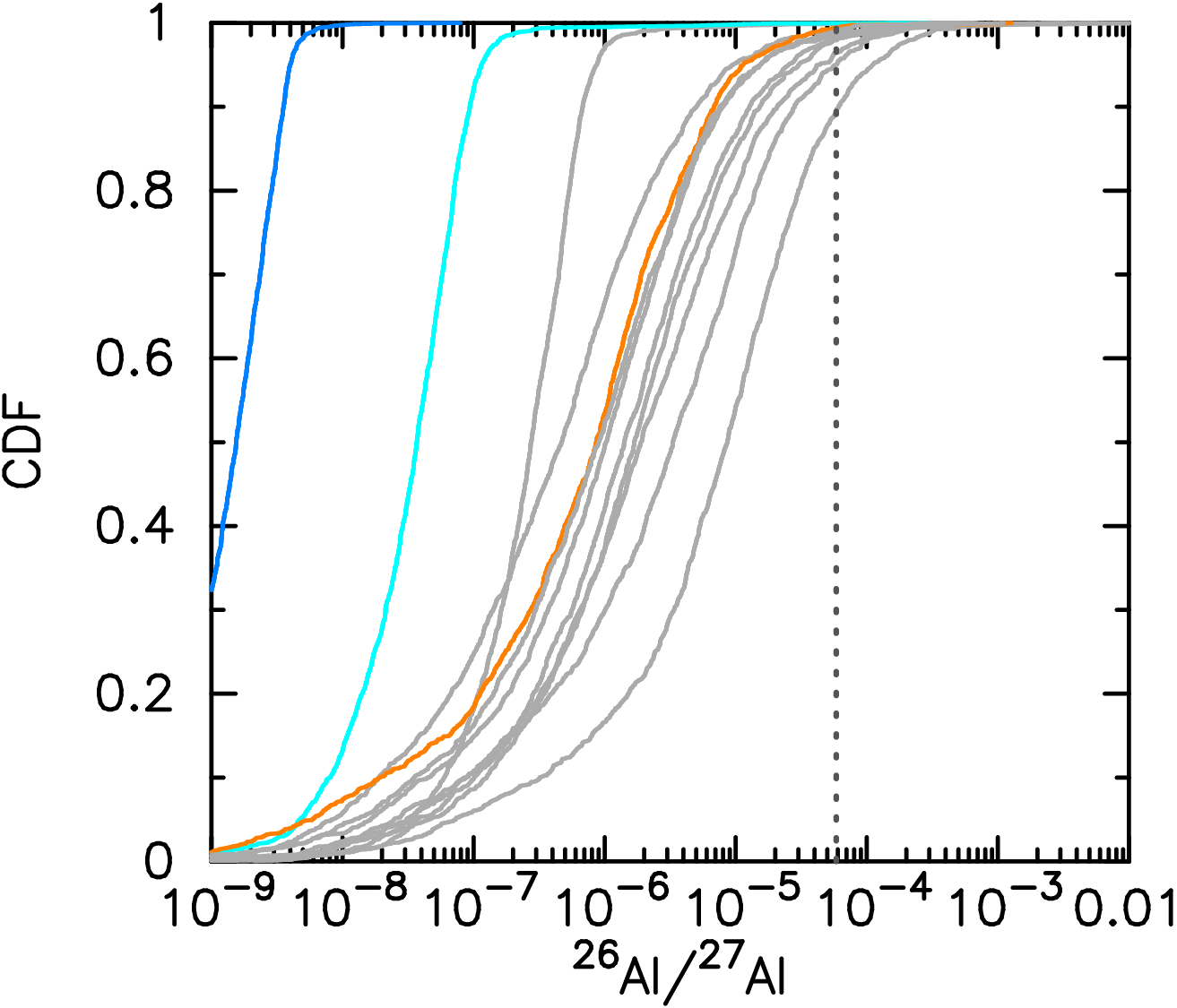}}}
\hspace*{0.3cm} 
\subfigure[$^{26}$Al, supernovae and local wind bubbles -- 10\,Myr]{\label{high_density-Al-LC18_r3-b}\rotatebox{270}{\includegraphics[scale=0.31]{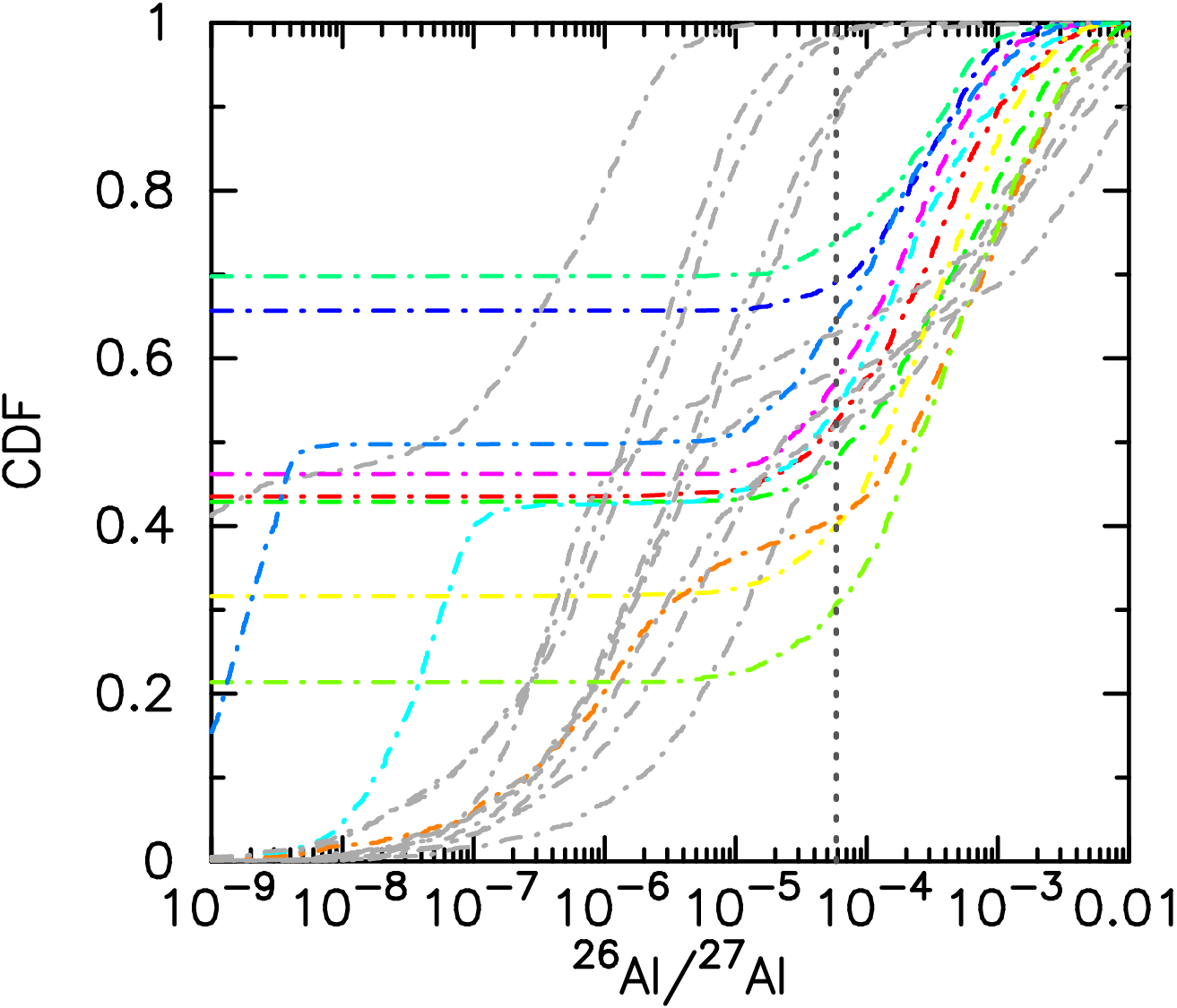}}}
\hspace*{-1.5cm} 
\subfigure[$^{26}$Al, supernovae and global wind bubbles -- 10\,Myr]{\label{high_density-Al-LC18_r3-c}\rotatebox{270}{\includegraphics[scale=0.31]{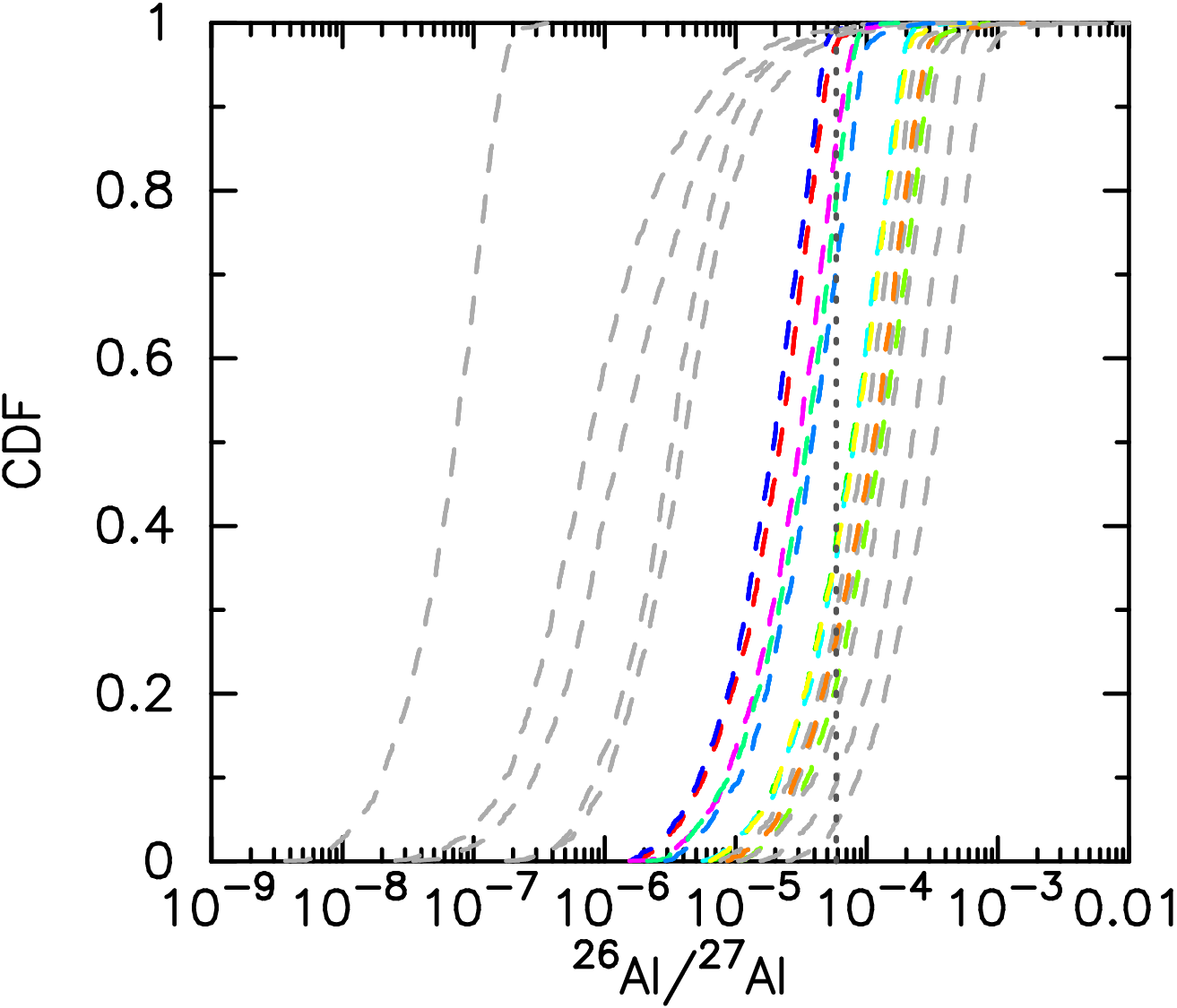}}}
\hspace*{0.3cm}\subfigure[$^{60}$Fe, supernovae only -- 10\,Myr]{\label{high_density-Al-LC18_r3-d}\rotatebox{270}{\includegraphics[scale=0.31]{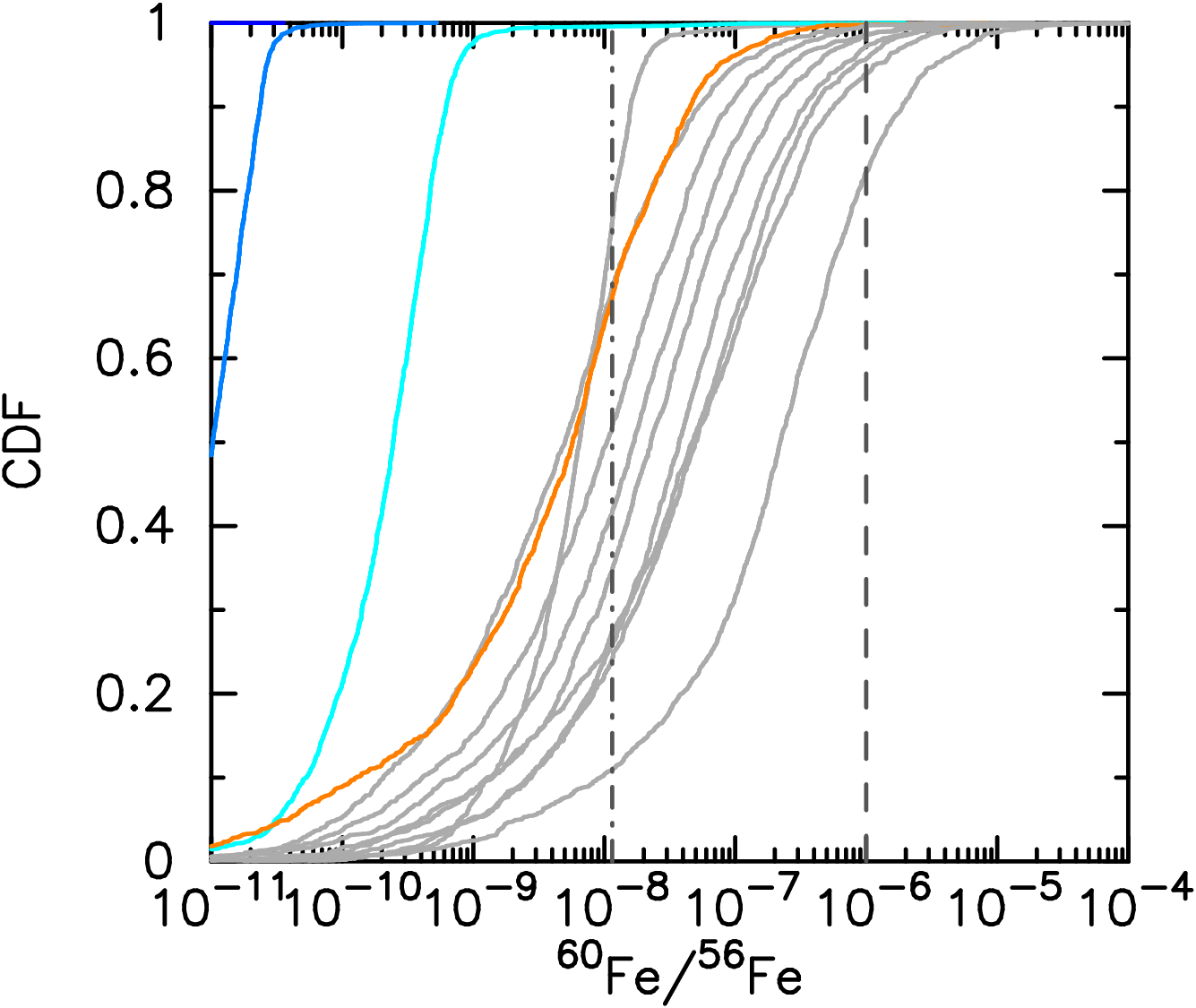}}}
\caption[bf]{Comparison between the  SLR ratios $Z_{\rm Al, Fe}$ calculated after 10\,Myr in our simulations with the \citet{Limongi06} models, versus the \citet{Limongi18} models. We show the Cumulative Distribution  Functions (CDFs) of the SLR ratios $Z_{\rm Al, Fe}$ in a star-forming region with an initially high stellar density ($\tilde{\rho} \sim 1000$M$_\odot$\,pc$^{-3}$). The initial disc radii are all $r_{\rm disc} = 100$\,au, and the discs do not evolve with time. The coloured lines in all panels are the yields calculated using the fast-rotating (300\,km\,s$^{-1}$) models from \citet{Limongi18}, whereas the grey lines are the yields calculated using the models in \citet{Limongi06}, which we show in Figs.~\ref{high_density-Al}~and~\ref{high_density-Fe}. Solid lines are ratios calculated from supernovae only. Dashed lines are global, or dispersed wind bubbles ($r_{\rm bub} = 2r_{1/2}$) plus any supernovae, dot-dashed lines are local wind bubbles ($r_{\rm bub} = 0.1$\,pc)  plus any supernovae. We show the $^{26}$Al/$^{27}$Al and  $^{60}$Fe/$^{56}$Fe ratios calculated at 10\,Myr. The vertical dotted lines in panels (a)--(c) indicate the measured Solar system value, $Z_{\rm Al,SS} = 5.85 \times 10^{-5}$ \citep{Thrane06}, whereas the vertical dot-dashed line in panel (d) indicates the Solar system value for Fe, $Z_{\rm Fe,SS} = 1.15 \times 10^{-8}$ \citep{Tang12} and the vertical dashed line indicates the alternative measurement, $Z_{\rm Fe,SS} = 1 \times 10^{-6}$ \citep{Mishra16}.}
\label{high_density-Al-LC18_r3}
  \end{center}
\end{figure*}

  Two major differences between the two sets of models are apparent in Fig.~\ref{yield_models}. The \citet{Limongi18} models produce almost no $^{60}$Fe in stars $>$25\,M$_\odot$, due to the direct collapse to a black hole. For stars $<$25\,M$_\odot$, the $^{60}$Fe production is completely dominated by the supernovae, and only at masses $>$50\,M$_\odot$ is there a non-negligible contribution from stellar winds (though still six orders of magnitude smaller than the contribution from supernovae from lower mass stars). We also note here that our simulations do not contain any stars $>$50\,M$_\odot$, so the  $^{60}$Fe is only produced in supernovae.

  The second major difference is that the models that include stellar rotatation produce far more $^{26}$Al and $^{60}$Fe for stars $<$25\,M$_\odot$.

We repeat the calculations for our first set of simulations (a high density star-forming region where the discs all have radii $r_{\rm disc} = 100$\,au, shown in Figs.~\ref{high_density-Al}~and~\ref{high_density-Fe} in the main paper) and use the yields from the non-rotating (0\,km\,s$^{-1}$) and fast rotating (300\,km\,s$^{-1}$) models in \citet{Limongi18} to calculate the the SLR ratios $Z_{\rm Al, Fe}$ for each star.

 We show the results for the non-rotating models in Fig.~\ref{high_density-Al-LC18_r1} and the fast rotating models in Fig.~\ref{high_density-Al-LC18_r3}. The coloured lines show the results using these \citet{Limongi18} models, whereas the grey lines show the results using the older models of \citet{Limongi06}, which we presented in Figs.~\ref{high_density-Al}~and~\ref{high_density-Fe}.

  The main differences between the two sets of models are in the lack of supernovae from stars $>$25\,M$_\odot$, which means that far fewer star-forming regions have SLR ratios commensurate with Solar system values, when only considering enrichment from supernovae. The reason for this is that stars with masses $<$20\,M$_\odot$ explode after the end time of our models (10\,Myr) but those with masses $>$20\,M$_\odot$ do not explode during the simulations. Interestingly, in the fast-rotating models (Fig.~\ref{high_density-Al-LC18_r3}) there are fewer supernovae than in the non-rotating models, which results in less enrichment from supernovae in the rotating models.

  The models with rotation produce far more $^{26}$Al via the winds of stars than supernovae, and in some cases produce \emph{more} enrichment than the older models in \citet{Limongi06}, especially when we assume the ejected wind material remains in local bubbles, rather than being dispersed throughout the star-forming region.

In summary, whilst the yield calculations using the data in \citet{Limongi18} are different from those in \citet{Limongi06}, the uncertainties in stellar rotation rates (and the other missing physics and approximations in our simulations) suggest using the older models is not overestimating the SLR yields in our simulated star-forming regions. \citet{deMink13} show that the rotation rates of massive stars are at least $>$200\,km\,s$^{-1}$, so the wind enrichment models using yields from fast-rotating stars are likely to be the most appropriate for our simulations.

\label{lastpage}

\end{document}